\documentclass[twocolumn, tighten, trackchanges, resetfootnote,twocolappendix]{aastex7}

\usepackage{color, xcolor, graphicx, amsmath, amssymb, mathrsfs, placeins, xspace}
\usepackage{appendix}
\usepackage{orcidlink}
\usepackage{hyperref}

\shorttitle{Data Quality Control and Multiple-star Modeling in Wide Binaries}
\shortauthors{Chae \& Yoon}

\begin{document}

\title{Revisiting Data Quality Control and Multiple-star Modeling in Wide Binary Gravity Tests: Confirmation of MOND-type Gravitational Anomaly at Low Acceleration}


\author[orcid=0000-0002-6016-2736, gname=Kyu-Hyun, sname=Chae]{Kyu-Hyun Chae}
\email{chae@sejong.ac.kr}\thanks{corresponding author: chae@sejong.ac.kr \\ kyuhyunchae@gmail.com}
\affiliation{Department of Physics and Astronomy, Sejong University, 209 Neungdong-ro Gwangjin-gu, Seoul 05006, Republic of Korea}
\author[orcid=0000-0002-0096-4702, gname=Youngsub, sname=Yoon]{Youngsub Yoon}\thanks{youngsuby@gmail.com}
\affiliation{Department of Physics and Astronomy, Sejong University, 209 Neungdong-ro Gwangjin-gu, Seoul 05006, Republic of Korea}
\email{youngsuby@gmail.com}

\begin{abstract}
Wide binary stars provide natural laboratories for directly probing gravity in the low-acceleration regime, as dark matter inferred from any viable gravity has negligible effects on their internal dynamics. Various recent studies including Bayesian 3D analyses have shown that wide binaries with separations greater than several thousand astronomical units experience MOND-type gravity with a boost factor of $\gamma\approx 1.3-1.6$. However, results claiming preference for, or no deviation from, standard gravity have also been published during the same period, particularly highlighting the roles of data quality control and realistic modeling of multiple-star (i.e., triple and higher-order) systems that host hidden companion stars. Here we carefully reexamine the issues of data quality control and modeling multiple-star systems in statistical gravity tests based on sky-projected 2D velocities of wide binary stars. Through extensive tests including the acceleration-plane test, the $\tilde v$-distribution test, and the median-$\tilde v$-profile test (where $\tilde v$ is the sky-plane 2D relative velocity normalized by the Newtonian circular velocity between the two stars), we show that proper data quality control or reasonable variation in multiple-star modeling cannot remove the low-acceleration gravitational anomaly but confirms the MOND-type gravitational anomaly, particularly consistent with recent realistic MOND solutions of wide binary orbits. We find that studies claiming no evidence for the low-acceleration gravitational anomaly are consequences of bypassed calibration of the fraction of multiple-star systems using the Newtonian-regime data, bias-introduction in data quality control that is not taken into account in gravity tests, or insufficient statistics in the low-acceleration regime.
\end{abstract}

\keywords{Binary stars (154);  Wide binary stars (1801); Gravitation (661); Modified Newtonian dynamics (1069); Non-standard theories of gravity (1118)}

\section{Introduction}

Wide binary stars have recently emerged as a crucial testbed in the debate between Newton-Einstein standard gravity with dark matter and modified gravity demanding a major revision or extension of general relativity in the low-acceleration limit. Because internal dynamics of wide binaries cannot be affected by the negligible amount of hypothetical dark matter inferred within the orbits of size less than about $50$ kilo astronomical units (kau), any observationally verified anomalies will plainly mean that standard gravity is broken, in favor of nonstandard paradigms such as modified Newtonian dynamics \citep[MOND; ][]{Milgrom:1983}.

Using various statistical methods to analyze wide binaries selected from Gaia data release 3 \citep[DR3; ][]{Gaia:2023}, two groups led by Chae and Hernandez have independently shown that standard gravity is broken and the effective gravity is boosted in wide binaries with separation greater than several kau. \cite{Chae:2023} first obtained this boost factor $\gamma(\equiv G/G_{\rm N})$, where $G$ is the effective gravitational constant while $G_{\rm N}$ is Newton's constant, to be around $1.4-1.5$ with a strong statistical significance of $>5\sigma$ in the low internal acceleration regime ($< 10^{-9}\,{\rm m}\,{\rm s}^{-2}$), which is consistent with the generic prediction \citep{Banik:2018,ChaeMilgrom:2022} of MOND gravity such as AQUAL \citep{BekensteinMilgrom:1984} and QUMOND \citep{Milgrom:2010}. Many subsequent studies \citep{Hernandez:2023, Hernandez:2024b, Chae:2024a, Chae:2024b, HernandezKroupa:2025, Yoonetal:2025} have obtained similar gravitational boosts with varying statistical significance. These results are based on samples of various data qualities and properties ranging from samples of \emph{statistically} pure binaries to samples of apparent binaries that include up to about 50\% of hierarchical systems (triples and higher-order multiples) hosting unseen companion stars. 

Moreover, Bayesian analyses of 3D velocities \citep{Chae:2025, Chae:2026, Chaeetal:2026} have confirmed gravity boosts in the low-acceleration regime. \cite{Saglia:2025} also constructed Newtonian orbit solutions for 32 binaries of extreme data qualities, a quarter of which have $g_{\rm N}< 10^{-9}\,{\rm m}\,{\rm s}^{-2}$, and found that low-acceleration binaries exhibited overabundance near the periastron phase\footnote{Note, however, that the authors take a conservative view of their results considering the small sample size.} as clearly confirmed and demonstrated by \cite{Chae:2026}. Bayesian 3D analyses are expected to play more important roles in wide binary gravity research with more and better data in the future. 

In contrast to all the above results, several studies have claimed that the observed wide binaries statistically favor, or are consistent with, standard gravity (in particular, \citealt{Banik:2024, Pittordis:2025, Cookson:2026}).\footnote{\cite{Makarov:2026} also claimed that MOND effect was absent in a Gaia DR3 sample of wide binaries through a statistical analysis of a projected angular momentum parameter. This analysis considers a MOND model without the external field effect of the Milky Way and thus the claim is incorrect. See, e.g., Figure~1 of \cite{Chae:2024a} for the dramatic difference between the MOND predictions with and without the external field effect. Moreover, \cite{Makarov:2026} does not take into account multiples (i.e., apparent binaries with hidden additional star(s)) in the analysis, although his large sample is not pure even statistically. Thus, we do not further discuss \cite{Makarov:2026} in this work.} We note that \cite{Banik:2024} and \cite{Cookson:2026} emphasized quality control of the data, in particular regarding the normalized velocity parameter $\tilde{v}$ (to be defined below), while \cite{Pittordis:2025} emphasized realistic modeling of hierarchical systems in gravity tests. 

Here we carefully examine these wide binary studies that do not support the low-acceleration gravitational anomaly and carry out extensive gravity tests based on sky-plane 2D velocities, to independently check whether their key points of data quality control or realistic triple modeling can legitimately remove the gravitational anomaly reported by many studies mentioned above. From this work, we will show the following. 

First, the data quality cut based on the error of $\tilde v$ biases the data in a way to selectively remove accurate and \emph{valid} velocities so that the MOND-type scaling of $\tilde v$ appears weakened. Even for such biased data, the observed scaling of $\tilde v$ still deviates (less strongly) from the corresponding Newtonian prediction, as was also demonstrated by \cite{Chae:2024b}.

Second, the ``realistic'' triple model introduced by \cite{Pittordis:2025} for wide binaries satisfying Gaia's ${\tt ruwe} <1.2$ is applied to wide binaries from three samples recently used by \cite{Chae:2023}, \cite{Banik:2024}, and \cite{Pittordis:2025}. By implementing the \cite{Pittordis:2025} triple model in the \cite{Chae:2023} acceleration-plane test algorithm, we first calibrate the fraction of triples (i.e., apparent binaries hosting a hidden tertiary star) in each of the \cite{Chae:2023} and \cite{Pittordis:2025} samples using their Newtonian-regime data (in the case of the \cite{Banik:2024} sample that does not include Newtonian-regime data, fractions estimated from the other samples are used under similar conditions). Then, we investigate gravity in the low-acceleration regime through the acceleration-plane test and the $\tilde v$ distribution (histogram) test with the three samples, paying particular attention to the latest sample by \cite{Pittordis:2025}, based on the calibrated fractions of triples. 

We will show that the breakdown of standard gravity in the low-acceleration regime is robustly confirmed with statistical significance $>5\sigma$, and the behavior of the gravitational anomaly agrees well with the recent realistic numerical QUMOND solution of wide binary orbits by \cite{Pflamm-Altenburg:2025}. The preference for Newton over a control MOND model obtained by \cite{Pittordis:2025} was significantly affected by the lack of a reliable calibration of the triple fraction with a small-separation Newtonian subsample free of chance-alignment/flyby pairs, and the lack of investigation of the median trend over a sufficiently broad dynamic range from deep Newtonian regime to MOND regime. Moreover, as they discussed the caveat of their analysis, another key issue is that the control MOND model is not accurate in the transition regime according to realistic numerical solutions by \cite{Pflamm-Altenburg:2025}, so that the distinction between Newton and MOND appeared stronger in the transition regime where it should have been smaller.

Finally, we will show that the low acceleration gravitational anomaly is not affected by the data quality framework introduced by \cite{Cookson:2026} including a distance limit, a {\tt ruwe} cut, correction of the perspective effect \citep{Shaya:2011}, a color-magnitude diagram (CMD) cut, requirement of Gaia's parameter ${\tt ipd\_frac\_multi\_peak} = 0$, the \cite{Banik:2024} $\tilde v$ error cut, etc. The $<2\sigma$ claim of `no evidence for MOND' by \cite{Cookson:2026} was largely a consequence of low number statistics (their sample includes only 61 wide binaries in the MOND regime) along with other factors such as the \cite{Banik:2024} biased cut, a biased CMD cut, and incorrect masses for a large portion of the stars. A much larger sample including 8 times as many wide binaries as their sample in the MOND regime with a correct data quality framework confirms the low-acceleration gravitational anomaly with $>3\sigma$ significance.

The organization of this paper is as follows. In Section~\ref{sec:WBTnature}, we first carefully examine the nature of statistical tests with wide binaries, the error properties of the parameters used in statistical tests, and how the error cuts affect the scaling of the parameters in the context of testing gravity. For this purpose, we use three samples from \cite{Chae:2023}, \cite{Banik:2024}, and \cite{Pittordis:2025} that are available to us. In Section \ref{sec:tildev}, we describe the observed profiles of $\tilde v$ with various data quality cuts and show in detail how only the error cut on $\tilde v$ invented by \cite{Banik:2024} and followed by \cite{Cookson:2026} can bias the data by removing some selective fraction of high-precision velocities. In Section~\ref{sec:mondgravity}, we summarize existing predictions of classical MOND gravity paying particular attention to the recent realistic QUMOND solutions of wide binary orbits by \cite{Pflamm-Altenburg:2025}. In Section~\ref{sec:triple_model}, we investigate how the triple model proposed by \cite{Pittordis:2025} affects gravity tests by performing statistical tests with their triple model. In Section~\ref{sec:nearby_WBs}, we investigate how the quality framework proposed by \cite{Cookson:2026} can affect gravity tests using much larger samples within their distance limit of 150~pc and a realistic numerical MOND prediction. We summarize our key findings and give future prospects in Section~\ref{sec:conclusion}. The Python code implementing the \cite{Pittordis:2025} triple model will be made public on Zenodo.

\section{The nature of statistical tests with wide binaries: quality control versus bias introduction in data} \label{sec:WBTnature}

An observed wide binary system refers to a set of (mostly incomplete) measurements for the instantaneous state of the relative motion described by $(\mathbf{r},\mathbf{v})$, where $\mathbf{r}$ and $\mathbf{v}$ are the 3D displacement and relative velocity between the two stars. If all six components are well measured (save the radial separation which can only be roughly measured with currently available telescopes), a 3D analysis can be performed through the Bayesian approach \citep{Chae:2025,Chae:2026}. However, for general samples without accurate 3D velocities because of imprecise or unavailable radial velocities (RVs), gravity tests must rely on statistical methods. 

The Bayesian 3D approach \citep{Chae:2025,Chae:2026} takes into account all measurement uncertainties, and consequently individual gravity inferences are naturally weighted in the combined gravity inference for a sample. Thus, the only issue regarding data in 3D analysis is to select a random sample of wide binaries whose measured velocities are not contaminated \citep{Chaeetal:2026}.\footnote{Here, note that when RVs are not precise enough as in Gaia DR3 alone, some systematic biases must be estimated and taken into account \citep{Chae:2025}.} However, data issues have to be dealt with very carefully for statistical methods that rely on the 2D scalar data $r_p(\equiv |\mathbf{r}_p|)$ and $v_p(\equiv |\mathbf{v}_p|)$, where ($\mathbf{r}_{p},\mathbf{v}_{p}$) are the projections of $(\mathbf{r},\mathbf{v})$ onto the sky plane. In this section, we (re)examine the data issues relevant to the gravity tests proposed in the literature.

For a circular orbit, the magnitude of the kinematic acceleration $g$ is given by  
\begin{equation}
    g=g(r) \equiv \frac{v^2}{r}, 
    \label{eq:g_r}
\end{equation}
which in Newtonian gravity is equal to
\begin{equation}
    g_{\rm N} = g_{\rm N}(r) \equiv G_{\rm N} \frac{M_{\rm tot}}{r^2}, 
    \label{eq:gN_r}
\end{equation}
where $r\equiv|\mathbf{r}|$, $v\equiv|\mathbf{v}|$, and $M_{\rm tot}$ is the total mass of the binary system. For an ideal circular system with perfect measurements of $r$ and $v$ (along with $M_{\rm tot}$), a comparison between Equation~(\ref{eq:g_r}) and Equation~(\ref{eq:gN_r}) can provide a direct test of Newtonian gravity and a quantification of the deviation if present.

For an elliptical orbit, Newtonian gravity predicts 
\begin{equation}
    \frac{g}{g_{\rm N}(r)}= 2 - \frac{r}{a}= \frac{1 + e^2 + 2 e \cos\nu}{1+e\cos\nu},
    \label{eq:gratio}
\end{equation}
where $a$ is the semi-major axis, $e$ is the eccentricity, and $\nu\equiv\phi-\phi_0$ is the orbit true anomaly with the phase  $\phi$ and the argument of the periastron $\phi_0$. This Newtonian prediction of $g$ can be used to test Newtonian gravity through the Monte Carlo (MC) deprojection of the sky-plane 2D data as introduced by \cite{Chae:2023}. 

With the definition of the Newtonian circular velocity
\begin{equation}
    v_c(r) = \sqrt{G_{\rm N} \frac{M_{\rm tot}}{r}} = 941.9\text{\,m\,s}^{-1} \sqrt{\frac{M_{\rm tot}/{\rm M}_\odot}{r/{\rm kau}}},
    \label{eq:vc}
\end{equation}
the acceleration ratio (Equation~(\ref{eq:gratio})) can be related to the normalized 3D velocity
\begin{equation}
    \tilde{v}_{\rm 3D} \equiv \frac{v}{v_c(r)} \equiv \sqrt{\frac{g}{g_{\rm N}(r)}}.
    \label{eq:vratio_3D}
\end{equation}
In principle, one may use this 3D velocity $v$ through the MC deprojection, although the acceleration $g$ is more convenient because it can be analyzed as a function (or in bins) of $g_{\rm N}$.

However, most groups other than Chae's have analyzed the 2D data without deprojection to the 3D space, and in particular worked with the parameter $\tilde{v}$ defined by
\begin{equation} 
  \tilde{v} \equiv \frac{v_p}{v_c(r_p)}. 
  \label{eq:vtilde}
\end{equation} 
What would be the proper requirement on the quality of the data in testing gravity based on the 2D scalar data $(r_p, v_p)$? Since $r_p$ is measured with negligibly small errors by Gaia, the main issue is with $v_p$. Also, the uncertainty of the total mass ($M_{\rm tot}$) for the binary system needs to be considered. Specifically, the question is as to which binaries should be selected or excluded from a raw sample by what standards. In this respect, it is illuminating to investigate mock Newtonian binaries with controlled errors of $v_p$ and $M_{\rm tot}$.

\begin{figure*}[t!]
    \centering
   \includegraphics[width=0.8\textwidth]{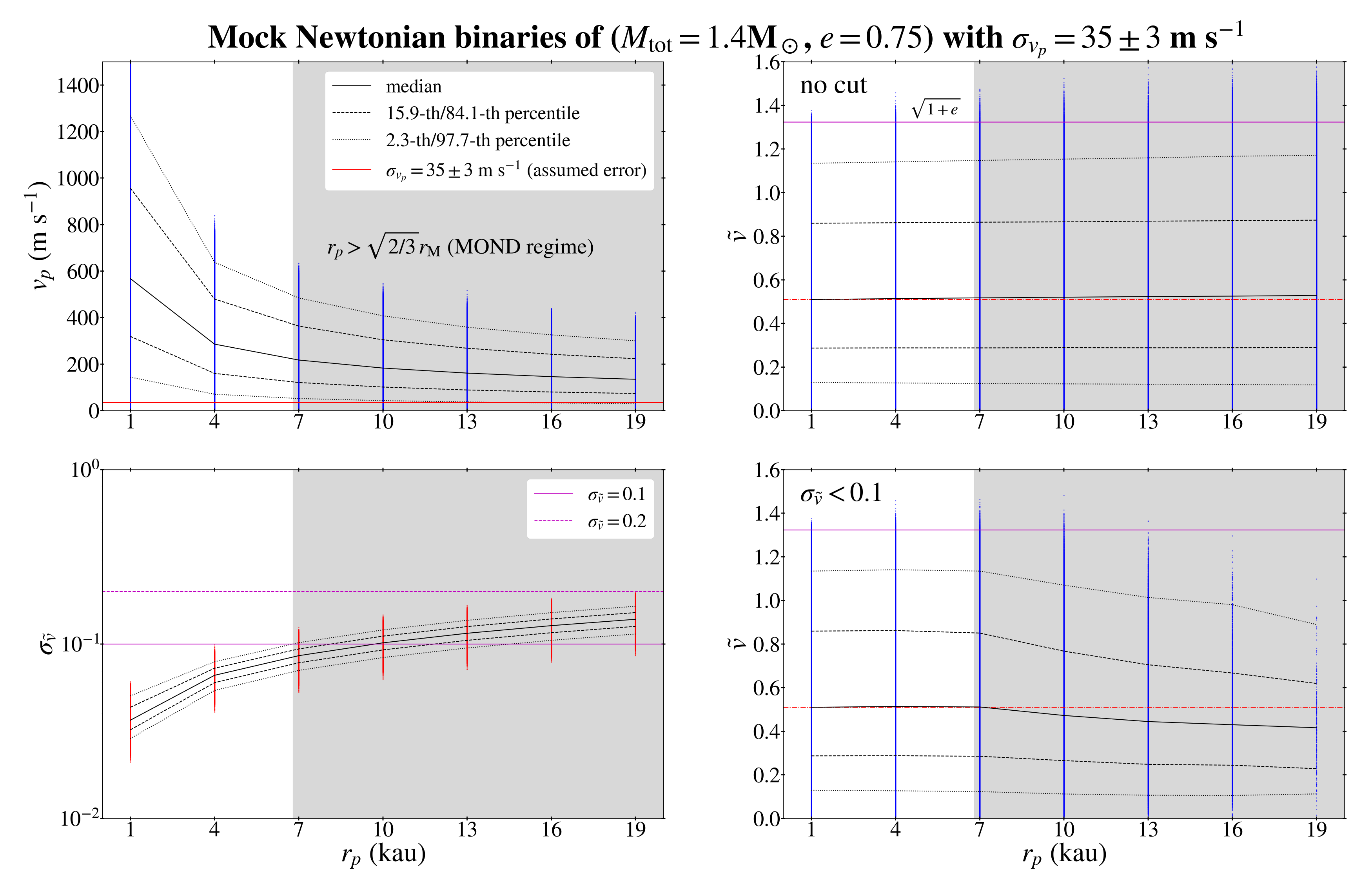}
    \caption{This figure illustrates the statistical properties of mock wide binaries generated assuming Newtonian gravity. For clarity, all wide binaries have the same eccentricity $e=0.75$ and similar total masses with $M_{\rm tot}=1.4\times(1.000\pm 0.068){\rm M}_\odot$. Random scatters of $\sigma_{v_p}=35\pm 3\,{\rm m}\,{\rm s}^{-1}$ are added to Newton-predicted sky-projected velocities $v_p$ as described in the text. The top left panel shows the predicted values of $v_p$ as a function of $r_p$. The parameter $\tilde v (= v_p/v_c$ (Equation~(\ref{eq:vtilde})) is expected to be flat as a function of $r_p$ in Newtonian gravity. The mock data agree well with this expectation as shown in the upper right panel. The lower left panel shows the statistical properties of the uncertainties of $\tilde v$ (Equation~(\ref{eq:vterr})). If the Banik cut (Equation~(\ref{eq:vtcut})) is used to remove some portions of the data, the median of $\tilde v$ gets biased to a lower value at large values of $r_p$. }
    \label{fig:vpvt_illustration_M14_vperr35}
\end{figure*}

For a binary with mass $M_{\rm tot}$ and sky-projected separation $r_p$, Newtonian gravity predicts the two components of $\mathbf{v}_p$ as
\begin{equation}
    \left. \begin{array}{ccl}
        v_{x^\prime} & = & kv|\cos\psi|\\
       v_{y^\prime}  & =  & kv \cos i \tan\psi|\cos\psi|
    \end{array}  \right\}
  \label{eq:vpcomps_Newton}  
\end{equation}
with
\begin{equation}
\begin{array}{ccl}
    v & =  & 941.9\text{\,m\,s}^{-1} \sqrt{\frac{M_{\rm tot}/{\rm M}_\odot}{r_p/{\rm kau}}}  \\
  &\times & (1-\sin^2i\sin^2\phi)^{1/4} \sqrt{\frac{1+e^2+2e\cos\nu}{1+e\cos\nu}} ,
  \label{eq:v_Newton}
  \end{array}
\end{equation}
and 
\begin{equation}
    \psi = \tan^{-1}\left(- \frac{\cos\phi + e\cos\phi_0}{\sin\phi + e\sin\phi_0}\right),
    \label{eq:psi}
\end{equation}
where $i$ is the inclination, $\nu=\phi-\phi_0$ as before, and $k=\pm 1$ (a sign function of $\phi$ and $\phi_0$) is irrelevant in the present analysis. Then, we have
\begin{equation}
  v_p  = \sqrt{v_{x^\prime}^2+v_{y^\prime}^2}.
  \label{eq:vp_Newton}  
\end{equation}
Here the two components (Equation~(\ref{eq:vpcomps_Newton})) of $\mathbf{v}_p$ (rather than the worked-out expression for $v_p$) are used to properly add a scatter to $v_p$. Specifically, for an assumed error of $\sigma_{v_p}$ for $v_p$, $v_{x^\prime}$ and $v_{y^\prime}$ are given, respectively, a normal scatter of $\sigma_{v_p}/\sqrt{2}$.

\begin{figure*}
    \centering
   \includegraphics[width=0.8\linewidth]{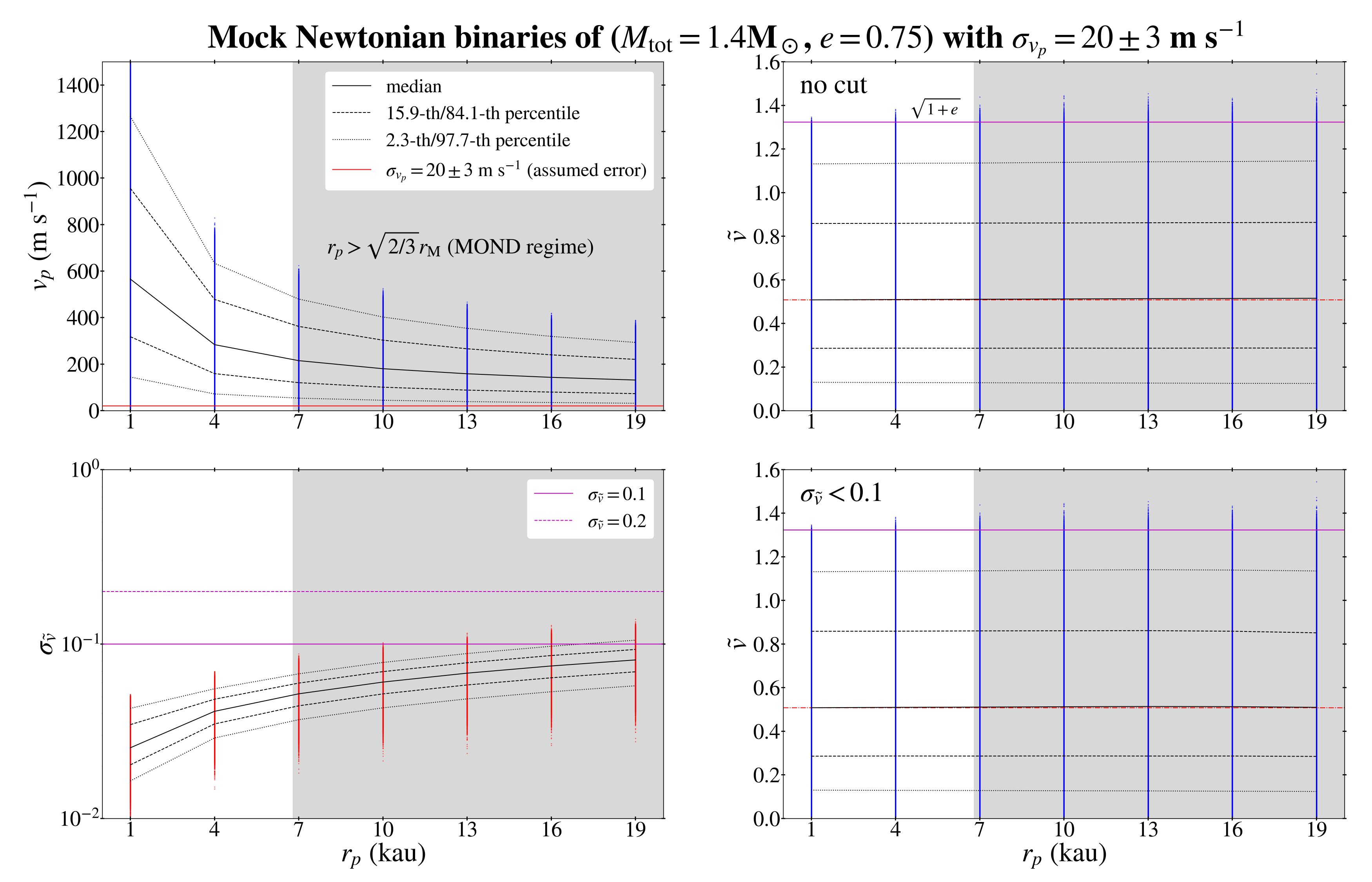}
    \caption{Same as Figure~\ref{fig:vpvt_illustration_M14_vperr35} but with $\sigma_{v_p}=20\pm 3\,{\rm m}\,{\rm s}^{-1}$. In this case, the Banik cut removes little of the data and $\tilde v$ is not biased.}
    \label{fig:vpvt_illustration_M14_vperr20}
\end{figure*}

Figure~\ref{fig:vpvt_illustration_M14_vperr35} exhibits Newtonian predictions of $v_p$ and $\tilde v$ at various values of projected separation $r_p$ for mock binaries. For this simulation, we take a total mass of $M_{\rm tot}=1.4{\rm M}_\odot$ (the median mass for the \cite{Banik:2024} sample) with a 6.8\% scatter (a mass uncertainty to be used for the purpose of illustration: see below) and eccentricity $e=0.75$ (a typical value for wide binaries under consideration). Here, the probability distributions of the parameters follow from the randomness of the mock observations: $f_{\rm pr}(i)=\sin i$, $\phi_0$ is uniform, and $f_{\rm pr}(\nu)=(1-e^2)^{3/2}(2\pi)^{-1}(1+e\cos\nu)^{-2}$. To the Newtonian predictions of $v_p$ are added mock measurement scatters with $\sigma_{v_p}=35\pm 3\,{\rm m}\,{\rm s}^{-1}$ (which is a typical value: see below). The uncertainties of $\tilde v$ are given by 
\begin{equation}
    \sigma_{\tilde v} = {\tilde v} \sqrt{\left(\frac{\sigma_{v_p}}{v_p}\right)^2 + \left(\frac{\sigma_{M_{\rm tot}}}{2M_{\rm tot}}\right)^2}
    \label{eq:vterr}
\end{equation}
with $\sigma_{M_{\rm tot}}/M_{\rm tot}=0.068$, which is chosen so that the cut based on Equation~(\ref{eq:vterr}) mimics the \cite{Banik:2024} quality control cut to be described below.

The top left panel of Figure~\ref{fig:vpvt_illustration_M14_vperr35} shows that most of the values of $v_p$ are significantly higher than the adopted uncertainty even at large separations $r_p > 10$~kau. This means that in most occurrences of the mock observation, the signal-to-noise ($S/N$) $v_p/\sigma_{v_p}$ is higher than 1. Consequently, the predicted values of $\tilde v$ closely follow the expected flatness of the median as a function of $r_p$, as shown in the top right panel. Now consider the uncertainties of $\tilde v$ shown in the bottom left panel. Most values are smaller than $0.2$, which is sufficiently low. Moreover, since $\tilde{v}/\sigma_{\tilde v} \simeq v_p/\sigma_{v_p}>1$ is already satisfied, no additional cut would be warranted. 

\begin{figure*}
    \centering
   \includegraphics[width=0.8\linewidth]{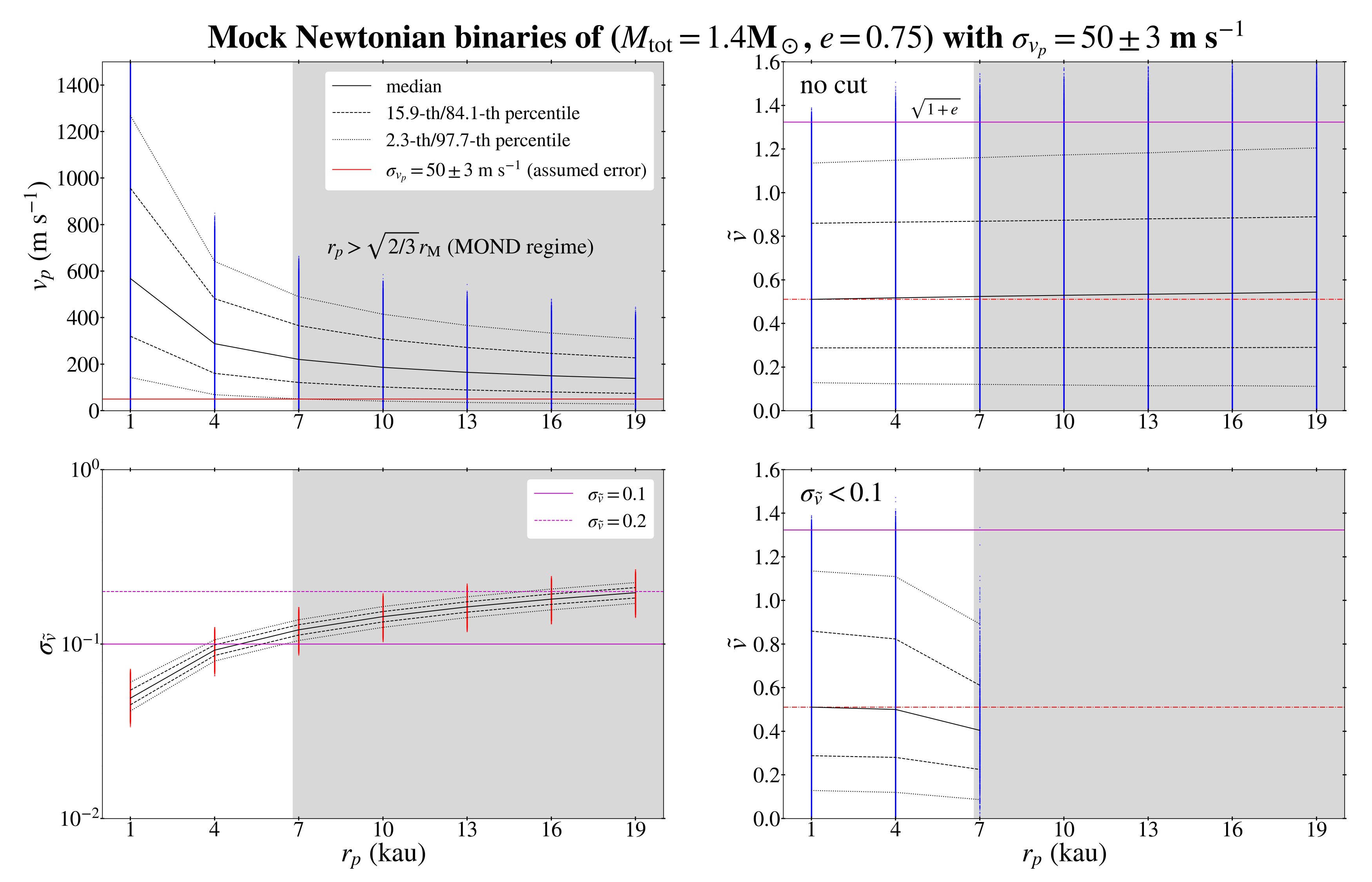}
    \caption{Same as Figure~\ref{fig:vpvt_illustration_M14_vperr35} but with $\sigma_{v_p}=50\pm 3\,{\rm m}\,{\rm s}^{-1}$. In this case, the Banik cut removes much of the data and $\tilde v$ is biased even at $r_p=7$~kau, while the whole data without the cut is unbiased (i.e.\ nearly flat). }
    \label{fig:vpvt_illustration_M14_vperr50}
\end{figure*}

\begin{figure*}
    \centering
   \includegraphics[width=0.8\linewidth]{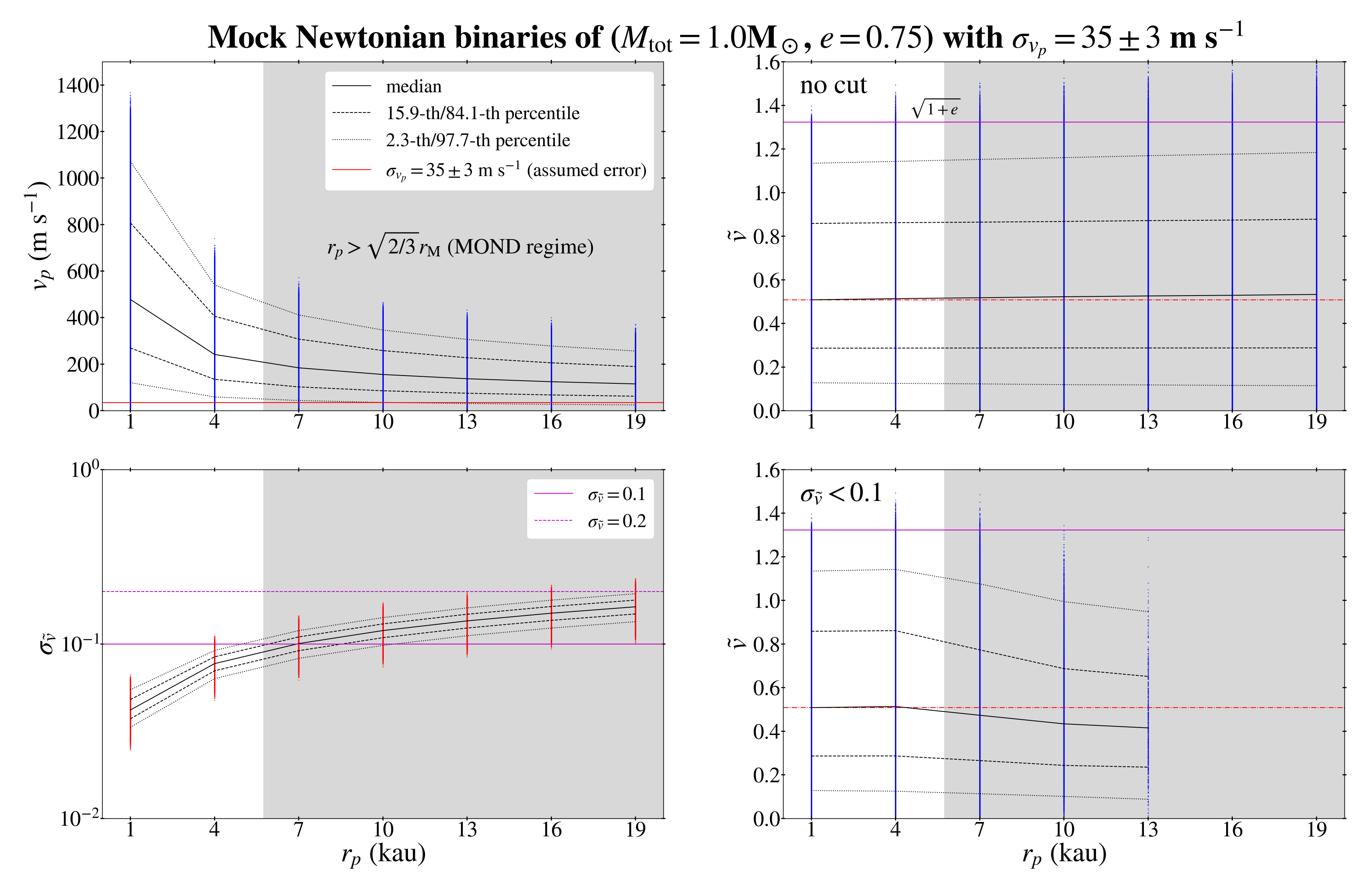}
    \caption{Same as Figure~\ref{fig:vpvt_illustration_M14_vperr35} but with $M_{\rm tot}=1.0\times(1.000\pm 0.068){\rm M}_\odot$. In this case, the Banik cut removes more of the data compared with the larger mass case and thus $\tilde v$ is more biased.}
    \label{fig:vpvt_illustration_M10_vperr35}
\end{figure*}

However, \cite{Banik:2024} suggested a cut (to be referred to as the Banik cut throughout)
\begin{equation}
    \sigma_{\tilde v} < 0.1 ~\mathrm{max} \left(1,\frac{\tilde v}{2}\right)
    \label{eq:vtcut}
\end{equation}
as a data ``quality control''. Because $\tilde v<2$ is satisfied in pure binaries with small uncertainties of $v_p$, this cut is identical to $\sigma_{\tilde v} < 0.1$, which is shown in the bottom left panel of Figure~\ref{fig:vpvt_illustration_M14_vperr35}. In passing, we note that $\tilde v<2$ is also satisfied in the majority of the actual data that include impure binaries. This cut removes large portions of the mock data for $r_p>r_{\rm M}$ where $r_{\rm M} \equiv \sqrt{G_{\rm N}M_{\rm tot}/a_0}$ (with $a_0=1.2\times 10^{-10}\,{\rm m}\,{\rm s}^{-2}$) is the MOND radius. The effect of this cut on $\tilde v$ is shown in the bottom right panel. The median scaling of $\tilde v$ with $r_p$ is no longer flat but goes downward as $r_p$ increases, in contrast to the case where the cut was not applied. 

The effects of the Banik cut vary depending on the combination of $M_{\rm tot}$ and $\sigma_{v_p}$. Figures~\ref{fig:vpvt_illustration_M14_vperr20} and Figures~\ref{fig:vpvt_illustration_M14_vperr50} show cases with varied values of $\sigma_{v_p}$ at the same value of $M_{\rm tot}$. In the case of $\sigma_{v_p}=20\pm 3\,{\rm m}\,{\rm s}^{-1}$, there is no bias in the median behavior of $\tilde v$ as the Banik cut can remove little of the data. On the other hand, in the case of $\sigma_{v_p}=50\pm 3\,{\rm m}\,{\rm s}^{-1}$, the median $\tilde v$ is significantly lowered at $r_p = 7$~kau and all data at $r_p\ge 10$~kau are completely removed by the Banik cut. However, as the upper right panel of Figures~\ref{fig:vpvt_illustration_M14_vperr50} shows, the removed data follow the Newtonian flat line to a good approximation (the slight rising trend of $\tilde v$ with $r_p$ due to scatters is minor). 

Figure~\ref{fig:vpvt_illustration_M10_vperr35} shows a case with the variation of $M_{\rm tot}$: $1.4\rightarrow 1.0\,{\rm M}_\odot$. For the smaller mass of $M_{\rm tot}=1.0{\rm M}_\odot$ at the given uncertainty, the bias is larger. This property implies that the Banik cut will selectively remove lower-mass systems in increasingly larger proportion at larger $r_p$. This was already noticed in \cite{Chae:2024b} for a sample from \cite{Chae:2023}. In the next section, we will show that this is also the case in the \cite{Banik:2024} sample. The above simulation results clearly show that when the Banik cut removes a significant portion of precise $v_p$ data, it introduces a bias in the statistical distribution of $\tilde v$ at the relevant $r_p$ in a way that the apparent median velocity is lower than the value implied by the assumed gravity.

\begin{figure*}
    \centering
   \includegraphics[width=0.9\linewidth]{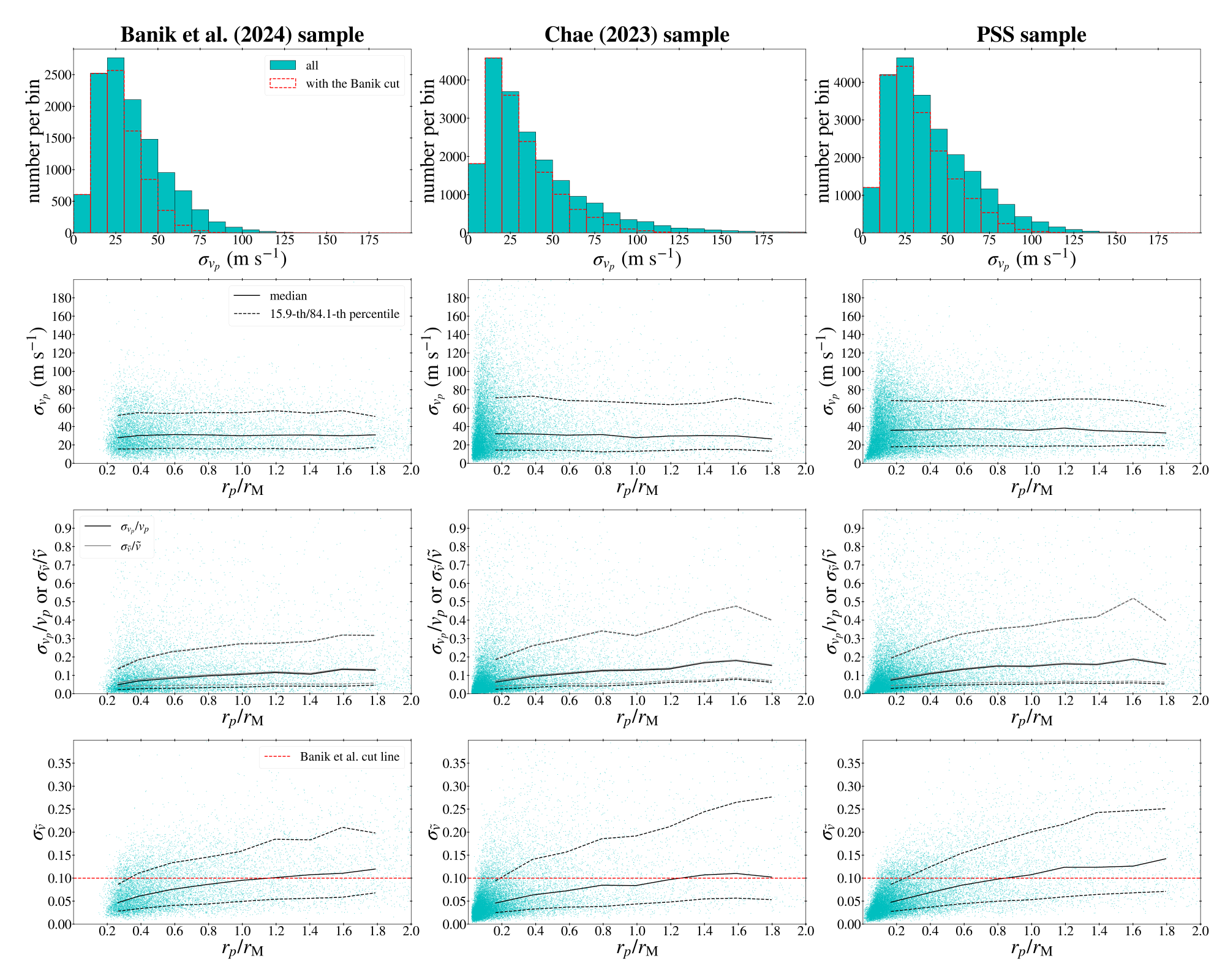}
    \caption{This figure shows error properties in three samples from the literature \citep{Banik:2024,Chae:2023,Pittordis:2025}. Errors of $v_p$ and $\tilde v$ as well as their relative errors (i.e., the inverse of their $S/N$) are shown. The three samples have similar error properties. The top row shows that the Banik cut starts to remove data from the third bin centered at $\sigma_{v_p} = 25\,{\rm m}\,{\rm s}^{-1}$. The third row shows that most data satisfy $S/N>2$ and almost all data satisfy $S/N>1$. The bottom row shows that the Banik cut selectively removes more proportions of data at larger $r_p/r_{\rm M}$.  }
    \label{fig:vpvt_error_distribution}
\end{figure*}

\begin{figure*}[tbh!]
    \centering
   \includegraphics[width=1.0\linewidth]{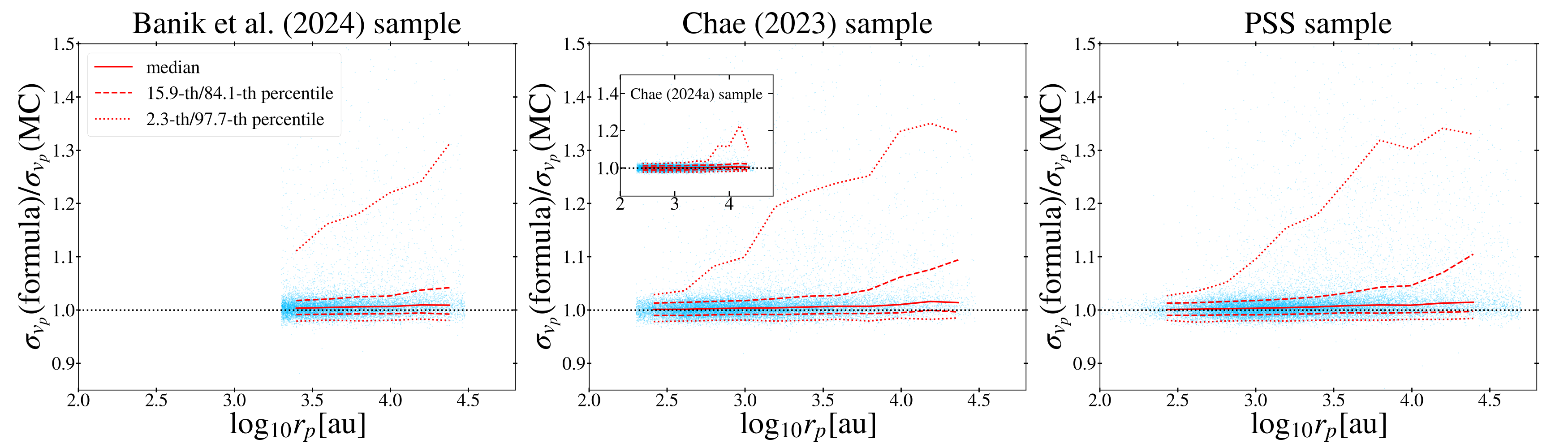}
    \caption{Distributions of the ratio of two estimates of the $v_p$ error are shown for the three samples used in this work.  }
    \label{fig:vperr_test}
\end{figure*}

Now we examine the error properties in wide binary samples recently used in the literature: 11865 systems within distance $250$~pc from \cite{Banik:2024}, 19716 systems within $200$~pc(the case with PM relative error $<0.005$) from \cite{Chae:2023}, and 23223 systems within $300$~pc from \cite{Pittordis:2025} (hereafter PSS). We consider the errors of $v_p$, $\tilde v$, and their relative errors (i.e., the inverse of their $S/N$). See below for the estimates of the error of $v_p$. For the relative error of $\tilde v$, we introduce a notation to be used throughout
\begin{equation}
    \beta \equiv \sigma_{\tilde v}/{\tilde v} = \sqrt{\left(\frac{\sigma_{v_p}}{v_p}\right)^2 + \left(\frac{\sigma_{M_{\rm tot}}}{2M_{\rm tot}}\right)^2},
    \label{eq:beta}
\end{equation}
where the last equality follows from Equation~(\ref{eq:vterr}). The numerical value of $\beta$ is similar to $\sigma_{v_p}/v_p$ as long as $\sigma_{M_{\rm tot}}/M_{\rm tot}$ is not unreasonably large. 

Figure~\ref{fig:vpvt_error_distribution} shows the distributions of the estimated errors. The top row shows that all three samples have similar distributions of $\sigma_{v_p}$, and most values are smaller than about $100\,{\rm m}\,{\rm s}^{-1}$. The second row shows that $\sigma_{v_p}$ does not depend on the separation between the two stars. The third row shows that virtually all data satisfy $\sigma_{v_p}/v_p<1$ (i.e.\ $S/N>1$) as well as $\beta <1$ and it has only a weak dependence on $r_p/r_{\rm M}$. The bottom row shows that $\sigma_{\tilde v}$ increases with $r_p/r_{\rm M}$, and thus the Banik cut selectively removes larger portions of the data in the MOND regime ($r_p/r_{\rm M}\ga 1$). As the top row shows, the Banik cut removes even very precise ($\sigma_{v_p}<50\,{\rm m}\,{\rm s}^{-1}$) data, although such data are consistent with the underlying gravity as demonstrated by simulations shown in the top right panels of Figures~\ref{fig:vpvt_illustration_M14_vperr35}, \ref{fig:vpvt_illustration_M14_vperr20}, \ref{fig:vpvt_illustration_M14_vperr50}, and \ref{fig:vpvt_illustration_M10_vperr35}. 

We note that the error of $v_p$ follows from the measurement errors of the sky-plane four PM components of the two stars. It can be conveniently calculated using the standard formula given by Equations~(7) through (9) of \cite{Chae:2024a}. For statistical purposes as in the present study, this estimate is sufficiently precise for the samples under consideration as can be verified through MC simulations. Figure~\ref{fig:vperr_test} compares the error based on the standard formula and a direct estimate based on MC simulations of the PM components. For all three samples used in this work, the ratio of the two estimates is close to unity, and the majority are within 10\% difference even at the largest $r_p$, which makes no difference in statistical gravity tests. The agreement between the two estimates is particularly good (see the inset of the middle panel of Figure~\ref{fig:vperr_test}) for the sample satisfying the selection criteria of \cite{Chae:2024a} that include parallax relative error $<0.005$ in addition to PM relative error $<0.005$. Thus, the claim by \cite{Makarov:2026} that the error estimate by \cite{Chae:2024a} was incorrect is baseless. The use of the standard formula can be too imprecise when uncertainties of PMs and parallaxes are large, but in the case of \cite{Chae:2024a} the uncertainties are sufficiently small. We further note that in the acceleration-plane test to be considered in this work, the uncertainties of the PM components are directly used (see \cite{Chae:2023,Chae:2024b}) and thus the error estimate of $v_p$ is irrelevant.

To sum up, simulations show that the Banik cut on $\sigma_{\tilde v}$ can bias the median trend of $\tilde v$ when significant portions of the data are selectively removed by the cut. The statistical properties of the existing wide binary samples, regardless of some differences in the details of how they were defined, show that $\sigma_{\tilde v}$ is \emph{not} independent of the separation between the pair and thus the Banik cut selectively removes larger portions of the data at larger separations. This indicates that the Banik cut will introduce bias in the data. The properties of the cut-introduced bias in the wide binary samples will be described in the following section.

\section{Profiles of the Normalized 2D Velocity with Data Quality Cuts}\label{sec:tildev}

\begin{figure*}[tbh!]
    \centering
   \includegraphics[width=1.0\linewidth]{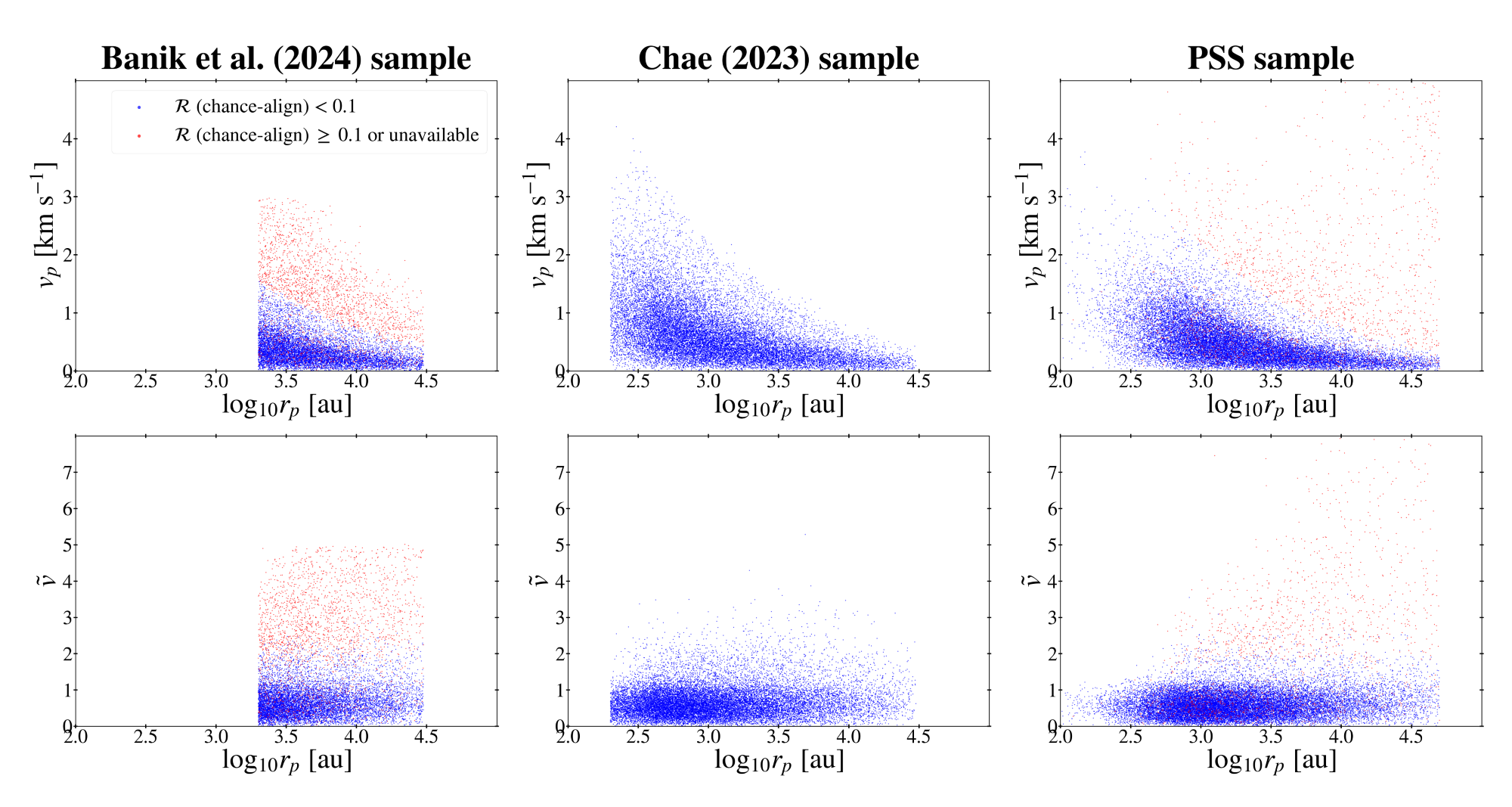}
    \caption{Distributions of $v_p$ and $\tilde v$ with respect to $r_p$ in three samples (introduced in Figure~\ref{fig:vpvt_error_distribution}) to be used for statistical analyses in this work. The \cite{Banik:2024} sample covers a relatively narrow dynamic range $2<r_p<30\,{\rm kau}$ excluding the Newtonian regime, while the other samples cover much broader ranges including the Newtonian regime. The \cite{Chae:2023} sample includes only systems that are likely to be gravitationally bound based on the \cite{El-badry:2021} $\mathcal{R}$ parameter. The \cite{Banik:2024} sample and the PSS sample include chance-alignment/flyby pairs that are indicated by red dots based on $\mathcal{R}$ (here those red dots include systems absent from the \cite{El-badry:2021} catalog, so not all of them may be chance-alignment/flyby). }
    \label{fig:vpvt_rp}
\end{figure*}

In this section, we examine in detail the profiles of the normalized 2D velocity $\tilde v$ (Equation~(\ref{eq:vtilde})) with respect to $r_p/r_{\rm M}$ in the three samples introduced in Figure~\ref{fig:vpvt_error_distribution} and investigate the effects of various quality cuts including the Banik cut. Figure~\ref{fig:vpvt_rp} shows the distributions of $v_p$ and $\tilde v$ with respect to $r_p$. \cite{Chae:2023} sample was derived from the \cite{El-badry:2021} catalog of 1.8 million pairs, while \cite{Banik:2024} and PSS samples were directly derived from the Gaia DR3 database. \cite{El-badry:2021} defined a chance-alignment (i.e., gravitationally-unbound pair) probability parameter $\mathcal{R}$ and estimated its value for every system in the catalog. 

The \cite{Chae:2023} sample includes only systems that are very likely to be gravitationally bound according to their $\mathcal{R}$ values. \cite{Banik:2024} and PSS samples include chance-alignment/flyby pairs without estimating their probabilities. However, the majorities of the \cite{Banik:2024} and PSS samples (81\% and 92\%, respectively)\footnote{The fraction of systems included in the \cite{El-badry:2021} catalog is much lower in the \cite{Banik:2024} sample in part because it includes only systems of relatively larger separations with $r_p>2\,{\rm kau}$.} are included in the \cite{El-badry:2021} catalog. In Figure~\ref{fig:vpvt_rp}, all pairs satisfying $\mathcal{R}<0.1$\footnote{The particular \cite{Chae:2023} sample taken from the previous study satisfies a stronger cut $\mathcal{R}<0.01$, but use of $\mathcal{R}<0.1$ makes no tangible difference.} are represented by blue dots while the rest are by red dots. The systems without $\mathcal{R}$ values are not included in the \cite{El-badry:2021} catalog because they did not pass the selection function used by \cite{El-badry:2021}. These systems are likely to have relatively higher probabilities of being gravitationally unbound, but some of them may be gravitationally bound and happen to not satisfy the complex selection function. 

\begin{figure*}[t!]
    \centering
   \includegraphics[width=1.0\linewidth]{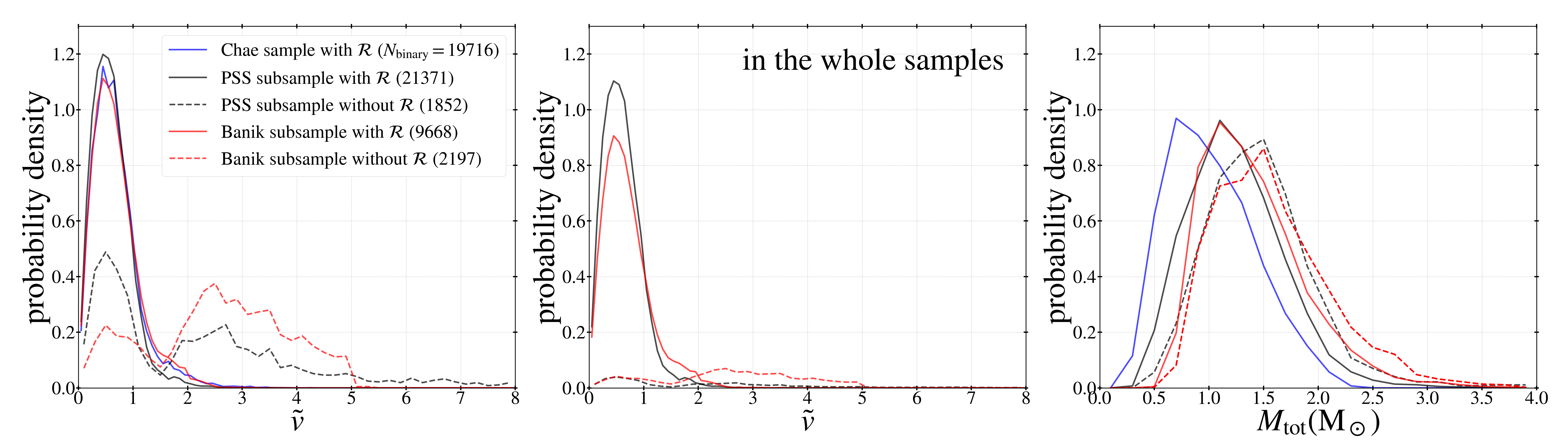}
    \caption{For the samples shown in Figure~\ref{fig:vpvt_rp}, distributions of $\tilde v$ and $M_{\rm tot}$ (total mass of the binary system) are compared for the subsamples with or without $\mathcal{R}$ (the \cite{El-badry:2021} parameter of chance-alignment/flyby probability). The \cite{Chae:2023} sample does not have a subsample without $\mathcal{R}$ by definition. For the \cite{Banik:2024} and \cite{Pittordis:2025} (PSS) subsamples without $\mathcal{R}$, the distributions of $\tilde v$ have relatively large probabilities at $\tilde v > 2$ and the distributions of $M_{\rm tot}$ are shifted to to the higher direction. The middle panel shows distributions of the subsamples in each whole sample, while the other panels show distributions within the subsamples. }
    \label{fig:vt_mass_hist}
\end{figure*}

\begin{figure*}[t!]
    \centering
   \includegraphics[width=1.0\linewidth]{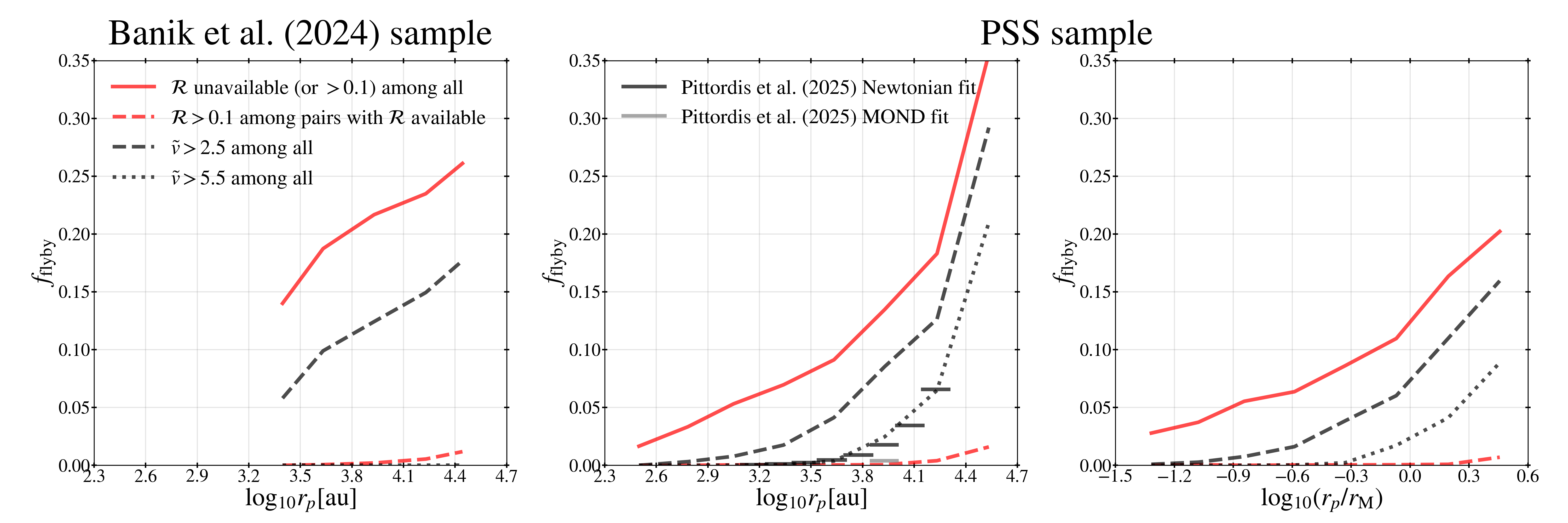}
    \caption{For the samples by \cite{Banik:2024} and \cite{Pittordis:2025} (PSS) shown in Figure~\ref{fig:vpvt_rp}, chance-alignment/flyby fractions are estimated in bins of $\log_{10}r_p$. Three criteria based on $\mathcal{R}$ (the \cite{El-badry:2021} parameter) and $\tilde v$ (Equation~(\ref{eq:vtilde})) are used to identify chance-alignment/flyby pairs. For the PSS sample, bins of $\log_{10}(r_p/r_{\rm M})$ are also considered. The short black solid lines in the middle panel represent the numbers of flybys in bins of $r_p$ estimated from the Newtonian fit by \cite{Pittordis:2025}. The short gray solid line represents the number of flybys in the benchmark bin of $7.1<r_p<10\,{\rm kau}$ from the MOND fit by \cite{Pittordis:2025}.  }
    \label{fig:f_flyby}
\end{figure*}

Figure~\ref{fig:vt_mass_hist} shows the distributions of $\tilde v$ and $M_{\rm tot}$ (total mass of the binary system) in the three samples. For the (sub)samples of systems having values of $\mathcal{R}$, the \cite{Chae:2023} and \cite{Banik:2024} distributions of $\tilde v$ (which do not have any limit on {\tt ruwe}) are similar although their distributions of $M_{\rm tot}$ are different, but the PSS systems, which satisfy the cut ${\tt ruwe}<1.2$, have a significantly different distribution of $\tilde v$. The \cite{Banik:2024} and PSS subsamples without $\mathcal{R}$ are systematically different from those with $\mathcal{R}$ both in the $\tilde v$ and $M_{\rm tot}$ distributions. The distributions of $\tilde v$ in the subsamples without $\mathcal{R}$ have double peaks, one at $\tilde v < 1$ and the other at $\tilde v > 2$, indicating that they are a mix of two populations. The population peaked at $\tilde v > 2$ has a long tail and is likely to be significantly contributed by chance-alignment/flyby pairs, while the population peaked at $\tilde v < 2$ may be significantly contributed by gravitationally-bound binaries. As the middle panel of Figure~\ref{fig:vt_mass_hist} shows, the latter population makes a tiny contribution to the distribution in the range $\tilde v < 1.5$, so that exclusion or inclusion of them makes little difference statistically.

For the \cite{Banik:2024} and PSS samples, about 99.9\% of binaries having $\mathcal{R}$ satisfy $\mathcal{R}<0.1$ indicating that systems simultaneously satisfying the selection function of \cite{El-badry:2021} and that of \cite{Banik:2024} or PSS are very likely to be gravitationally bound. Thus, the fractions of chance-alignment/flyby pairs in the \cite{Banik:2024} and PSS samples depend on how to treat the systems without $\mathcal{R}$. Since Figure~\ref{fig:vt_mass_hist} indicates that not all systems without $\mathcal{R}$ are gravitationally unbound, we consider a few possibilities to estimate the fraction of chance-alignment/flyby pairs in the given sample. Figure~\ref{fig:f_flyby} shows the fraction of chance-alignment/flyby pairs ($f_{\rm flyby}$) as a function of separation for the \cite{Banik:2024} and PSS samples. Chance-alignment/flyby pairs are identified by three criteria as indicated in the figure. The value of $f_{\rm flyby}$ selected with the unavailability of $\mathcal{R}$ or $\mathcal{R}>0.1$ is significantly higher than that with $\tilde v>2.5$ at a given bin. The cut $\tilde v > 5.5$ gives much lower values of $f_{\rm flyby}$, which are interestingly similar to the values of $f_{\rm flyby}$ from the Newtonian modeling results by \cite{Pittordis:2025}. For the PSS sample that covers a broad range of $r_p$, it is clear that $f_{\rm flyby}\rightarrow 0$ as $r_p\rightarrow 0$ regardless of which choice is used to identify $f_{\rm flyby}$. This property will be used throughout when we use the PSS sample for gravity tests. For simplicity of the notation, whenever we refer to a sample with $\mathcal{R}<0.1$, it is implicit that any pairs without $\mathcal{R}$ have been excluded.

The \cite{Banik:2024} sample has two peculiar properties compared with the other samples: (1) all pairs have $r_p>2\,{\rm kau}$ and thus do not include systems with high internal acceleration ($\ga 10^{-8.5}\,{\rm m}\,{\rm s}^{-2}$) that can be useful for calibration purposes; (2) the sample includes a significantly higher fraction of gravitationally-unbound pairs. 

The PSS sample satisfies ${\tt ruwe}<1.2$ and a narrow band in the color-magnitude diagram (CMD), and thus it includes a significantly lower fraction of hierarchical systems (i.e., triple and higher-order multiple systems including hidden stars) although it includes some fraction of chance-alignment cases. This is confirmed in this work as will be shown through various analyses.

\begin{figure*}[tbh!]
    \centering
   \includegraphics[width=0.95\linewidth]{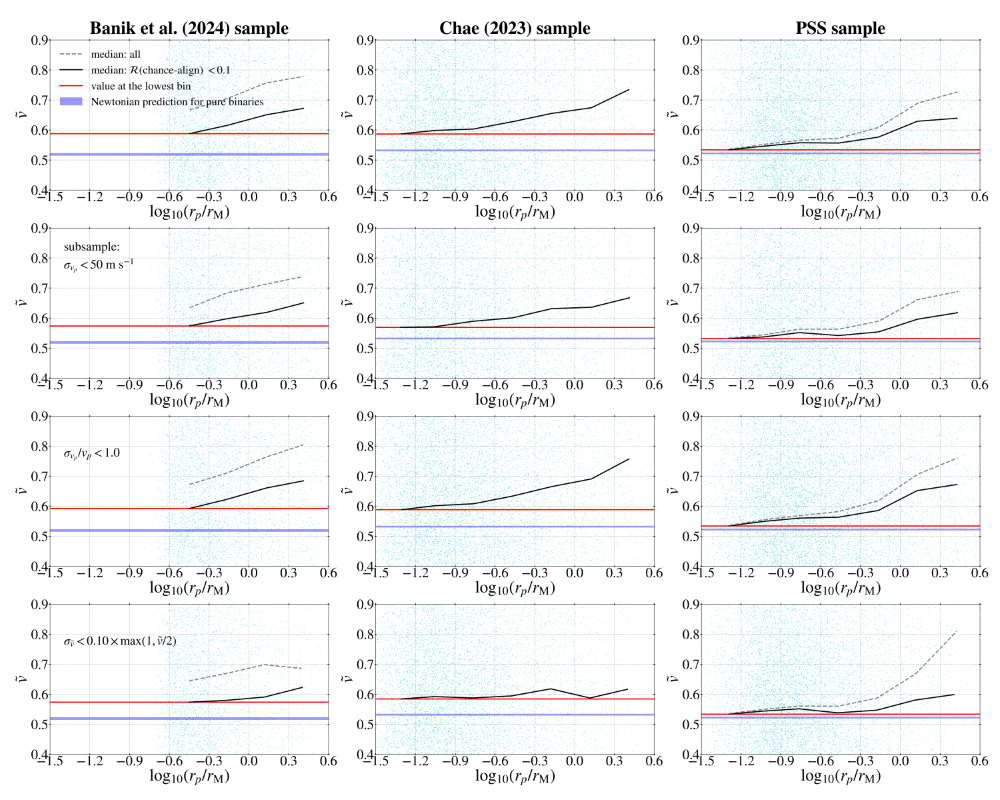}
    \caption{Scaling of the median of $\tilde v$ with respect to $r_p/r_{\rm M}$ in the three samples shown in Figure~\ref{fig:vpvt_rp}. The bins of $\log_{10}(r_p/r_{\rm M})$ are indicated by thin vertical lines. Gray dashed curves are for all systems including chance-alignment/flyby cases, while black solid curves are only for gravitationally-bound systems based on $\mathcal{R}<0.1$. The Newtonian prediction for pure binaries with eccentiricties shown in Figure~\ref{fig:eccentricity} is represented by the blue band. The top row shows the results for the entire samples, while the other rows show results for subsamples with the indicated quality cuts. Here the perspective effect (which is overall very minor and matters only at the last bin of $0.3<\log_{10}(r_p/r_{\rm M})<0.6$) has been corrected for the \cite{Banik:2024} and PSS samples.  }
    \label{fig:vt_scaling_3samples}
\end{figure*} 

\begin{figure}[tbh!]
    \centering
   \includegraphics[width=0.9\linewidth]{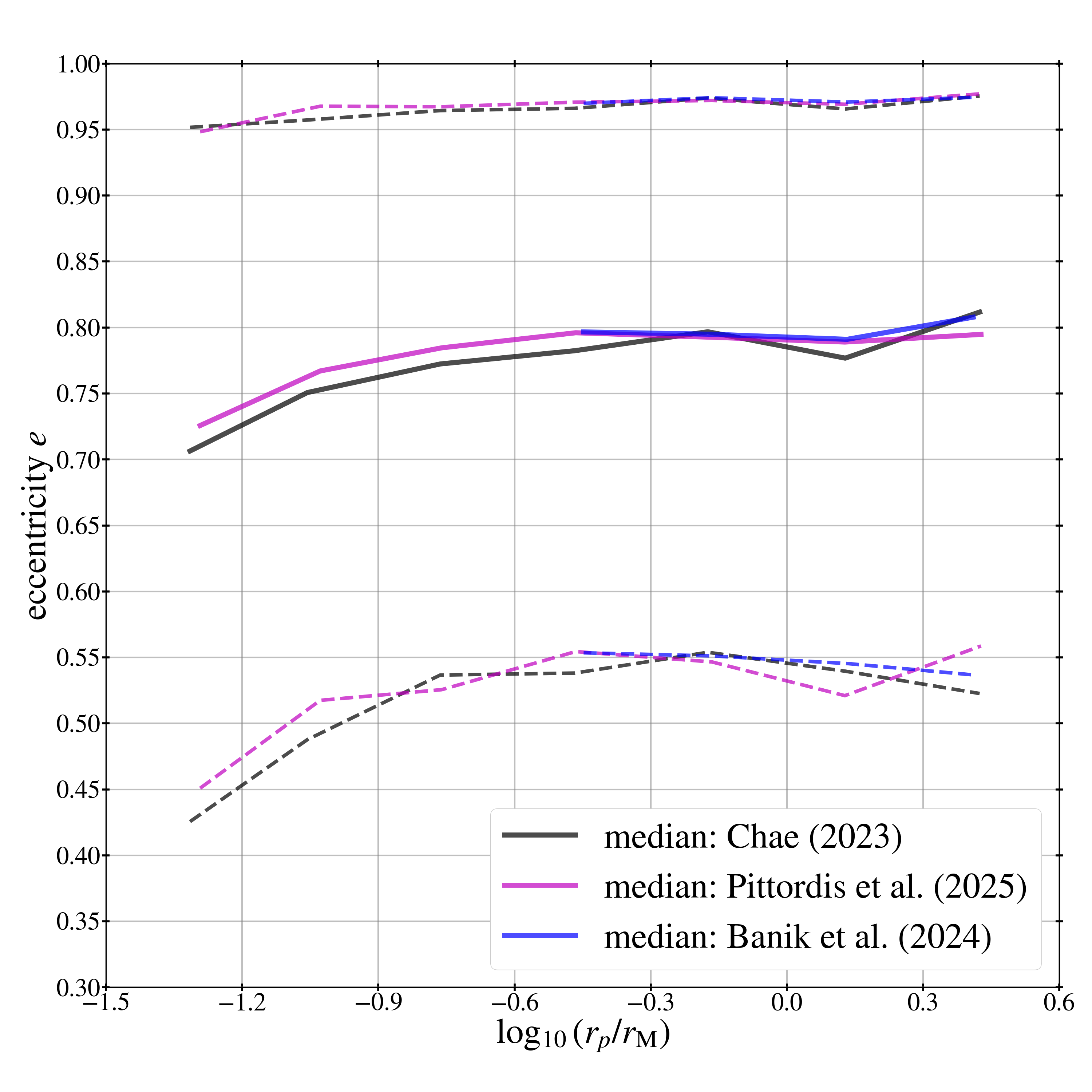}
    \caption{This figure shows the statistical properties of eccentricities for the three samples of wide binaries with $\mathcal{R}<0.1$ shown in Figure~\ref{fig:vt_scaling_3samples}. The horizontal bins are the same as Figure~\ref{fig:vt_scaling_3samples}. Solid curves show the medians while dashed curves show the 16-th and 84-th percentiles in the bins. All eccentricities are based on the Bayesian results by \cite{Hwang:2022}. }
    \label{fig:eccentricity}
\end{figure}

Figure~\ref{fig:vt_scaling_3samples} shows the profile of median $\tilde v$ with respect to $r_p/r_{\rm M}$ in the three samples. The top row shows the results for all pairs without or with the constraint $\mathcal{R}<0.1$. For the \cite{Banik:2024} and PSS samples, the results with $\mathcal{R}<0.1$ are lower than the corresponding results for all systems including chance-alignment/flyby pairs. The exceptions are the first few lowest-$r_p/r_{\rm M}$ bins in the PSS sample in which chance-alignment/flyby pairs are negligible. In all cases, the median rises with $r_p/r_{\rm M}$, as can be clearly seen compared to the flat red line that is set by the lowest-$r_p/r_{\rm M}$ bin with $\mathcal{R}<0.1$ in each sample. The medians of $\tilde v$ at a fiducial bin of $-0.6<\log_{10}(r_p/r_{\rm M})<-0.3$ (the lowest bin in the Banik sample) differ in the samples because of the differences in the implicit fraction of hierarchical systems. The median $\tilde v$ in the fiducial bin decreases in the order of the \cite{Chae:2023} sample, the \cite{Banik:2024} sample, and the PSS sample. The median of the PSS sample is significantly lower than the other samples because the cut ${\tt ruwe}<1.2$ removed a large fraction of hierarchical systems.

The height of the flat red line (the median of $\tilde v$ in the lowest bin) for each sample in Figure~\ref{fig:vt_scaling_3samples} is compared with the corresponding Newtonian prediction for pure binaries. The height of the flat red line in the PSS sample is only slightly above the Newtonian prediction for pure binaries, whereas those in the other samples are well above their Newtonian predictions. The Newtonian prediction depends on the underlying distribution of eccentricities ($e$) in the samples. The statistical properties of eccentricities in the samples can be found in Figure~\ref{fig:eccentricity}. The medians and 16-th and 84-th percentiles are derived from stacked distributions of all wide binaries in each bin based on Bayesian inferences by \cite{Hwang:2022}. The medians are in the range $0.71\la \langle e \rangle \la 0.78$ in all three samples. Throughout $\langle\cdots\rangle$ refers to the median whenever the symbol appears. 

\begin{figure}[tb!]
    \centering
   \includegraphics[width=1.\linewidth]{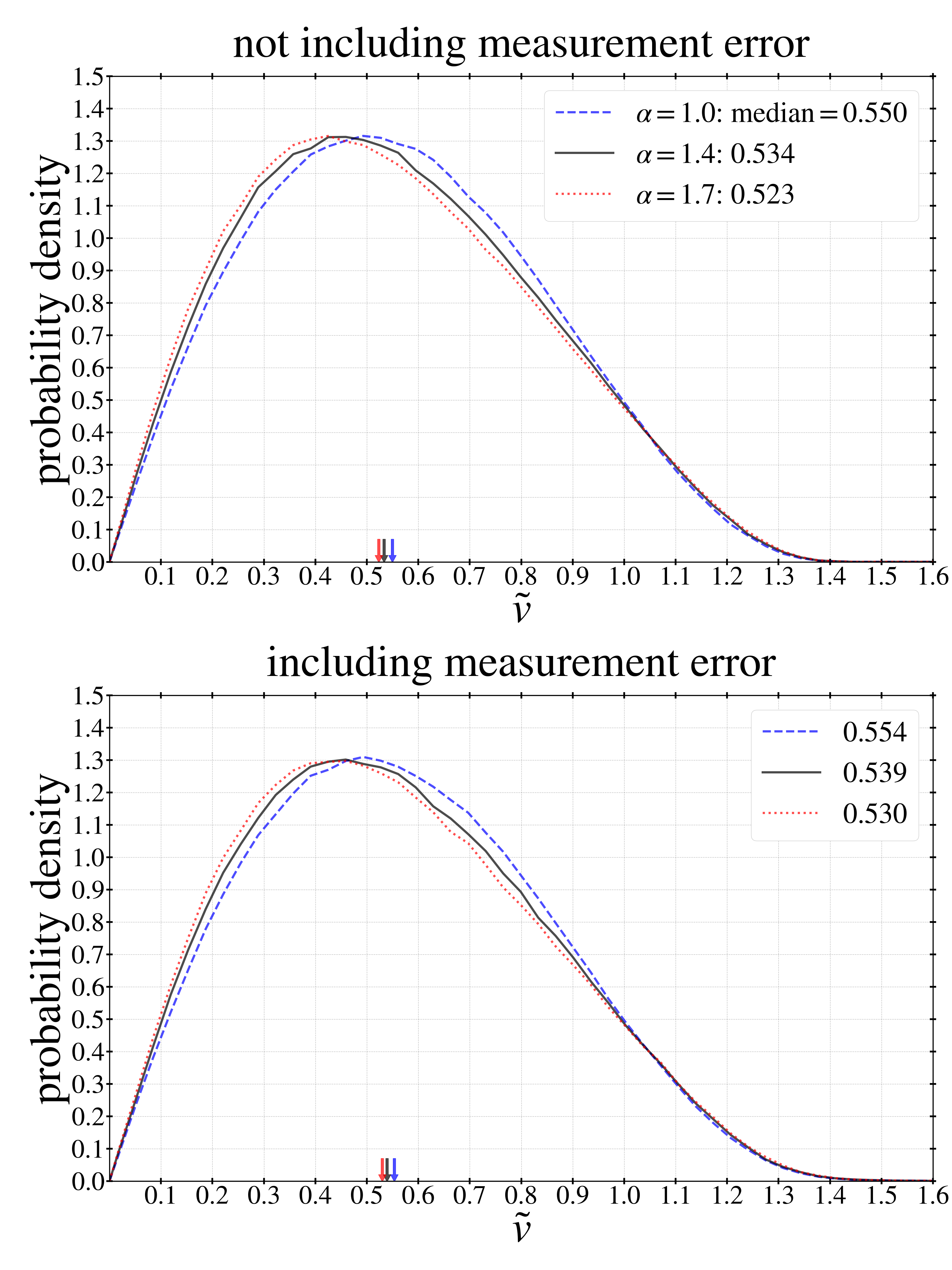}
    \caption{This figure shows the predicted distribution of $\tilde v$ (Equation~(\ref{eq:vtilde})) in Newtonian gravity at random phase and orientation of elliptical orbits whose eccentricities are sampled from the probability distribution given by Equation~(\ref{eq:powerlaw}). Three cases are shown with power-law index $\alpha=1.0,\,1.4,$ and $1.7$. The medians are indicated by arrows. The top panel shows results without any errors on $v_p$ or mass (so $v_c$), while the bottom panel shows the results including measurement errors on $v_p$ taken from the \cite{Banik:2024} sample as shown in left column of Figure~\ref{fig:vpvt_error_distribution} and an error of 6.8\% for mass. }
    \label{fig:vtdistribution_alpha}
\end{figure}

\begin{figure}[tbh!]
    \centering
   \includegraphics[width=1.0\linewidth]{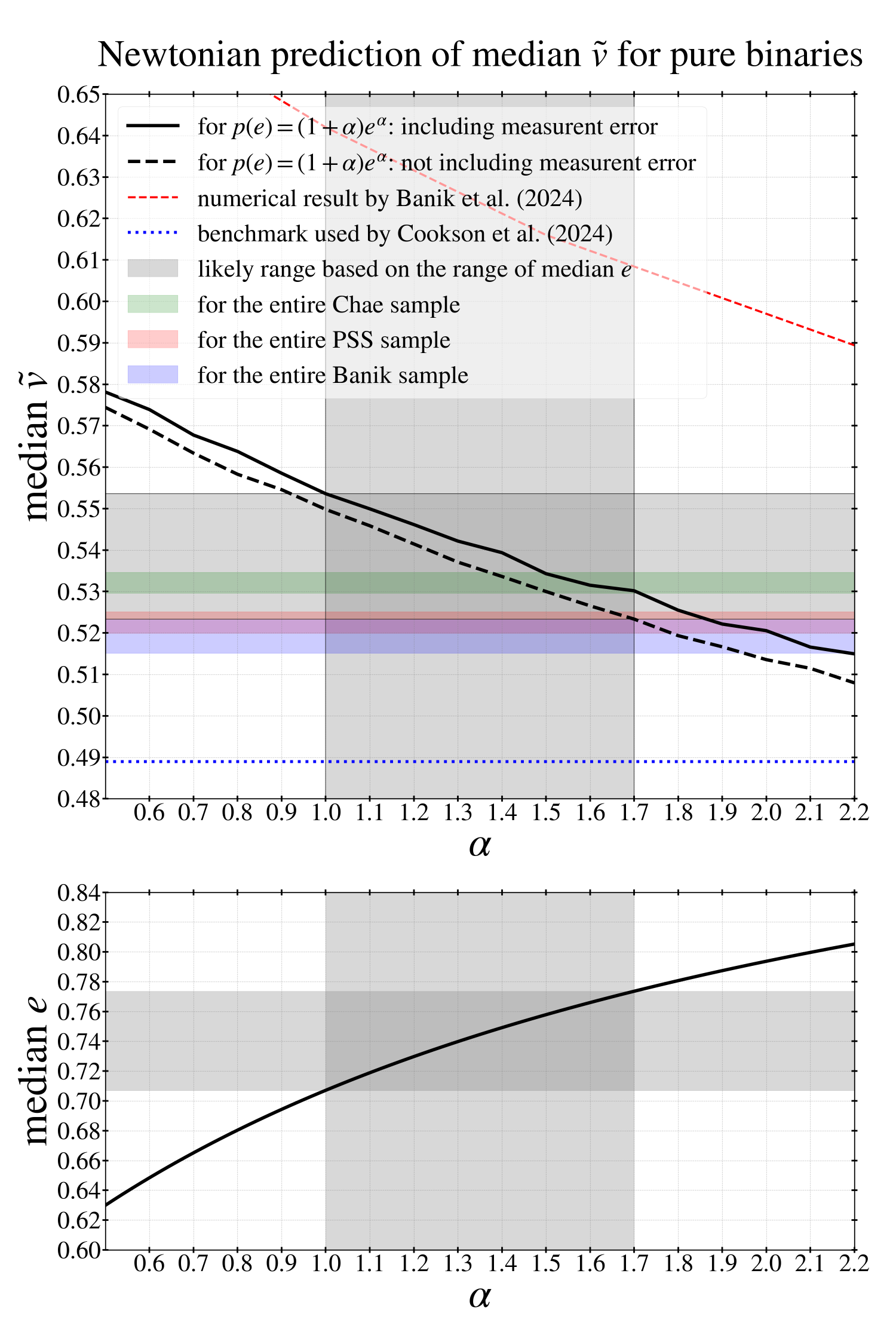}
    \caption{This upper panel shows how median $\tilde v$ varies with the exponent $\alpha$ of the power-law eccentricity distribution (Equation~(\ref{eq:powerlaw})) in Newtonian gravity. The black solid and dashed curves represent respectively the results with and without measurement errors as illustrated in Figure~\ref{fig:vtdistribution_alpha}. The gray band represents the likely range based on currently available empirical results on eccentricities shown in Figure~\ref{fig:eccentricity}. The erroneous numerical result by \cite{Banik:2024} (their Figure~15) is represented by the red dashed curve. It corresponds to a velocity boost factor of $\approx 1.17$. The biased Newtonian benchmark used by \cite{Cookson:2026} is represented by the blue dotted line. The colored bands indicate the ranges of median $\tilde v$ predicted for the three samples based on their individual eccentricities taken from \cite{Hwang:2022}. The medians for the three samples overlap with the gray band. The lower panel shows the scaling of median eccentricity with $\alpha$.}
    \label{fig:vtmedian_alpha}
\end{figure}

To investigate the Newtonian prediction on the $\tilde v$ distribution as a function of eccentricity distribution, we assume the widely used power-law probability distribution of $e$ given by
\begin{equation}
    p(e)=(1+\alpha)e^\alpha.
    \label{eq:powerlaw}
\end{equation}
Figure~\ref{fig:vtdistribution_alpha} shows numerical examples of the predicted $\tilde v$ distribution. The median of $\tilde v$ is a monotonically decreasing function of $\alpha$, and thus median $e$. When the measurement errors of $v_p$ and mass are taken into account, the distribution and median of $\tilde v$ will be affected given the definition of $\tilde v = v_p / v_c$. The effect will be more important when $v_p$ is smaller, i.e., $r_p$ is larger. Here, for illustration we use the entire \cite{Banik:2024} sample to estimate the statistical effect of the measurement errors of $v_p$ and $M_{\rm tot}$. For the uncertainty of $v_c$, we assume an uncertainty of 6.8\% in $M_{\rm tot}$ (see below). For the uncertainty of $v_p = \sqrt{v_{x^\prime}^2 + v_{y^\prime}^2}$, we add a scatter of $\sigma_{v_p}/\sqrt{2}$ to both $v_{x^\prime}$ and $v_{y^\prime}$ (Equation~(\ref{eq:vpcomps_Newton})). When the entire sample is used, the effect of measurement errors is minor because binaries of large $r_p$ are relatively few. In testing gravity with binaries of large $r_p$ through $\tilde v$ analyses, we will estimate the effect of measurement errors in the specific bin.

Figure~\ref{fig:vtmedian_alpha} shows $\langle\tilde v\rangle$ and $\langle e\rangle$ as functions of $\alpha$. The gray band corresponds to the likely range of $\langle e\rangle$ shown in Figure~\ref{fig:eccentricity}. The power-law index $\alpha$ has the range $1.0<\alpha<1.7$ which is somewhat superthermal and consistent with Figure~7 of \cite{Hwang:2022}. The likely range of median $\tilde v$ for $1.0<\alpha<1.7$ is given by $0.52\la \langle\tilde v\rangle \la 0.55$. Figure~\ref{fig:vtmedian_alpha} also shows the medians of $\tilde v$ predicted with individual eccentricities in the three samples. Here, individual eccentricities refer to the individual ranges reported by \cite{Hwang:2022} and they are implemented as described in Section~2.4 of \cite{Chae:2024b}. The medians for the three samples are on the lower side of the gray band.

We note that the numerical result shown in Figure~15 of \cite{Banik:2024} corresponds to a velocity boost factor of $\approx 1.17$ compared to the numerical result shown here. Specifically, \cite{Banik:2024} obtained $\langle\tilde v\rangle=0.64$ for $\alpha=1$ while the correct value is $0.55$, which has been (implicitly) shown multiple times in the literature (e.g., Figure~10 of \cite{Pittordis:2019}, Figure~3 of \cite{HernandezChae:2024}, and the right panel of Figure~13 of \cite{Chae:2024b}). If this error factor in \cite{Banik:2024} was present in all their calculations, what they called Newtonian gravity actually meant a gravity boosted by $\approx 1.17^2 \approx 1.37$. Interestingly, they concluded that their ``Newtonian gravity'' was preferred over their numerical representation of MOND gravity, which is also questionable in the acceleration range $10^{-10}\la g_{\rm N}\la 10^{-9}\,{\rm m}\,{\rm s}^{-2}$ (see Section~\ref{sec:mondgravity}). 

\cite{Cookson:2026} adopted a value of $0.489$ for their Newtonian benchmark (as indicated in their Figure~8). This value is out of the question for any conceivable distribution of eccentricities. We will discuss \cite{Cookson:2026} in more detail in Section~\ref{sec:nearby_WBs} where we analyze samples of wide binaries selected with a tight distance constraint.

\begin{figure*}[tbh!]
    \centering
   \includegraphics[width=0.95\linewidth]{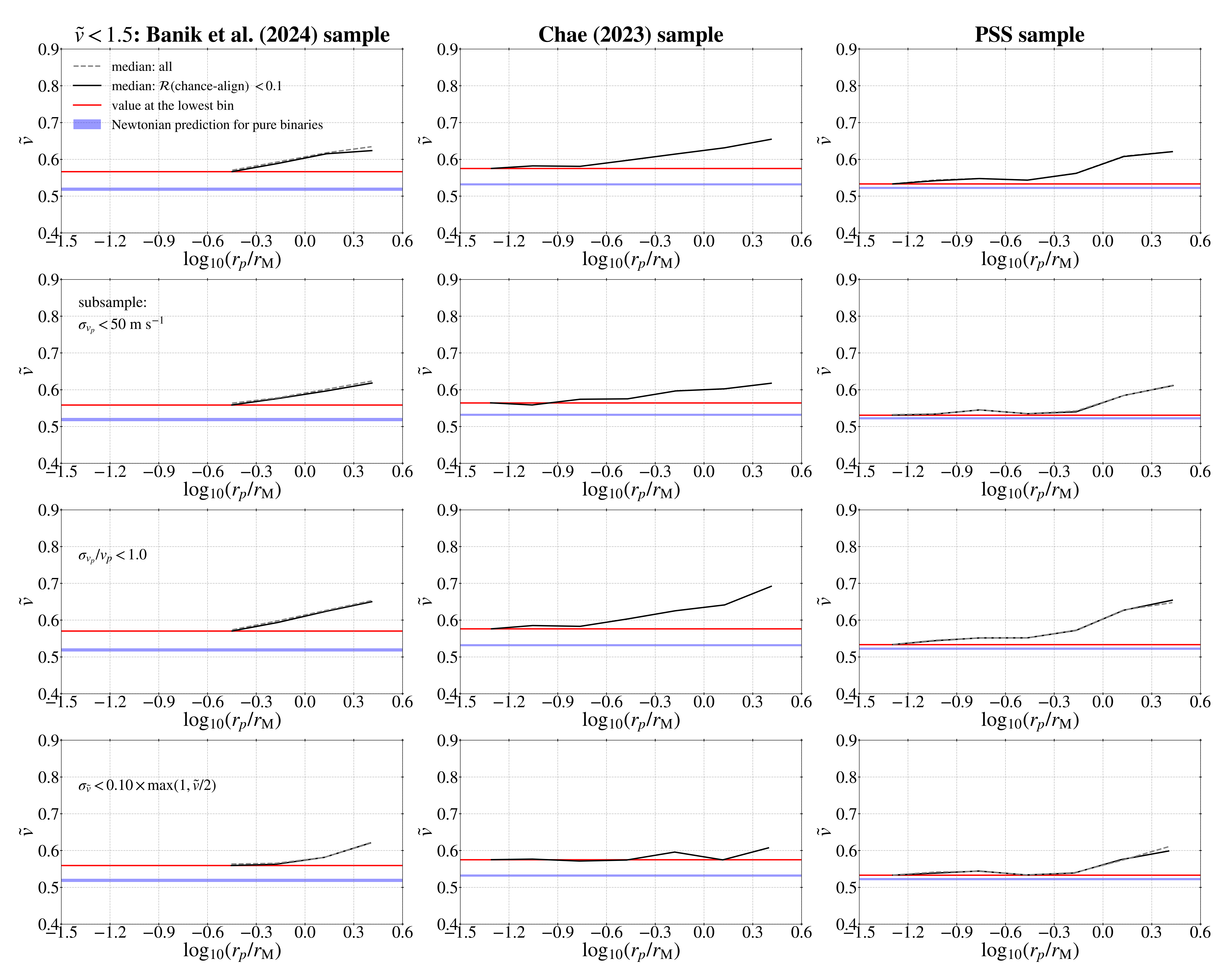}
    \caption{Same as Figure~\ref{fig:vt_scaling_3samples} but only for systems satisfying $\tilde v<1.5$. }
    \label{fig:vt_scaling_3samples_vtmax15}
\end{figure*}

\begin{figure*}[tbh!]
    \centering
   \includegraphics[width=0.95\linewidth]{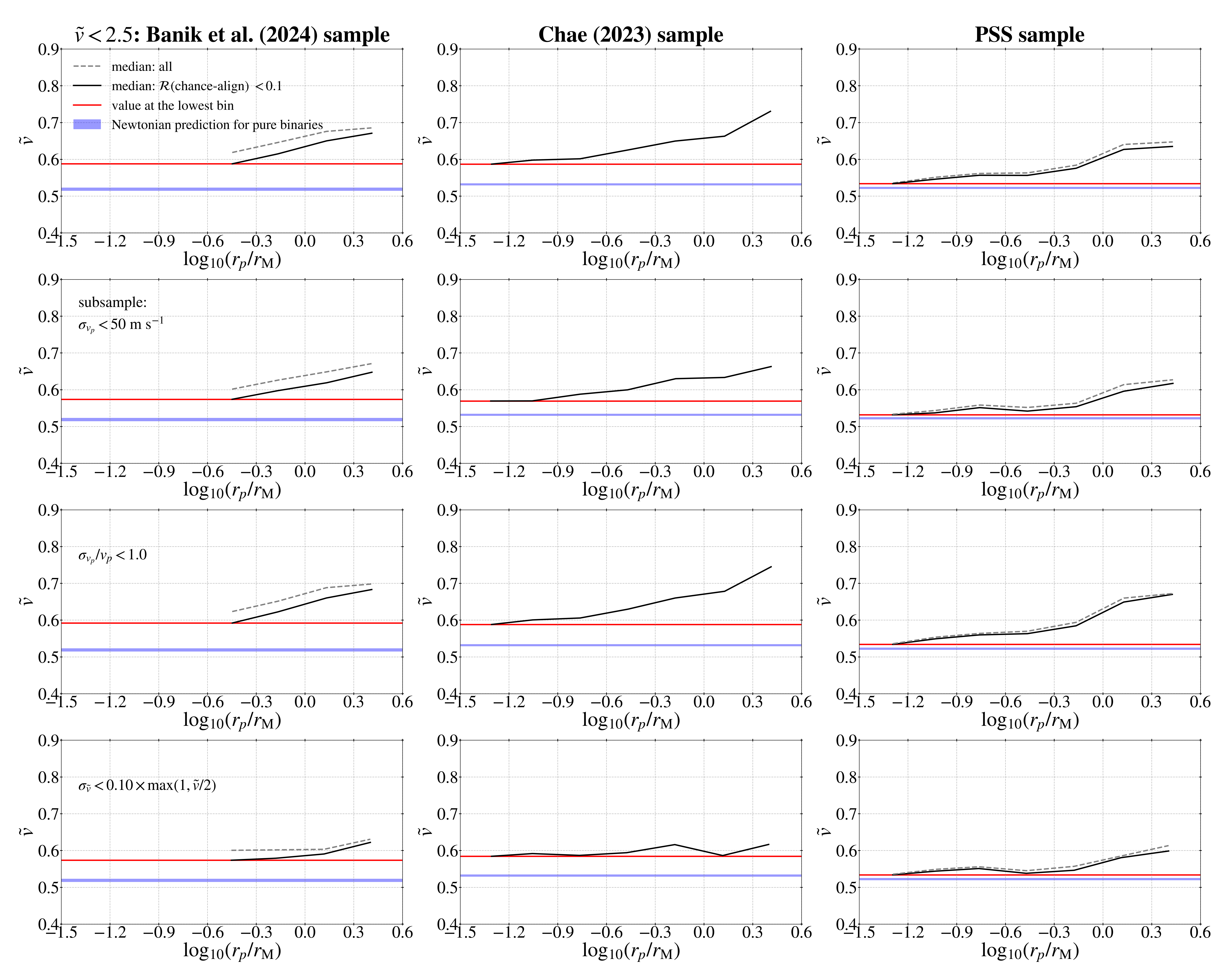}
    \caption{Same as Figure~\ref{fig:vt_scaling_3samples} but only for systems satisfying $\tilde v<2.5$.}
    \label{fig:vt_scaling_3samples_vtmax25}
\end{figure*} 

\begin{figure} 
    \centering
   \includegraphics[width=0.7\linewidth]{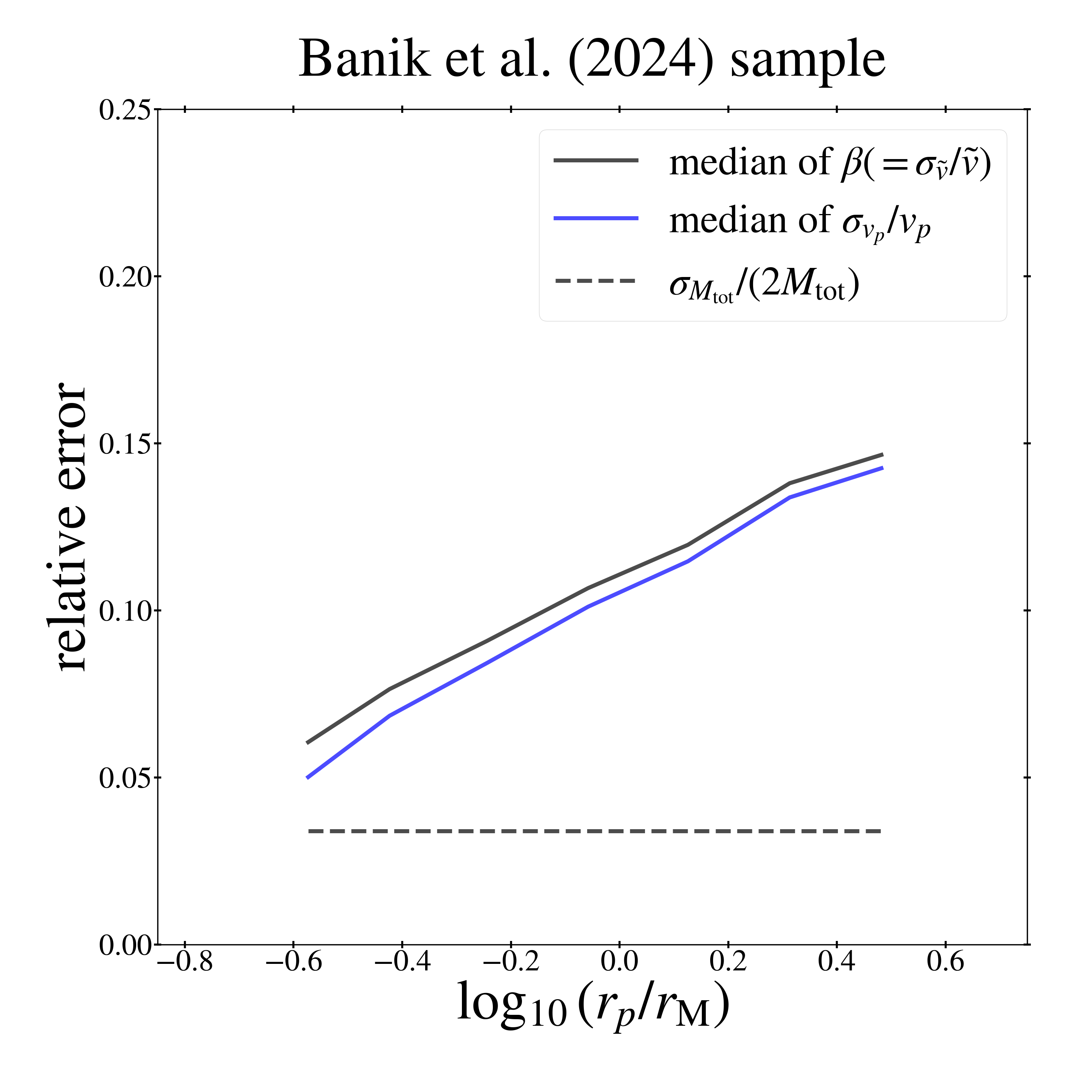}
    \caption{The black(blue) curve represents the scaling of the median relative error of $\tilde v$($v_p$) with respect to $r_p/r_{\rm M}$. The horizontal dashed line represents one half of the relative uncertainty of $M_{\rm tot}$ used for the Banik cut (see the text). }
    \label{fig:relerror}
\end{figure}
 
Because chance-alignment/flyby pairs are negligible in the lowest $r_p/r_{\rm M}$ bin of the PSS sample (Figure~\ref{fig:f_flyby}), the small difference between median $\tilde v$ and the Newtonian prediction for pure binaries indicates that the amount of hierarchical systems in the PSS sample will be small. We will show in Section~\ref{sec:test_acceleration} that this is indeed the case. 

The rows from the second to the bottom in Figure~\ref{fig:vt_scaling_3samples} show the results with various added cuts that are designed to further improve the quality of the data beyond the quality cuts used to define the samples. With $\sigma_{v_p}<50\,{\rm m}\,{\rm s}^{-1}$ (the second row), the median values of $\tilde v$ are overall lowered and the rising slopes are somewhat smaller, but the rising trends remain in all samples. For the \cite{Chae:2023} sample, the lowest-$r_p/r_{\rm M}$ bin with $\sigma_{v_p}<50\,{\rm m}\,{\rm s}^{-1}$ is now closer to the pure Newtonian line. With $\sigma_{v_p}/v_p<1$ (the third row), the overall trends of $\tilde v$ are similar to the cases of the top row with a minor indication that the rising slopes are slightly higher.

The bottom row of Figure~\ref{fig:vt_scaling_3samples} shows the results with the Banik cut. Unlike the cases with $\sigma_{v_p}<50\,{\rm m}\,{\rm s}^{-1}$ or $\sigma_{v_p}/v_p<1$, the rising trends of the median $\tilde v$ are significantly suppressed for gravitationally-bound binaries satisfying $\mathcal{R}<0.1$. Here it is particularly useful to compare the results with the Banik cut with those with $\sigma_{v_p}<50\,{\rm m}\,{\rm s}^{-1}$. The mock Newtonian binaries generated in Section~\ref{sec:WBTnature} show that the trend of median $\tilde v$ is not biased when $\sigma_{v_p}<50\,{\rm m}\,{\rm s}^{-1}$ is imposed, but the Banik cut can lower median $\tilde v$ in those binaries at large $r_p$ (e.g., $r_p/r_{\rm M}\ga 1$). These Banik cut-introduced biases are seen in all samples. The only difference between mock Newtonian binaries and real binaries is that the unbiased trend of median $\tilde v$ is flat (or shows a minor variation within the blue band) in pure Newtonian binaries while it is rising in real binaries. Finally, it is interesting to note that some rising trend still remains for $r_p/r_{\rm M}\ga 1$ with respect to the lowest-$r_p/r_{\rm M}$ bins in all three samples, indicating that the Banik cut does not completely remove the rising trend.

The rising profiles of $\tilde v$ visible in Figure~\ref{fig:vt_scaling_3samples} can be reliably interpreted only through detailed modeling taking into account hierarchical systems and chance-alignment pairs, which will be done in Section~\ref{sec:triple_model}. However, hierarchical systems and chance-alignment pairs can be understood qualitatively even without detailed modeling. Chance-alignment pairs can be controlled through two independent means, the \cite{El-badry:2021} $\mathcal{R}$ parameter and an upper bound on $\tilde v$. The effects of hierarchical systems can be controlled by considering samples selected with gradually more stringent criteria, as the three independent samples considered here. 

Thus, we repeat analyses of the profile of median $\tilde v$ for the three samples with upper bounds on $\tilde v$, for which we consider $\tilde v < 1.5$ and  $\tilde v < 2.5$. The cut $\tilde v < 1.5$ is expected to entirely remove chance-alignment pairs, but may also remove gravitationally-bound binaries with boosted gravity and/or hidden companions. Considering the scatter of $\tilde v$ arising from measurement errors of $v_p$ and $M_{\rm tot}$, the cut $\tilde v < 1.5$ may even remove some pure binaries at large $r_p$, i.e., in the low-acceleration regime. The cut $\tilde v < 2.5$ is less strict but may still effectively remove chance-alignment pairs while retaining the majority of gravitationally-bound binaries.

Figure~\ref{fig:vt_scaling_3samples_vtmax15} and Figure~\ref{fig:vt_scaling_3samples_vtmax25} show the $\tilde v$ profiles with the bounds $\tilde v < 1.5$ and $\tilde v < 2.5$, respectively. As expected, almost all systems with $\tilde v < 1.5$ (Figure~\ref{fig:vt_scaling_3samples_vtmax15}) in the \cite{Banik:2024} and PSS samples automatically satisfy low probability of chance-alignment/flyby $\mathcal{R}<0.1$. In all cases including the most reliable case of $\sigma_{v_p}<50\,{\rm m}\,{\rm s}^{-1}$ (the second row) the rising trend is clearly visible even with the strict requirement $\tilde v < 1.5$ for all samples. The lowest median of $\tilde v$ at $r_p/r_{\rm M}< 1$ is now closer to the pure Newtonian prediction in all three samples. Interestingly, the highest median (at $r_p/r_{\rm M}> 1$) with $\sigma_{v_p}<50\,{\rm m}\,{\rm s}^{-1}$ is similar to the Newtonian prediction multiplied by a MOND-inspired boost factor of $\approx 1.15-1.20$ in all three samples. 

With $\tilde v < 2.5$ (Figure~\ref{fig:vt_scaling_3samples_vtmax25}), the PSS sample appears to be almost free of chance-alignment pairs, while the \cite{Banik:2024} sample appears to include some fraction as indicated by the difference between all pairs and the pairs with $\mathcal{R}<0.1$. The profiles of median $\tilde v$ are intermediate between the case without any limit on $\tilde v$ and the case with $\tilde v < 1.5$.

\begin{figure*}[tbh!]
    \centering
   \includegraphics[width=1.0\textwidth]{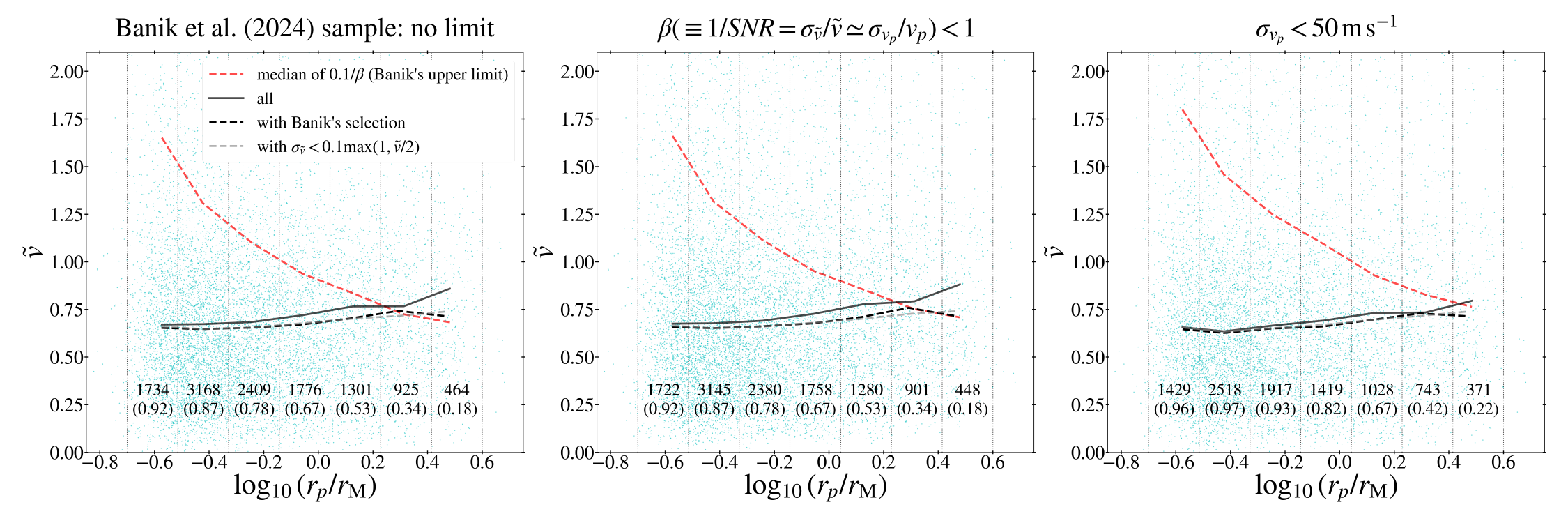}
    \caption{This figure exhibits the effect of the Banik cut on the scaling of $\tilde v$ with respect to $r_p/r_{\rm M}$. In each panel, the displayed curves are based on the medians in the 7 bins of $\log_{10}(r_p/r_{\rm M})$ indicated by the vertical gray dotted lines. Black solid curve is for the entire sample without applying the Banik cut. Black dashed curve is for the systems that are identified by \cite{Banik:2024} to satisfy their cut, while gray dashed curve is based on Equation~(\ref{eq:vtcut}) with $0.068$ for the relative uncertainty of the binary mass. The number given in each bin indicates the number of systems within the bin while the number in parentheses indicates the fraction satisfying the Banik cut (see the text for discussion on the dramatic decrease of the fraction with increasing $r_p/r_{\rm M}$). Red dashed curve shows the scaling of the median upper bound on $\tilde v$ implied by the Banik cut. The left panel is for the entire sample while the other panels are for the subsamples satisfying the quality cuts indicated in the panels. }
    \label{fig:vt_scaling_Banik}
\end{figure*}

\begin{figure}[tbh!]
    \centering
   \includegraphics[width=0.8\linewidth]{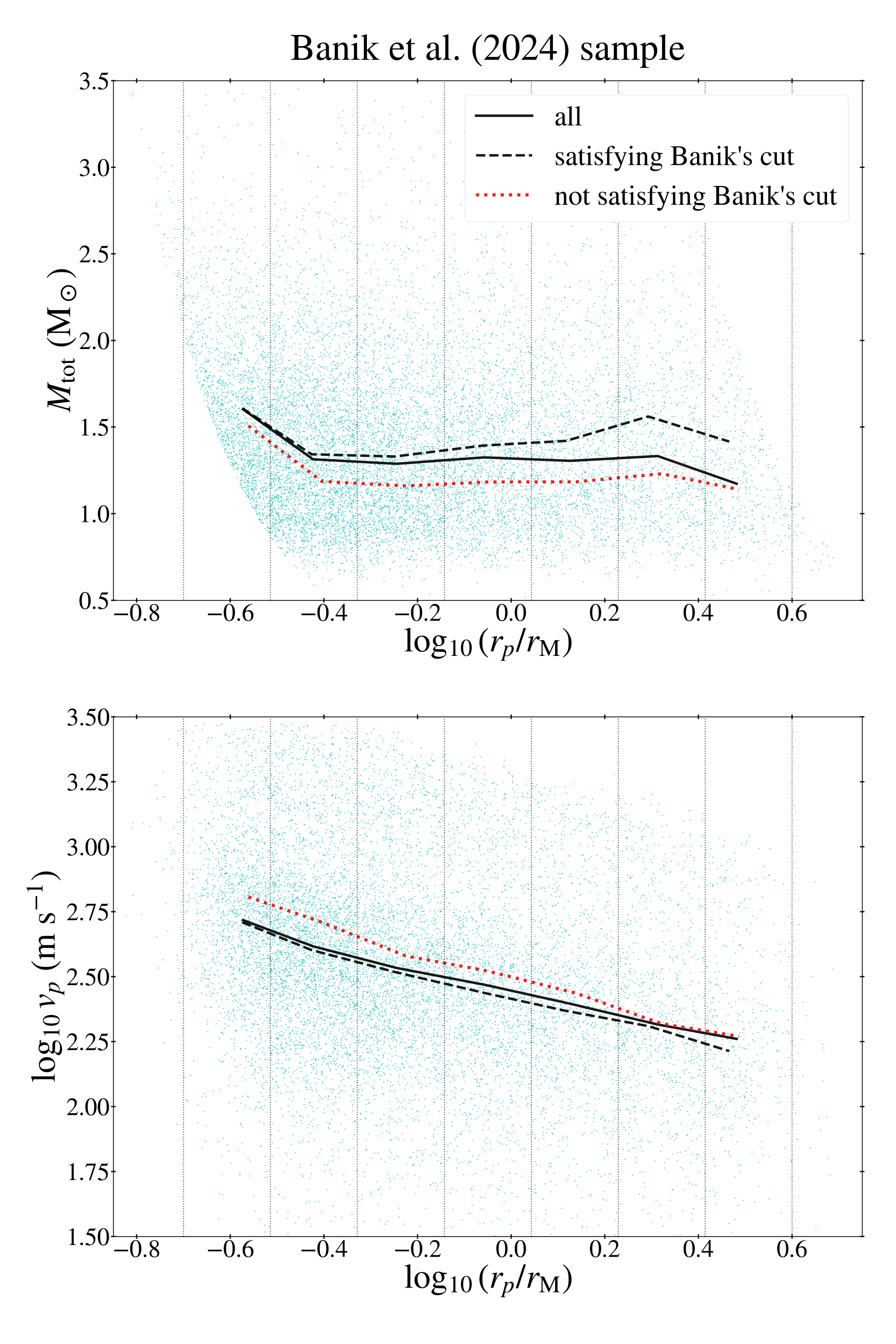}
    \caption{The upper panel shows the scaling of $M_{\rm tot}$ (total mass of the binary) for the \cite{Banik:2024} sample using the same bins of $\log_{10}(r_p/r_{\rm M})$ defined in Figure~\ref{fig:vt_scaling_Banik}. The lower panel is same for $\log_{10}v_p$.}
    \label{fig:mass_vp_Banik}
\end{figure}

We now look further into how the Banik cut can bias the data when it is naively imposed on a sample. As most data satisfy $\tilde v<2$, the Banik cut (Equation~(\ref{eq:vtcut})) becomes, to a good approximation, 
\begin{equation}
    \sigma_{\tilde v} < 0.1,
    \label{eq:vterrlt01}
\end{equation}
which can be rewritten as
\begin{equation}
    \tilde v < \frac{0.1}{\beta}
    \label{eq:vt01beta}
\end{equation}
with the definition of $\beta$ given by Equation~(\ref{eq:beta}).

Equation~(\ref{eq:vt01beta}) indicates that the Banik cut is equivalent to imposing an upper bound on $\tilde v$. However, as $\beta$ increases with $r_p/r_{\rm M}$ (Figure~\ref{fig:relerror}), the upper bound on $\tilde v$ decreases biasing the data. The relative error $\sigma_{v_p}/v_p$ (and thus $\beta$) increases with $r_p/r_{\rm M}$ because $v_p$ decreases with $r_p/r_{\rm M}$ while $\sigma_{v_p}$ is independent of $r_p/r_{\rm M}$.

Figure~\ref{fig:vt_scaling_Banik} shows how the upper bound on $\tilde v$ induced by the Banik cut depends on $r_p/r_{\rm M}$ using the \cite{Banik:2024} sample. The left panel exhibits the total sample. We find that the \cite{Banik:2024} selection of individual systems provided by the authors is statistically in agreement with Equation~(\ref{eq:vtcut}) when $\sigma_{M_{\rm tot}}/M_{\rm tot}=0.068$. This is the reason why we have been using the value throughout. The median upper bound indicated by the red dashed curve dramatically decreases as $r_p/r_{\rm M}$ increases. In particular, in the MOND regime with $r_p/r_{\rm M} \ga 1$, the median upper bound is well below unity, biasing the intrinsic distribution of $\tilde v$. We consider logarithmic bins of $r_p/r_{\rm M}$ and indicate the number of systems as well as the fraction selected by the Banik cut in each bin. The fraction decreases dramatically with $r_p/r_{\rm M}$. In the highest-$r_p/r_{\rm M}$ bin, the fraction meeting the Banik cut is just 18\%. 

The middle and right panels of Figure~\ref{fig:vt_scaling_Banik} exhibit results for subsamples with the requirement $\beta < 1$ or $\sigma_{v_p}<50\,{\rm m}\,{\rm s}^{-1}$. The subsample with $\beta < 1$ is similar to the total sample because almost all data already satisfy $S/R>1$. The subsample with $\sigma_{v_p}<50\,{\rm m}\,{\rm s}^{-1}$ is of particular interest because the median $\tilde v$ profile is not expected to be biased. In this subsample, the effect of the Banik cut is less severe but still present. In detail, for the three lowest-$r_p/r_{\rm M}$ bins, the fractions are above 93\% and the bias is almost absent. However, the bias is present in bins with higher $r_p/r_{\rm M}$, with the overall trend of the bias increasing with $r_p/r_{\rm M}$. 

The effects of the Banik cut on $\tilde v$ can also be seen from the profiles of $M_{\rm tot}$ and $v_p$ as shown in Figure~\ref{fig:mass_vp_Banik}. As the top panel of Figure~\ref{fig:mass_vp_Banik} shows, the median mass is flat from the bin centered at $\log_{10}(r_p/r_{\rm M})\approx -0.4$ to the bin at $\approx 0.3$ for the original sample without the Banik cut, but it rises steeply for systems satisfying the Banik cut because the cut selectively removes systems with lower masses, as simulations performed in Section~\ref{sec:WBTnature} show and \cite{Chae:2024b} noticed. However, as the bottom panel shows, the Banik cut selectively removes systems with large values of $v_p$ making the subsample satisfying the Banik cut has a larger median mass but a lower $v_p$ at a given $r_p$. Thus, a sample satisfying the Banik cut will have a statistically lowered median of $v_p$ for a statistically boosted median of mass. Because a statistical gravity test is performed essentially by comparing the observed median of $v_p$ with the predicted median at a given median mass, a sample with biased medians of $v_p$ and $M_{\rm tot}$ will return a biased result on gravity.

\begin{figure}[tb!]
    \centering
   \includegraphics[width=1.0\linewidth]{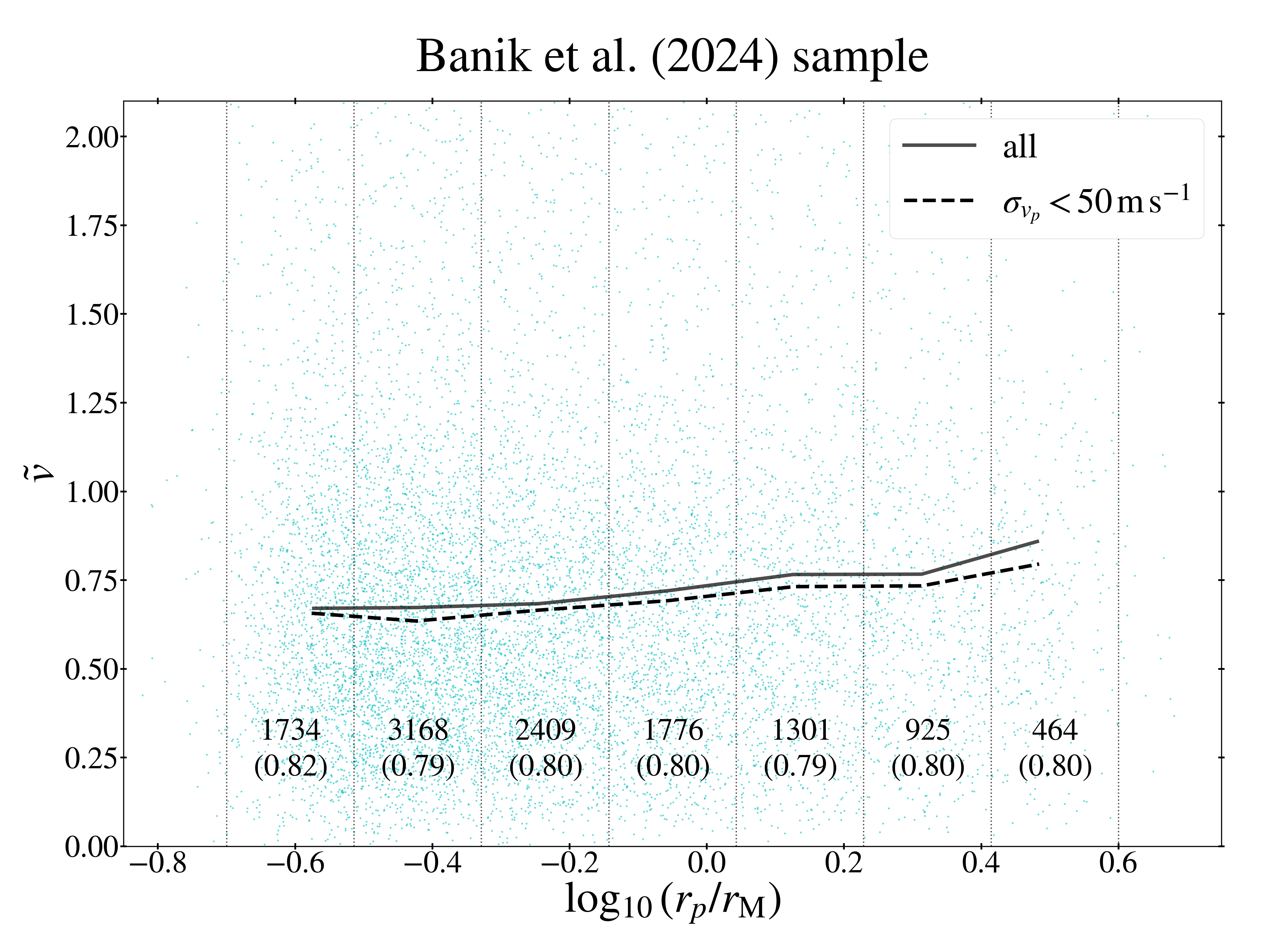}
    \caption{This figure shows the effect of the cut $\sigma_{v_p}<50\,{\rm m}\,{\rm s}^{-1}$ on the scaling of $\tilde v$ using the same bins of $\log_{10}(r_p/r_{\rm M})$ defined in Figure~\ref{fig:vt_scaling_Banik}. Unlike the Banik cut, the fractions satisfying this cut are nearly the same in all the bins. }
    \label{fig:vt_vperr50}
\end{figure}

Unlike the Banik cut, the cut $\sigma_{v_p}<50\,{\rm m}\,{\rm s}^{-1}$ removes less precise data in an overall way and does not introduce any bias in mass or $v_p$. This is because the measurement uncertainty of $v_p$ (the relative velocity between the pair measured from Gaia's high-precision PMs) is largely independent of the binary mass and separation. Figure~\ref{fig:vt_vperr50} shows that the cut $\sigma_{v_p}<50\,{\rm m}\,{\rm s}^{-1}$ removes a uniform fraction of about 20\% in all bins of $\log_{10}(r_p/r_{\rm M})$. 

We note that the subsample satisfying $\sigma_{v_p}<50\,{\rm m}\,{\rm s}^{-1}$ has somewhat lowered medians of $\tilde v$ in all bins compared to the original sample. This can be understood as follows. When an apparent binary has a kinematic contaminant (e.g., a hidden close companion star), $v_p$ will be boosted and its measurement uncertainty will be larger. Removing systems with measurement uncertainties larger than $50\,{\rm m}\,{\rm s}^{-1}$ will selectively remove systems with larger $v_p$ in a given bin of $r_p$, and thus the subsample with $\sigma_{v_p}<50\,{\rm m}\,{\rm s}^{-1}$ will have a lower median of $v_p$ in the same bin.

The lesson we get from the above comparison between the Banik cut and the cut $\sigma_{v_p}<50\,{\rm m}\,{\rm s}^{-1}$ is that any quality control of data should not bias the data with respect to a gravity-dependent parameter such as $r_p/r_{\rm M}$ but must be done in an overall and unbiased way. This was already pointed out in \cite{Chae:2024b}, and the cut $\sigma_{v_p}<50\,{\rm m}\,{\rm s}^{-1}$ along with ${\tt ruwe} < 1.2$ was used in \cite{Yoonetal:2025}. The useful unbiased quality control ${\tt ruwe} < 1.2$ was also adopted by \cite{Pittordis:2025}. We note that recent studies including \cite{Pittordis:2025} and \cite{Yoonetal:2025} did not adopt the Banik cut. However, \cite{Cookson:2026} advocated adopting the Banik cut. We will address issues with \cite{Cookson:2026} in Section~\ref{sec:nearby_WBs} where we analyze samples with a tight distance limit.

\begin{figure}[tbh!]
    \centering
   \includegraphics[width=1.0\linewidth]{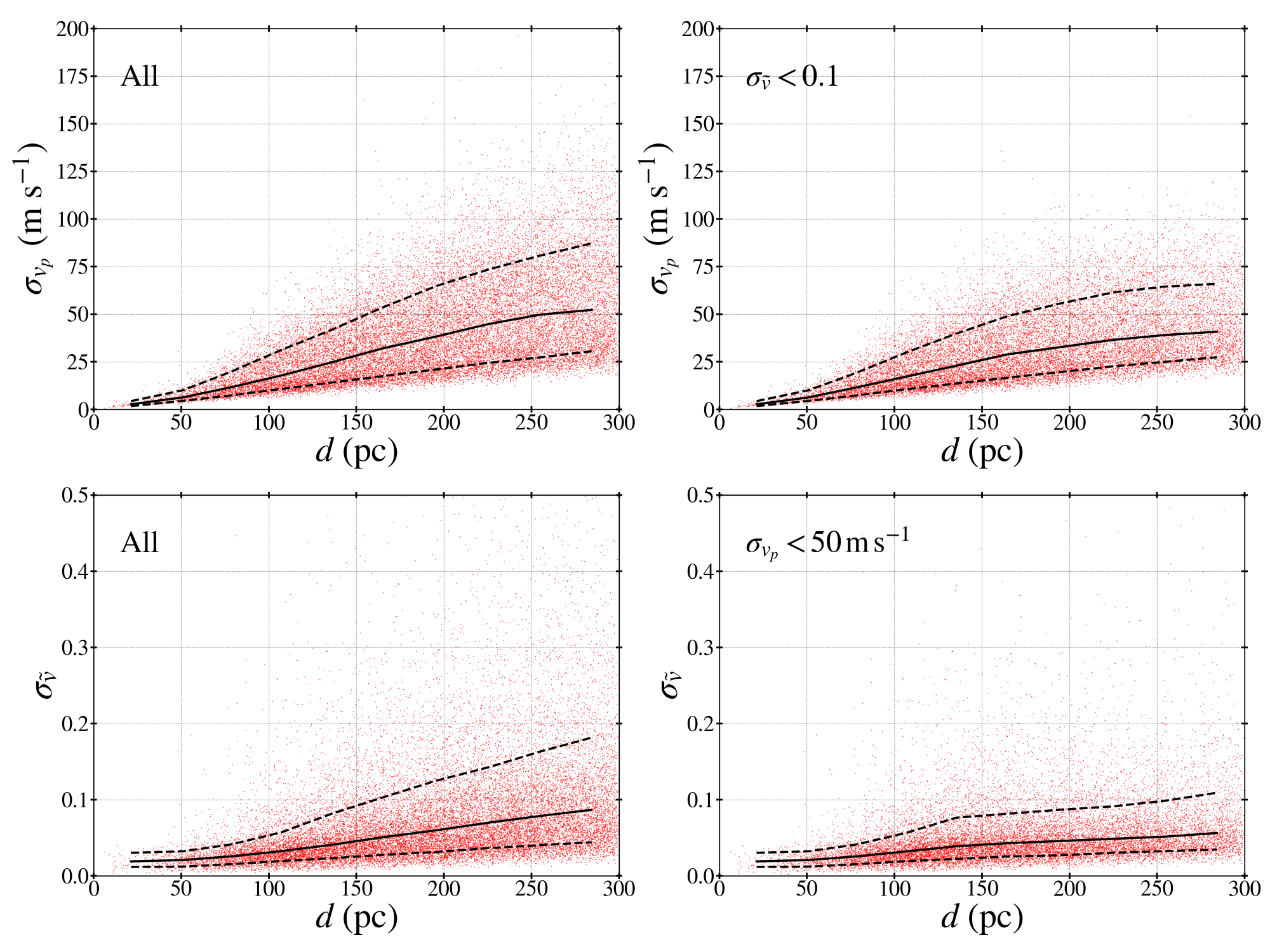}
    \caption{This upper left panel shows the distribution of $\sigma_{v_p}$ (the measurement error of $v_p$) as a function of distance ($d$) from the Sun using the PSS sample, while the upper right panel shows a subsample with $\sigma_{\tilde v}<0.1$. The lower left panel shows $\sigma_{\tilde v}$ for the PSS sample while the lower right panel shows a subsample with $\sigma_{v_p}<50\,{\rm m}\,{\rm s}^{-1}$. The solid curve shows the scaling of the median with $d$ while the dashed curves show the 16th and 84th percentiles. }
    \label{fig:vperr_vterr_distance}
\end{figure}

We note that the Banik cut can be statistically satisfied in an unbiased manner through various cuts. One such sample was constructed by \cite{Chae:2024a} by requiring relatively precise RVs for both components. Here we find that the quality cut $\sigma_{v_p}<50\,{\rm m}\,{\rm s}^{-1}$ or a distance limit (along with some basic quality cuts) can also statistically satisfy the Banik cut, as can be seen in Figure~\ref{fig:vperr_vterr_distance}. The distance limit $d<150\,{\rm pc}$ satisfies both $\sigma_{v_p}<50\,{\rm m}\,{\rm s}^{-1}$ and $\sigma_{\tilde v}<0.1$. (In passing, we note that the good quality of nearby data led \cite{Chae:2023} to define a benchmark sample with $d<80\,{\rm pc}$ and used it to define a general sample.) The cut $\sigma_{v_p}<50\,{\rm m}\,{\rm s}^{-1}$ satisfies $\sigma_{\tilde v}<0.1$ up to about 300\,pc, but the vice versa is not true. This is another point that the $\sigma_{v_p}<50\,{\rm m}\,{\rm s}^{-1}$ is better than $\sigma_{\tilde v}<0.1$.

\begin{figure*}[tbh!]
    \centering
   \includegraphics[width=0.95\linewidth]{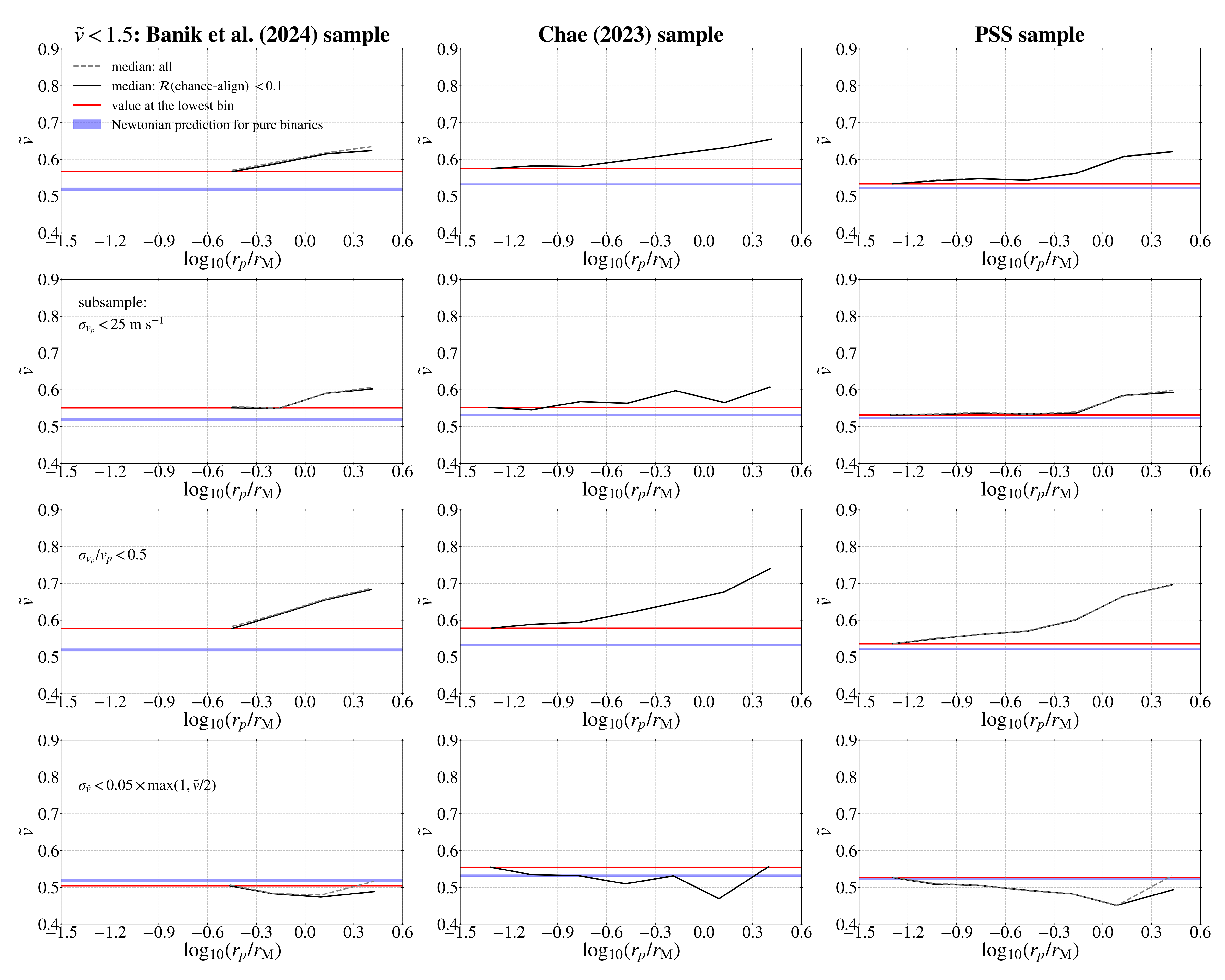}
    \caption{Same as Figure~\ref{fig:vt_scaling_3samples_vtmax15} but with tighter cuts in defining the subsamples.}
    \label{fig:vt_scaling_3samples_vtmax15_tighter}
\end{figure*}

The nature of the quality cut based on $\sigma_{v_p}$, $\sigma_{v_p}/v_p$, or $\sigma_{\tilde v}$ can be further investigated by investigating the effect of tightening the cut. If a quality or precision cut is correct and unbiased, better or improved quality/precision may be desired as long as the statistical power is not sacrificed significantly. Thus, we investigate the effect of tightening each cut by a factor of two. Figure~\ref{fig:vt_scaling_3samples_vtmax15_tighter} shows the results on the median $\tilde v$ with the tighter cuts for the samples with the limit $\tilde v<1.5$. Comparison of this figure with  Figure~\ref{fig:vt_scaling_3samples_vtmax15} shows that the trend of $\tilde v$ with the tighter cut $\sigma_{v_p}<25\,{\rm m}\,{\rm s}^{-1}$ is similar to that with the cut $\sigma_{v_p}<50\,{\rm m}\,{\rm s}^{-1}$. 

However, the tighter cut on $\sigma_{\tilde v}$ dramatically changes the trend of $\tilde v$. As the cut on $\sigma_{\tilde v}$ is tightened, $\tilde v$ gets more biased and it now clearly declines with $r_p/r_{\rm M}$. This confirms the biased nature of the cut. The tighter cut on $\sigma_{v_p}/v_p$ also biases $\tilde v$ as $\tilde v$ increases more steeply. This can be understood by the fact that a tight cut will selectively remove lower $v_p$ at a given $\sigma_{v_p}$. Thus, a tight cut based on $\sigma_{v_p}/v_p$ can bias $\tilde v$ in the opposite direction. 

As \cite{Chae:2024b} showed, even with the Banik cut or something similar, the gravity boost cannot be completely removed and can still be quantified if the corresponding Newtonian prediction is properly calculated and compared with the data. The estimated gravity boost from the Banik cut-imposed sample is somewhat weakened, as demonstrated by \cite{Chae:2024b}. This can be understood as follows. Both $\tilde v$ of the observed sample and that of the mock Newtonian control sample are subject to Equation~(\ref{eq:vterrlt01}), or Equation~(\ref{eq:vt01beta}). However, the Newtonian sample is less affected than the observed sample, as $\tilde v$ of the former is generally smaller than that of the latter, which means that more of the Newtonian sample satisfy Equation~(\ref{eq:vt01beta}) than the observed sample.

\section{Numerical Predictions of Classical MOND Gravity} \label{sec:mondgravity}

The Newtonian prediction on $\tilde v$ presented in the previous section (see Figures~\ref{fig:vtdistribution_alpha} and \ref{fig:vtmedian_alpha}) serves as the theoretical benchmark with respect to which wide binary data are compared. Since the observed scalings of $\tilde v$ with $r_p/r_{\rm M}$ (e.g., Figure~\ref{fig:vt_scaling_3samples_vtmax15}) indicate deviations from the Newtonian benchmark, it is warranted to consider nonstandard gravity models beyond the Newtonian benchmark. Here we summarize recent numerical results on the prediction of classical MOND gravity on wide binary kinematics that will be compared with the observed deviations through statistical modeling in the following sections. 

\begin{figure}[tbh!]
    \centering
    \includegraphics[width=0.9\linewidth]{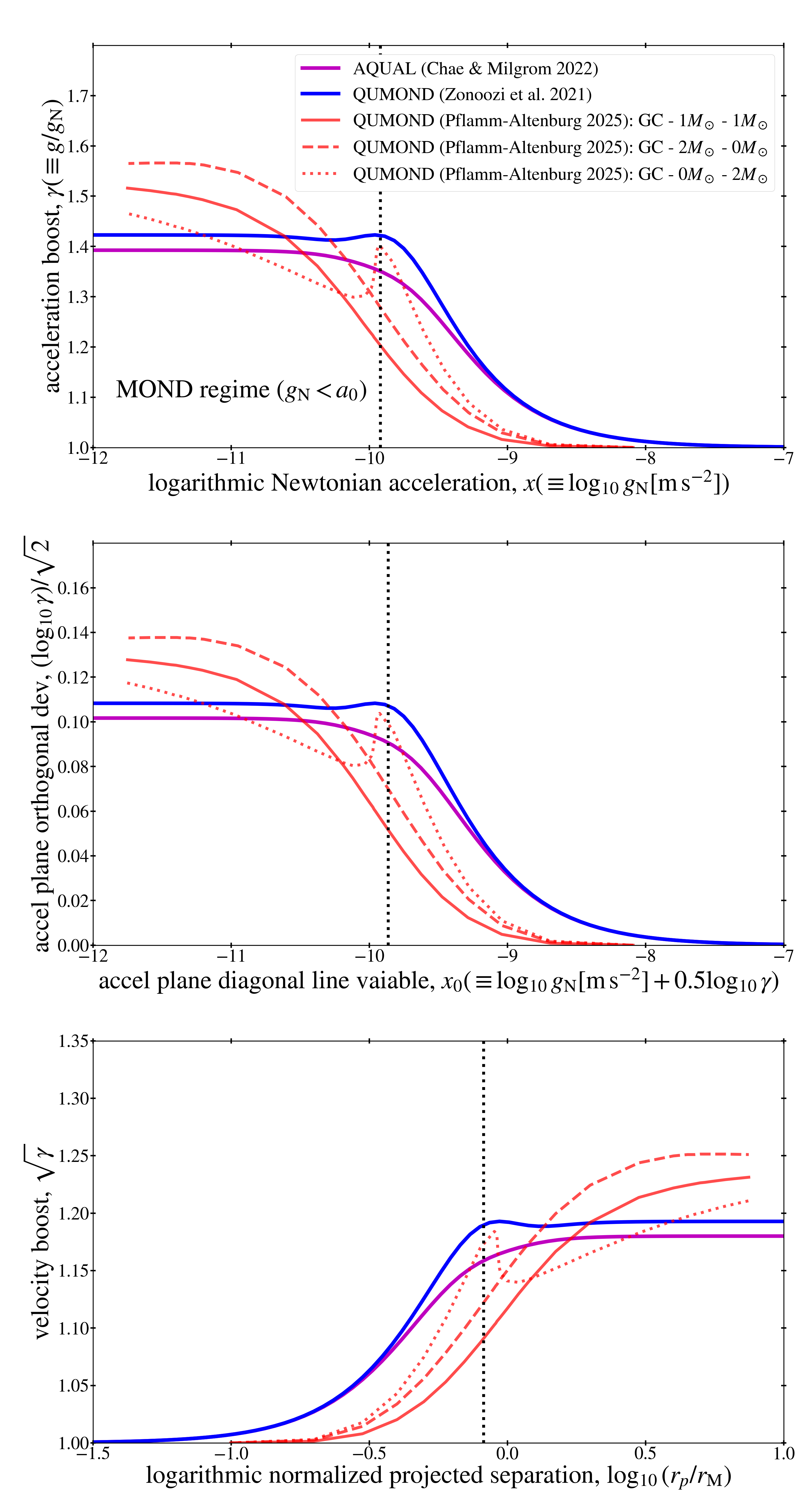}
    \caption{Various numerical predictions of AQUAL and QUMOND on wide binary kinematics are shown. The numerical results by \cite{ChaeMilgrom:2022} and \cite{Zonoozi:2021} are based on the assumption that two-body dynamics in MOND gravity can be treated as test-particle dynamics. The numerical results by \cite{Pflamm-Altenburg:2025} are based on actual orbit solutions for two-body dynamics in QUMOND: three configurations for the Galactic center (GC), one star, and the other star are shown. The configuration GC-$1M_\odot$-$1M_\odot$ is the one that is applicable to observed wide binaries. See the text for further details. Top panel shows the acceleration boost factor $\gamma(\equiv g/g_{\rm N})$ as a function of $x(\equiv \log_{10}g_{\rm N}[{\rm m}\,{\rm s}^{-1}])$. Middle panel shows the orthogonal deviation as a function of $x_0(\equiv x +0.5 \log_{10}\gamma)$ in the logarithmic acceleration plane introduced by \cite{Chae:2023}. Bottom panel shows the prediction on velocity boost, i.e., $\sqrt{\gamma}$, as a function of $\log_{10}(r_p/r_{\rm M})$ where $r_p$ and $r_{\rm M}$ are sky-projected separation and the MOND radius, and $\log_{10}(g_{\rm N}/a_0)\approx-2\log_{10}(1.22r_p/r_{\rm M})$.}
    \label{fig:theory_prediction}
\end{figure}

In the literature, two classical MOND models of nonrelativistic gravity, AQUAL \citep{BekensteinMilgrom:1984} and QUMOND \citep{Milgrom:2010}, have been used to make numerical predictions on the kinematic properties of wide binary stars orbiting each other under the external field of the Galaxy, i.e., taking into account a strong external field effect (EFE). The conventional approach was to derive solutions for a (massless) test particle around a point mass under a constant external field. For example, \cite{Zonoozi:2021} presented a functional form for $\gamma(g_{\rm N})$ based on the ring library for QUMOND obtained by \cite{Banik:2018} assuming that the external field is aligned with the orbit axis, while \cite{ChaeMilgrom:2022} presented an azimuthally-averaged result (see their Figure~5) at the typical tilt angle of $60^\circ$ for AQUAL. While these results are more realistic than analytical approximations, they may not be fully realistic because kinematic properties of comparable-mass stars orbiting each other under MOND or other nonlinear gravities may not be equivalently described by a test particle as in the Newtonian case. 

To deal with the limitation of the test-particle description of orbital motions, \cite{Pflamm-Altenburg:2025} obtained realistic orbital solutions for the two-body dynamics under QUMOND. Representing the Galaxy as a point mass at the Galactic center (GC), \cite{Pflamm-Altenburg:2025} derived the properties of the relative motion of the two stars in various configurations for the three bodies, i.e.\, GC, star1, and star2. \cite{Pflamm-Altenburg:2025} found that the configuration GC-$2M_\odot$-$0M_\odot$ or GC-$0M_\odot$-$2M_\odot$ (corresponding to the test-particle motion) returned results different from the cases of GC-$1M_\odot$-$1M_\odot$, GC-$1.5M_\odot$-$0.5M_\odot$, and GC-$0.5M_\odot$-$1.5M_\odot$. Because the observed mass ratio is in the range $0.3\la M_{\rm secondary}/M_{\rm primary} \le 1$ (see, e.g., Figure~3 of \cite{Chae:2024a}), the latter cases cover the likely cases. Since it turns out that the results for GC-$1.5M_\odot$-$0.5M_\odot$ and GC-$0.5M_\odot$-$1.5M_\odot$ are similar to that for GC-$1M_\odot$-$1M_\odot$, we will consider only the latter case.

Figure~\ref{fig:theory_prediction} summarizes the numerical results presented by \cite{Zonoozi:2021}, \cite{ChaeMilgrom:2022}, and \cite{Pflamm-Altenburg:2025}. All these results are based on the simple transition (or interpolating) function \citep{Famaey:2005} or the function by \cite{McGaugh:2008}, which are broadly similar to each other and are supported by rotationally-supported galaxies \citep{McGaugh:2016} and pressure-supported galaxies \citep{Chae:2019}. In the low-acceleration limit ($< 10^{-11}\,{\rm m}\,{\rm s}^{-2}$), the results by \cite{Zonoozi:2021} and \cite{ChaeMilgrom:2022} predict $\gamma\approx 1.4$ while that by \cite{Pflamm-Altenburg:2025} predicts $\gamma\approx 1.5$. Thus, the difference between the two-body solution and the test-particle case is minor in the low-acceleration regime. However, there is a significant difference between the two cases in the transition regime ($10^{-10}\la g_{\rm N} \la 10^{-9}\,{\rm m}\,{\rm s}^{-2}$). For the same QUMOND gravity, the test-particle case by \cite{Zonoozi:2021} significantly overpredicts $\gamma$ compared with the two-body solution by \cite{Pflamm-Altenburg:2025}. In this work, we will use the results by \cite{Pflamm-Altenburg:2025} and \cite{ChaeMilgrom:2022} to test classical MOND gravity.

\section{Testing Gravity with the PSS triple model} \label{sec:triple_model}

Testing gravity with a general sample of wide binaries/pairs requires proper modeling of, or taking into account, the two key unknowns, namely, hierarchical systems (i.e., triples and higher-order multiples) and chance-alignment/flyby systems that are observed as apparent pairs/binaries. We will deal with chance-alignment/flyby pairs based on the \cite{El-badry:2021} parameter $\mathcal{R}$ or an upper limit on $\tilde v$ (see Figure~\ref{fig:f_flyby}). As for hierarchical systems, \cite{Chae:2023} introduced an algorithm to model triples and quadruples in general samples, which have been used for various samples so far. More recently, \cite{Pittordis:2025} suggested a new numerical procedure to ``realistically'' model triples in a sample satisfying certain selection criteria including ${\tt ruwe}<1.2$, and presented numerical models of triples without considering quadruples. 

The new triple model by \cite{Pittordis:2025} is interesting because it is applicable to a sample satisfying well-defined criteria. Here we devise an algorithm to accommodate the numerical results on triples by \cite{Pittordis:2025}, and carry out new statistical analyses of the wide binary samples (defined and investigated in the previous sections) with the algorithm. Then, we compare our results with the gravity test results by \cite{Pittordis:2025}. We introduce the following consistent notation to represent the occurrence rate of hierarchical systems. We use $f_{\rm trip}$ to represent the fraction of triples among apparent binaries if all hierarchical systems are assumed to be triples as in \cite{Pittordis:2025}. The symbol $f_{\rm multi}$ (introduced in \citealt{Chae:2023}) is reserved for the fraction that includes both triples and quadruples.

\subsection{An effective algorithm for the PSS triple model} \label{sec:effectivePSS}

The statistical kinematic effect of a hidden companion can be modeled through a well-developed procedure as in \cite{Chae:2023} and \cite{Pittordis:2025} because the hidden companion is in a fully Newtonian regime. The statistical effect depends on the details of the statistical properties of the hidden companion, such as the probability distributions of companion-to-host mass ratio $M_c/M_h$ and physical separation between the host and the companion $r_{hc}$, which is on average similar to the semi-major axis of the inner orbit $a_{\rm inn}$. 

The kinematic effect is particularly sensitive to $r_{hc}$ (and thus $a_{\rm inn}$). This can be understood as follows. The kinematic effect for a host star of mass $M_h$ is proportional to $\sqrt{M_h(1 + M_c/M_h)/r_{hc}}$. The factor $1 + M_c/M_h$ varies between 1 and 2, but the factor $r_{hc}$ can vary from sub au to hundreds of au. A further important complication is that the kinematic effect is proportional to the offset between the barycenter and the photocenter of the two stars. If the offset is negligible or averaged out over the observation period, the net kinematic effect is negligible. Let the offset of the photocenter from the barycenter normalized by $r_{hc}$ be $\eta_{\rm phot}$ (or $f_{pb}$ according to \cite{Pittordis:2025}). Depending on whether the observed image of the $M_h+M_c$ inner system represents the combined light (the two are unresolved) or just $M_h$ (the two are resolved, but $M_c$ is too faint to be detected), the factor $\eta_{\rm phot}$ is given by
\begin{equation}
   \eta_{\rm phot} = \left\{ \begin{array}{l}
         \frac{M_c}{M_h + M_c} -  \frac{L_c}{L_h + L_c}\text{ (if unresolved)} \\
         \frac{M_c}{M_h + M_c} \text{ (if resolved)}
    \end{array} \right. ,
    \label{eq:etaphot}
\end{equation}
where $L_h$ and $L_c$ are the luminosities.

\begin{figure*}[tbh!]
    \centering
   \includegraphics[width=0.8\linewidth]{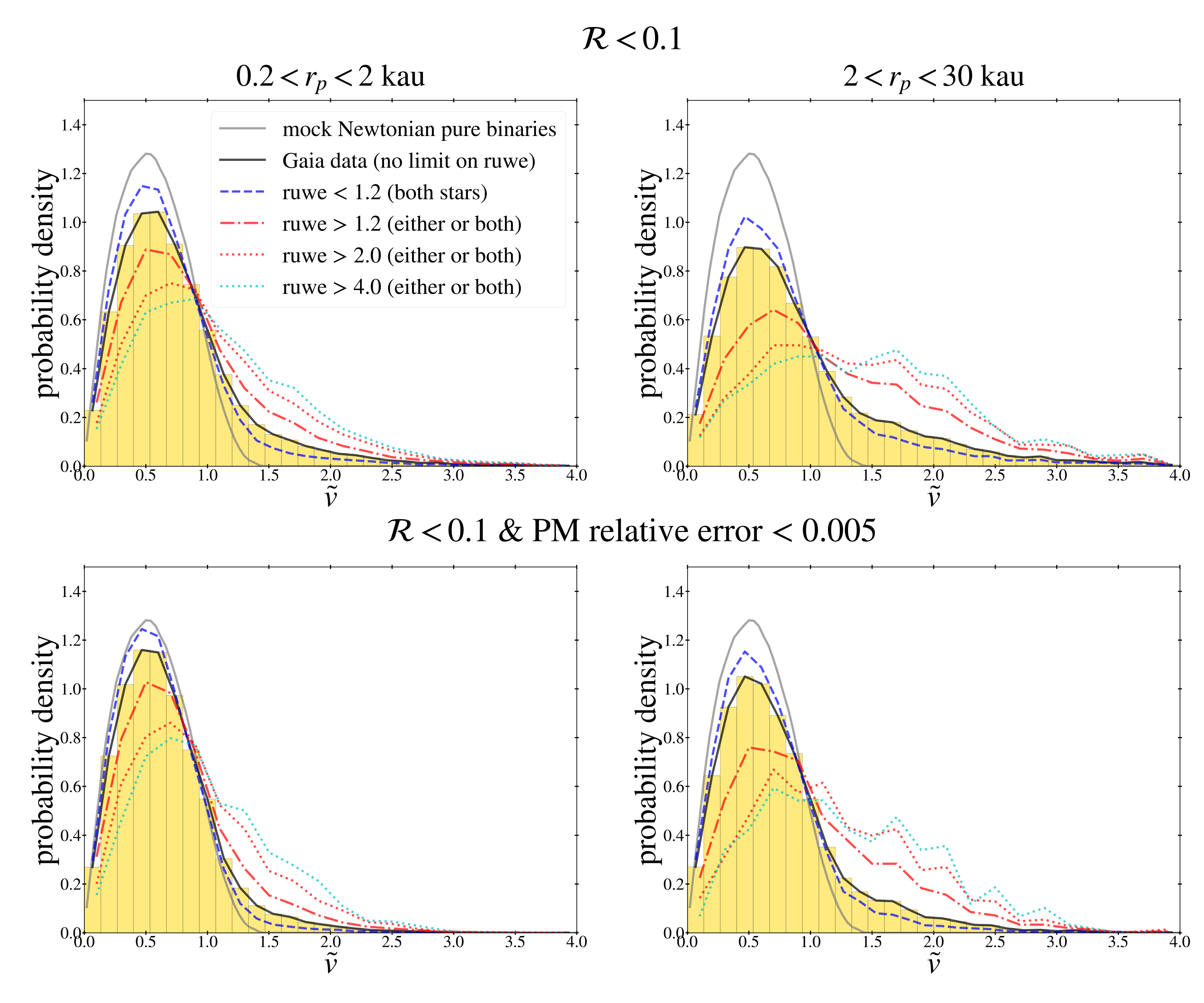}
    \caption{This figure investigates how the $\tilde v$ distribution in a sample depends on the limit on Gaia's {\tt ruwe} parameter using wide binaries from \cite{Chae:2023}. The left and right columns show distributions in two specific ranges of $r_p$. The top row shows all wide binaries that are likely to be gravitationally bound based on the \cite{El-badry:2021} $\mathcal{R}$ parameter, while the bottom row shows its subsample with relative PM error $<0.005$. All observed distributions shown here are broadened to varying degrees compared with the expectation of pure binaries in Newtonian gravity based on the generic thermal distribution of eccentricities.}
    \label{fig:vt_vt_hist_ruwe}
\end{figure*}

For a given sample of binaries, the range and distribution of $a_{\rm inn}$ and the effective value of $\eta_{\rm phot}$ depend on how the sample is defined. For example, if any resolved hierarchical systems have already been excluded in defining the sample down to a certain angular resolution limit, that puts a distance-dependent upper limit on (the sky-projected value of) $a_{\rm inn}$. For a general sample of host stars (e.g., including stars with large {\tt ruwe} values), the lower limit for $a_{\rm inn}$ is quite small (e.g., $0.01$~au: see \cite{Chae:2023} and references therein). When $a_{\rm inn}$ is small (such as a sub-au value), the instant kinematic effect is quite large (remember that the relative speed at 1 au for a total mass of $1M_\odot$ is $30\,{\rm km}\,{\rm s}^{-1}$). However, the observed PMs from Gaia DR3 represent the values averaged over the observation period of 34 months. This means that the net kinematic effect may be averaged out when the orbital period of the inner orbit is smaller than the observation period ($P_{\rm inn}<3\,{\rm yr}$ or $a_{\rm inn}<2\,{\rm au}$). In that case, the only effect of the hidden star would be increasing the effective mass of the host star, which will (moderately but not greatly) boost the relative velocity in the outer orbit between the two apparent binary stars. 

Thus, due to the time-averaging effect during the Gaia astrometric scan period, the \emph{observed} kinematic effect of a short-period inner binary will be much smaller than the effect at a certain instant (or for a very short period). For such a system, either star (or both stars) will have large {\tt ruwe} value(s) and the measurement uncertainties of the relative PM components will also be relatively large due to the wobbling effect. This means that binary systems with large {\tt ruwe} values or large relative PM errors will exhibit a velocity boost effect, but it may not be necessarily large. To check this expectation, we investigate the distribution of $\tilde v$. We use binaries from \cite{Chae:2023} because only it provides a sample without an upper limit on {\tt ruwe} while covering a broad range of $r_p$ among the samples considered in this study. 

Figure~\ref{fig:vt_vt_hist_ruwe} reveals the effects of cuts based on {\tt ruwe} or PM relative error. In the top row, we consider only gravitationally-bound systems based on $\mathcal{R}<0.1$ (precluding chance-alignment cases) and split the sample into the two ranges of $0.2<r_p<2\,{\rm kau}$ (left) and $2<r_p<30\,{\rm kau}$ (right). The bottom row is the same as the top, except that PM relative error $<0.005$ is imposed. In each panel, binaries with varied lower limits of {\tt ruwe} are compared with those with the upper limit {\tt ruwe} $<1.2$. Clearly, values of $\tilde v$ are statistically larger in binaries with {\tt ruwe} $>1.2$ than those with {\tt ruwe} $<1.2$, but the boost effect is moderate. Binaries with {\tt ruwe} $>2.0$ have values of $\tilde v$ that are statistically further boosted compared with those with {\tt ruwe} $>1.2$.

\begin{figure*}[tbh!]
    \centering
   \includegraphics[width=0.7\linewidth]{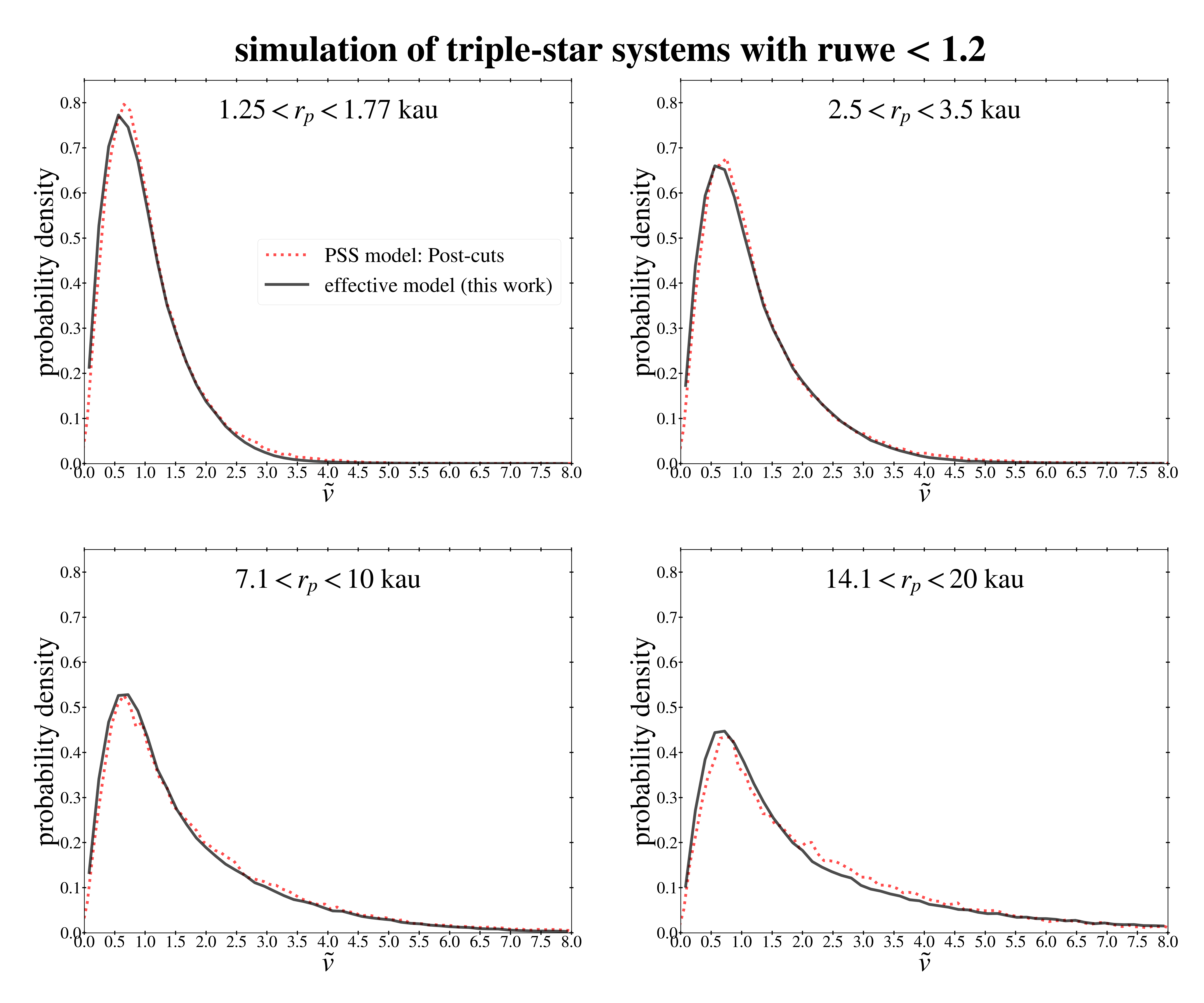}
    \caption{Distributions of $\tilde v$ from the \cite{Pittordis:2025} triple modeling in some bins of $r_p$. Red dotted curves are the same as those (`Post-cuts' option) shown in Figure~8 of \cite{Pittordis:2025}. The curves are reproduced here using numerical data provided by the authors. The black solid curves show our effective model as described in the text. Our effective model agrees well with the \cite{Pittordis:2025} simulation results. }
    \label{fig:vt_triple_PSS}
\end{figure*}

However, binaries with {\tt ruwe} $>4.0$ are similar to those with {\tt ruwe} $>2.0$. These binaries with large {\tt ruwe} values do not show dramatic boost compared with those with {\tt ruwe} $<1.2$. Most of such binaries are likely to have relatively short-period inner binaries and yet the boost effect is moderate, as would be the case when the time-averaging effect is at work. Binaries with $r_p > 2$~kau (the right column of Figure~\ref{fig:vt_vt_hist_ruwe}) appear to have relatively larger boost effect compared with those with $r_p < 2$~kau for the same {\tt ruwe} limit. This is well expected because the outer binary relative velocity in wider binaries is smaller for the same inner orbit kinematic effect.

For a general sample without a limit on {\tt ruwe}, to take into account the kinematic effect of an inner binary time-averaged over $\approx 3$ years, \cite{Chae:2023} employed a rather simple prescription that the effective value of $\eta_{\rm phot}$ is zero if the inner orbit period is less than 3 years (for an inner binary with a longer period Equation~(\ref{eq:etaphot}) was used). This prescription is not ideal, but the model provides adequate overall descriptions of the observed kinematic properties of wide binaries, e.g., through the distributions of $\tilde v$ \citep{Chae:2024b}.  

To take into account the kinematic effect of inner binaries more realistically than earlier studies including \cite{Chae:2023}, \cite{Pittordis:2025} introduced a new procedure. They suggested working with the limit {\tt ruwe} $<1.2$\footnote{They used additional cuts such as the ``Lobster cut'' in the CM diagram, but the {\tt ruwe} cut played the major role.} and then constructed a numerical model appropriate for the limit through a forward modeling approach based on empirical properties of binaries. As Figure~\ref{fig:vt_vt_hist_ruwe} shows, a sample with the {\tt ruwe} limit exhibits some discrepancy with mock Newtonian pure binaries, but the discrepancy is smaller compared to a sample without the limit. 

We will carry out various statistical tests, including the acceleration-plane test \citep{Chae:2023} and the $\tilde v$-distribution test, implementing the \cite{Pittordis:2025} approach of sample selection and triple modeling. For this, we use samples with the limit {\tt ruwe} $<1.2$ from \cite{Chae:2023} and \cite{Pittordis:2025} that cover broad dynamic ranges including both Newtonian and MOND regimes. \cite{Pittordis:2025} provide the statistical distributions of $\tilde v$ in bins of $r_p$ for a mock triple population, some of which are reproduced in Figure~\ref{fig:vt_triple_PSS} based on the data files provided by the authors. 

Because the statistical distribution of $\tilde v$ for triple systems cannot be used to statistically assign an inner orbit to a specific binary system, we devise an effective modeling procedure that closely reproduces $\tilde v$ distributions from the PSS modeling procedure in any bin of $r_p$ as shown in Figure~\ref{fig:vt_triple_PSS}. Our effective model makes use of the \cite{Pittordis:2025} finding that the {\tt ruwe} $<1.2$ cut preferentially removes inner orbits with period in a ``broad window around 3 years''. We control the exclusion window by varying the lower limit of $a_{\rm inn}$ (the inner orbit semi-major axis) from their adopted distribution of $\log_{10}a_{\rm inn}$, i.e., a Gaussian distribution with mean $\log_{10} (40\,{\rm au})$ and standard deviation 1.5 (see \cite{Pittordis:2025} for further details). 

We numerically find using the \cite{Pittordis:2025} sample that the following lower limit of $a_{\rm inn}$ reproduces the PSS modeling result: 
\begin{equation}
    a_{\rm inn,min}/{\rm au}=11.7-5.5\log_{10}(r_p/{\rm kau}),
    \label{eq:ainn_min}
\end{equation}
where $r_p$ is the sky-projected separation of the outer orbit (i.e., the observed binary). For the upper limit of $a_{\rm inn}$ we use the one used by \cite{Pittordis:2025}, i.e.,
\begin{equation}
    a_{\rm inn,max}=a_{\rm out}\times{\rm max}(0.342(1-e_{\rm out})^2,0.01),
    \label{eq:ainn_max}
\end{equation}
where $a_{\rm out}$ and $e_{\rm out}$ refer to the semi-major axis and eccentricity of the observed binary which is the outer orbit in the present context. In this work, there is a minor exception to Equation~(\ref{eq:ainn_max}). Unlike \cite{Pittordis:2025}, we consider binaries of small $r_p<1\,{\rm kau}$. In rare cases of very small $r_p$, $a_{\rm inn,max}$ can be smaller than $a_{\rm inn,min}$ given by Equation~(\ref{eq:ainn_min}). In those cases, we use $a_{\rm inn,max}=1.5\times a_{\rm inn,min}$ (where the multiplication factor $1.5$ is chosen to allow some width, but its value can be varied without significantly affecting gravity tests). With the lower and upper limits, $\log_{10}(a_{\rm inn}/{\rm au})$ is sampled from a normal distribution with mean $\log_{10}(40)$ and standard deviation $1.5$. For numerical calculations of the inner orbit and its kinematic effect, we use the code introduced in \cite{Chae:2023} with the adjustment of the range of $a_{\rm inn}$. The remarkable agreement between the PSS result for the $\tilde v$ distribution and our result for any bin of $r_p$ indicates that the two independent codes for the inner orbit calculation are consistent and the {\tt ruwe} cut can be mainly controlled by the range of $a_{\rm inn}$ as noted by \cite{Pittordis:2025}. 

\begin{figure*}[tbh!]
    \centering
   \includegraphics[width=1.0\linewidth]{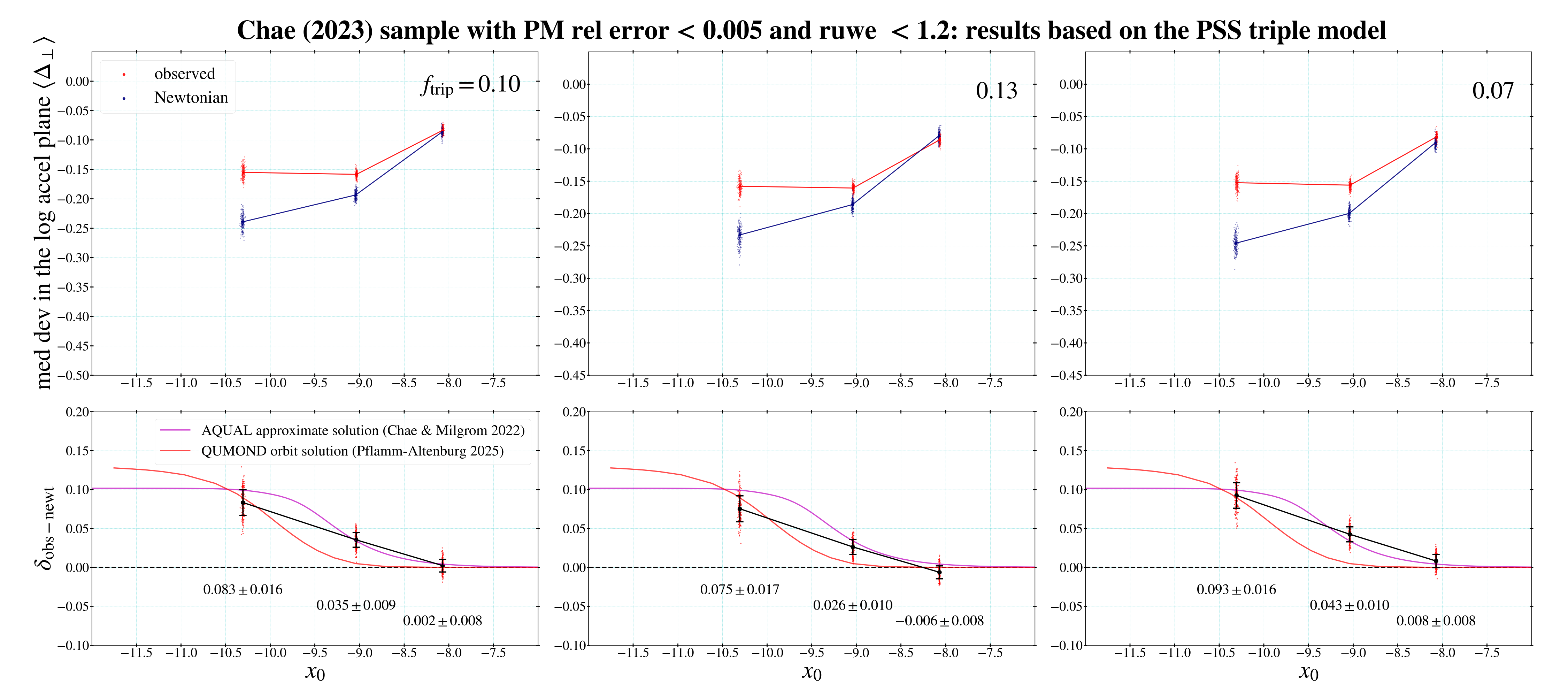}
    \caption{Acceleration-plane test results for the \cite{Chae:2023} sample with {\tt ruwe} $<1.2$ and PM relative error $<0.005$ based on the PSS triple model shown in Figure~\ref{fig:vt_triple_PSS}. (Upper panels) For each binary, $g_{\rm N}\equiv GM_{\rm tot}/r^2$ and $g\equiv v^2/r$ are calculated based on Monte Carlo (MC) deprojected 3D separation $r$ and relative velocity $v$. We define $x\equiv \log_{10}g_{\rm N}$ and $y\equiv \log_{10}g$, and then calculate $x_0 \equiv x+(y-x)/2$, so that $(x_0,x_0)$ corresponds to the projection of $(x,y)$ on the diagonal line, and $\Delta_\perp \equiv (y-x)/\sqrt{2}$, which represents the orthogonal deviation of $g$ from $g_{\rm N}$ in the logarithmic acceleration plane and is expected to be zero only in the special case of Newtonian circular orbits. The ordinate $\langle\Delta_\perp\rangle$ refers to the median of $\Delta_\perp$ in one MC set. Red points represent distributions of $\langle\Delta_\perp\rangle$ from 200 MC sets of the Gaia data in three bins of $x_0$. Blue points represent the corresponding mock Newtonian data. In each panel, $f_{\rm trip}$ is determined by matching data with the Newtonian prediction in the rightmost bin of $x_0$ center on $\approx -8.1$. The left panel is for the best-fit value of $f_{\rm trip}$ while the other panels show the variations due to the uncertainty of $f_{\rm trip}$. (Lower panels) The ordinate is defined by $\delta_{\rm obs-newt}\equiv \langle\Delta_\perp\rangle_{\rm obs}- \langle\Delta_\perp\rangle_{\rm newt}$, where $\langle\Delta_\perp\rangle_{\rm obs}$ and $\langle\Delta_\perp\rangle_{\rm newt}$ represent the points from the upper panels. Then, the gravitational boost factor ($\gamma_g$) is given by $\gamma_g=10^{\sqrt{2}\delta_{\rm obs-newt}}$. The magenta curve is the \cite{ChaeMilgrom:2022} numerical prediction in AQUAL gravity \citep{BekensteinMilgrom:1984}, while the red curve is the \cite{Pflamm-Altenburg:2025} numerical prediction in QUMOND gravity \citep{Milgrom:2010}, as described in Section~\ref{sec:mondgravity}. See \cite{Chae:2023} for further details of the methodology.}
    \label{fig:residual_3cols_Chaesample}
\end{figure*}

\subsection{Acceleration-plane test}  \label{sec:test_acceleration}

The acceleration-plane test of \cite{Chae:2023} is a straightforward and reliable way of investigating the median trend of data as a function of acceleration. The main advantage of the method is that data at high enough accelerations can be used to reliably calibrate(fit) $f_{\rm trip}$ (or $f_{\rm multi}$ in general).  

We perform the acceleration-plane test for samples with the limit {\tt ruwe} $<1.2$ with the adaptation of Equation~(\ref{eq:ainn_min}) to accommodate the properties of triples simulated by \cite{Pittordis:2025}. Because \cite{Pittordis:2025} assumed that all hierarchical systems are triples, we are also assuming it in the present study. We allow PM scatters in the Monte Carlo procedure for Gaia data and the counterpart mock Newtonian data (see \citealt{Chae:2024b}). For each binary, we use the individual Bayesian information on eccentricity derived by \cite{Hwang:2022} as described by Section~2.4 of \cite{Chae:2024b}. When an individual information is not available, we use the $r_p$-dependent power-law distribution as given by Equation~(18) of \cite{Chae:2024a}. Individual eccentricity information is available for every system in the \cite{Chae:2023} sample and for 92\% of the PSS sample.

\begin{figure*}[tbh!]
    \centering
   \includegraphics[width=0.8\linewidth]{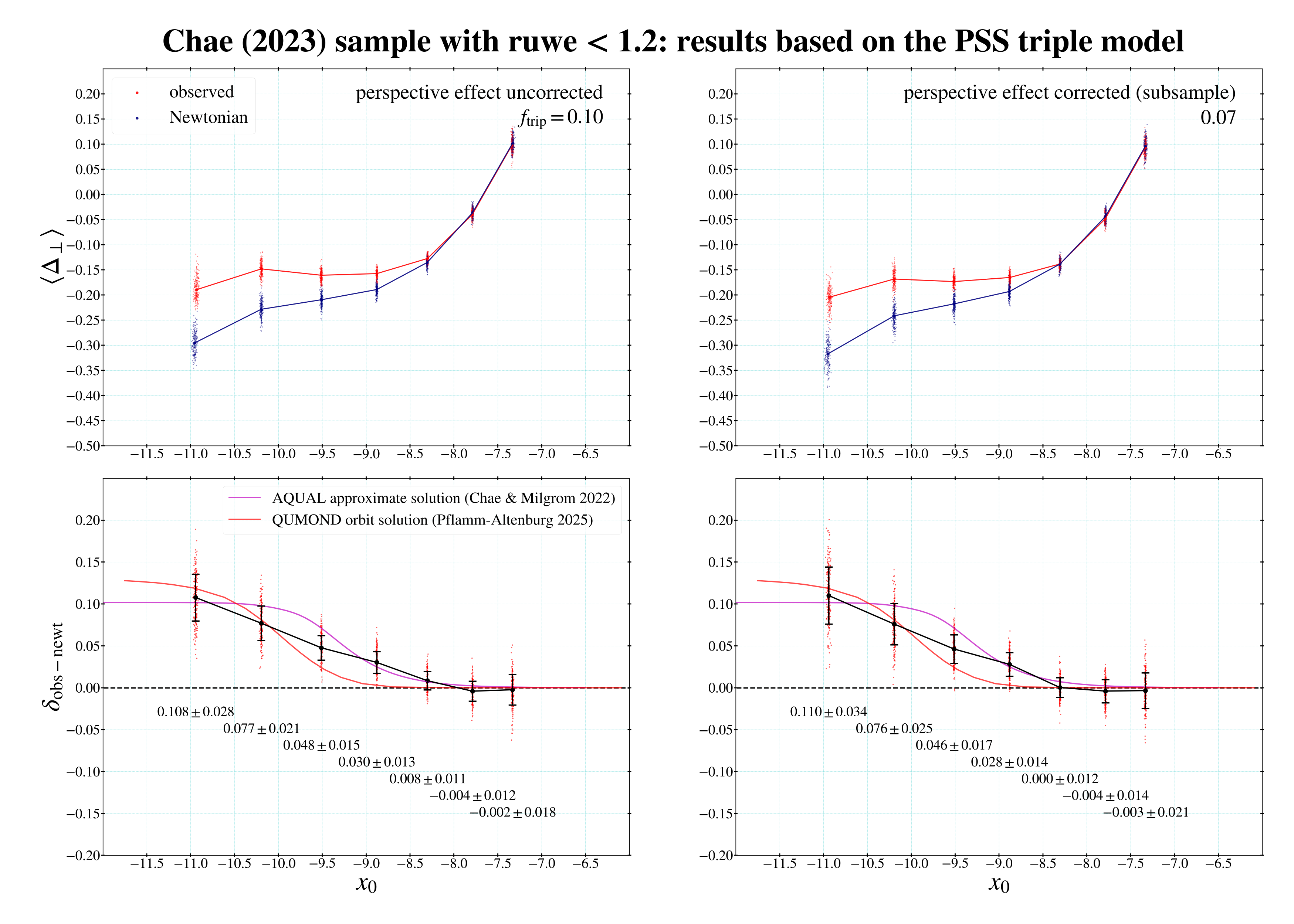}
    \caption{The left column is the same as the left column of Figure~\ref{fig:residual_3cols_Chaesample} except that 7 bins of $x_0$ are used. The right column is for a subsample of wide binaries for which RVs are available and the perspective effect is taken into account. The good agreement between the two results demonstrates that the perspective effect is very minor as most signal comes from data with $r_p<20\,{\rm kau}$ and $d>50\,{\rm pc}$. }
    \label{fig:residual_7bins_Chaesample}
\end{figure*}

\subsubsection{For the Chae (2023) sample with ${\tt ruwe}<1.2$}  \label{sec:acceleration_chaesample}

 We consider the sample of 19716 binaries with PM relative error $<0.005$ from \cite{Chae:2023}. If the cut {\tt ruwe} $<1.2$ is applied, 13602 binaries (69\%) remain. We note that the excluded 31\% has a much larger proportion of hierarchical systems than the remaining 69\%. 

Figure~\ref{fig:residual_3cols_Chaesample} shows the results for the sample of 13602 binaries with the {\tt ruwe} cut. As usual, we calibrate $f_{\rm trip}$ using the high-acceleration bin ($\ga 10^{-8.3}\,{\rm m}\,{\rm s}^{-2}$). The fitted value of $f_{\rm trip}=0.10\pm 0.03$ with the limit {\tt ruwe} $<1.2$ is dramatically reduced from $f_{\rm multi}=0.43\pm 0.05$ (see Appendix~A of \cite{Chae:2024b}) for the sample without the {\tt ruwe} cut (note here that \cite{Chae:2024b} allowed both triples and quadruples, but the allowance of quadruples was not a significant factor). The much lower fitted value of $f_{\rm trip}$ is roughly consistent with the reduction due to the {\tt ruwe} cut, but an exact match is not expected because the exact fraction of hierarchical systems among the excluded sample is unknown and both the \cite{Chae:2023} model without a {\tt ruwe} limit and the PSS effective model with the specific {\tt ruwe} limit have some uncertainties.

With the PSS sampling and triple modeling, the \cite{Chae:2023} sample clearly shows a gravitational anomaly at low acceleration $\la 10^{-10}\,{\rm m}\,{\rm s}^{-2}$ consistent with the result for the general sample and Chae's multiple modeling. The only significant difference is the fitted value of the fraction of hierarchical systems. The agreement about the low-acceleration gravitational anomaly is remarkable considering the variations in sample definition and modeling of hierarchical systems.

To further examine the detailed behavior of the parameter $\delta_{\rm obs-newt}$ (see the caption of Figure~\ref{fig:residual_3cols_Chaesample}), we consider finer bins of $x_0$, and the results are shown in Figure~\ref{fig:residual_7bins_Chaesample}. The left column is the result for the same sample shown in Figure~\ref{fig:residual_3cols_Chaesample}. In the right column, we show a result with the perspective effect included for a subsample of binaries for which measured RV(s) are(is) available for both stars (or one of the two). For this sample, the fitted value of $f_{\rm trip}$ is somewhat lower because binaries with measured RVs are less likely to be subject to kinematic contamination, as previously noted \citep{Chae:2024a,Chae:2024b,Chae:2025}. The result with the perspective effect is very similar to the one without it, but the statistical error is somewhat larger due to the reduced sample size.

Figure~\ref{fig:residual_3cols_Chaesample} and Figure~\ref{fig:residual_7bins_Chaesample} show that the inferred functional behaviors of $\delta_{\rm obs-newt}$ are consistent with the existing numerical predictions of MOND gravity models, while clearly ruling out Newtonian gravity. The inferred values of $\delta_{\rm obs-newt}$ for $x_0 \la -9$ are in between the two numerical results by \cite{ChaeMilgrom:2022} for AQUAL and \cite{Pflamm-Altenburg:2025} for QUMOND. Thus, the particular sample of wide binaries does not distinguish well the two numerical predictions. 

\begin{figure}[tbh!]
    \centering
   \includegraphics[width=0.9\linewidth]{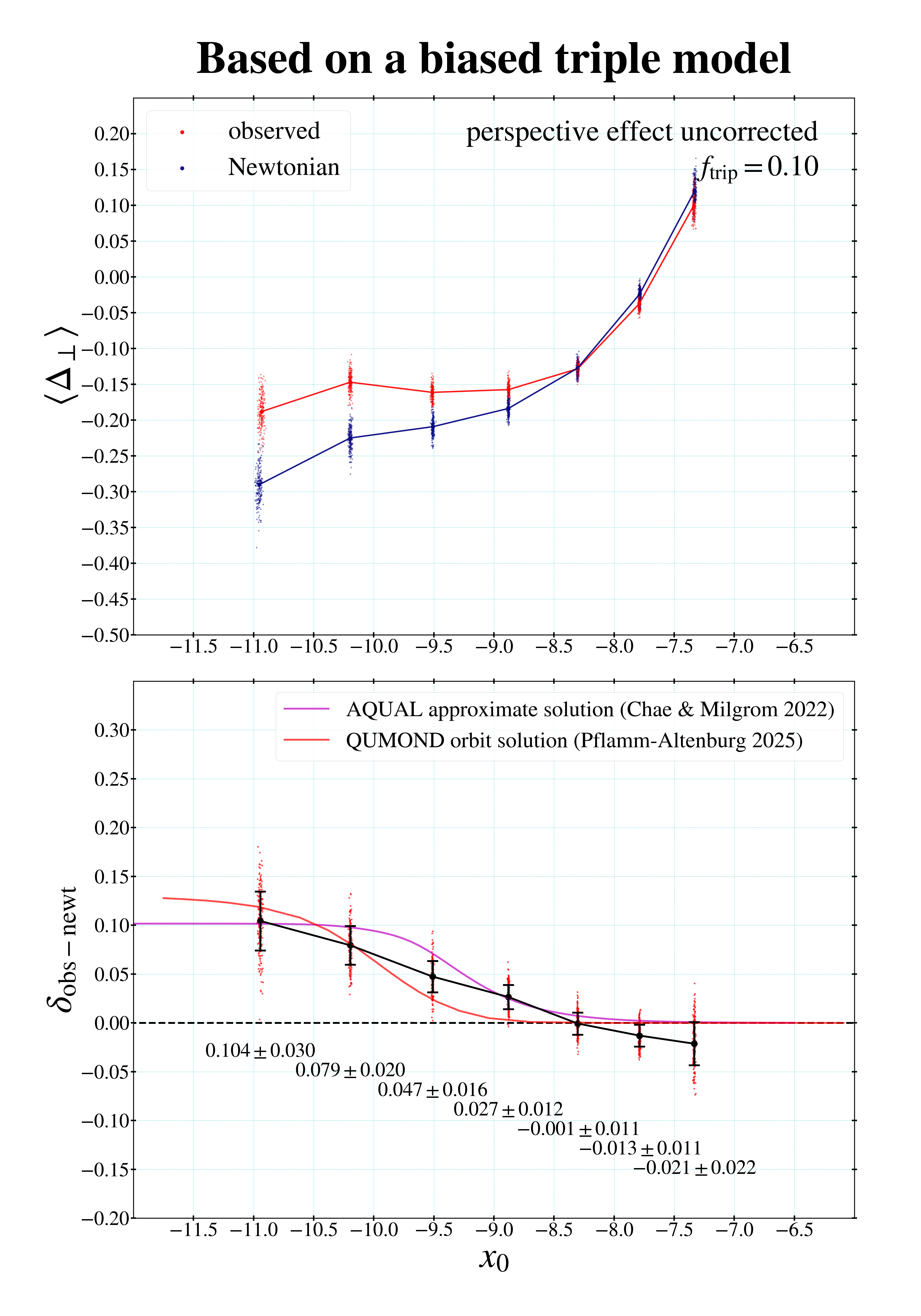}
    \caption{Same as the left column of Figure~\ref{fig:residual_7bins_Chaesample} but with a biased triple model, in which $a_{\rm inn}$ is constant and thus inconsistent with the correct model based on Equation~(\ref{eq:ainn_min}). }
    \label{fig:residual_7bins_PSSdistort}
\end{figure}

It is worth noting that the remarkable agreement of the data with the Newtonian prediction in the three bins with $x_0\ga -8.3$ is a non-trivial result. It means that the triple model is a good description of the data in the high-acceleration Newtonian regime. To check this, we have considered a biased triple model with a fixed $a_{\rm inn,min}=5\,{\rm au}$ (which is different from Equation~(\ref{eq:ainn_min})) that does not agree with the \cite{Pittordis:2025} triple model at small separation $r_p \la 3\,{\rm kau}$. Figure~\ref{fig:residual_7bins_PSSdistort} shows the acceleration-plane test result with the inaccurate triple model. While the values of $\delta_{\rm obs-newt}$ for $x_0<-9$ are nearly indistinguishable from those in the left column of Figure~\ref{fig:residual_7bins_Chaesample}, the values at $x_0>-8$ (the rightmost two bins) tend to deviate from the Newtonian prediction when the bin at $x_0\approx -8.3$ (the third from the right) was used to calibrate $f_{\rm trip}=0.10$.

\subsubsection{For the PSS sample}  \label{sec:acceleration_PSSsample}

\begin{figure*}[tbh!]
    \centering
   \includegraphics[width=1.0\linewidth]{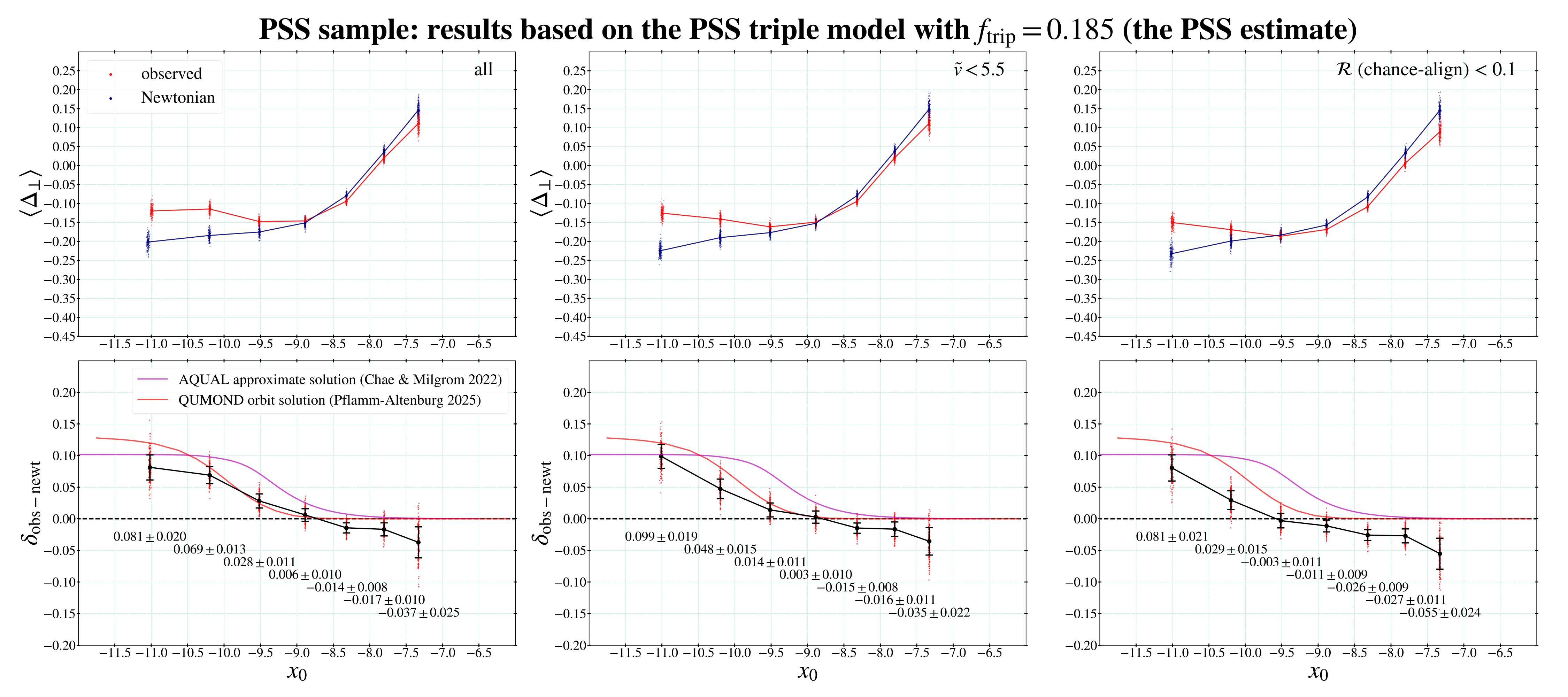}
    \caption{Each column is similar to the right column (with the perspective effect) of Figure~\ref{fig:residual_7bins_Chaesample} but for the PSS sample with a fixed value of $f_{\rm trip}=0.185$ from \cite{Pittordis:2025}. The left column is the result for the `no flyby' option while the right column is for the `max flyby' option. The middle column is for the PSS-like option of flybys (see Figure~\ref{fig:f_flyby}).}
    \label{fig:residual_7bins_PSSsample_ftripfixed}
\end{figure*}

We also perform the acceleration-plane test for the PSS sample as it includes binaries with relatively small separations $r_p<1$~kau (see Figure~\ref{fig:vpvt_rp}). However, \cite{Pittordis:2025} did not use those small-separation binaries for calibration purposes, but used the sample in the range $1.25<r_p<30\,{\rm kau}$ to simultaneously fit $f_{\rm trip}$ and the fraction of chance-alignment/flyby $f_{\rm flyby}$ in each assumed gravity model. For binaries in the MOND regime ($r_p>$ several kau) with $\tilde v > 1.5$, there is some uncertainty in distinguishing between gravitationally-bound apparent binaries with hidden additional star(s) and gravitational-unbound pairs. 

When both $f_{\rm trip}$ and $f_{\rm flyby}$ are unknowns that need to be determined from the data, it will be much more advantageous if one of them can be reliably determined regardless of the other. Two parameters $f_{\rm trip}$ and $f_{\rm flyby}$ have different properties and meanings. Parameter $f_{\rm trip}$ represents the probability for a member star to have a hidden close companion, and thus it depends on the kinematic properties of the star such as {\tt ruwe}, PM relative error, and RV error, but is independent of the size of the outer orbit of the wide binary system. We can determine it reliably using a subsample of small $r_p$ from a consistently defined sample by using the fact that when $r_p$ is sufficiently small, (1) $f_{\rm flyby}$ becomes negligibly small as long as the two stars have consistent distances from the Sun (see Figure~\ref{fig:f_flyby}), and (2) the binary system is in the Newtonian regime so that we can use Newtonian dynamics to determine $f_{\rm trip}$. This is precisely the approach introduced by \cite{Chae:2023} and will be followed here.  

Unlike $f_{\rm trip}$, $f_{\rm flyby}$ is a strong function of $r_p$ because cross-section of chance-alignment is proportional to $r_p^2$. Thus, while $f_{\rm flyby}$ refers to the fraction in the entire sample, it is proportionately contributed by wide binaries with larger $r_p$. However, chance-alignment/flyby pairs can be statistically studied and flagged using an empirically-assigned probability as done by \cite{El-badry:2021}. Of course, such a probability assignment has an uncertainty, and thus one needs to be careful in using empirical probabilities such as the \cite{El-badry:2021} $\mathcal{R}$ parameter. 

When the scientific question is whether there exists a gravitational anomaly at large $r_p$ or not, one should be careful not to take chance-alignment/flyby pairs for gravitationally-bound systems, as those pairs can give rise to a false gravitational boost. In this respect, it may be preferable to be conservative in statistically removing chance-alignment/flyby pairs. In other words, it may be better to remove all suspicious cases at the risk of removing some gravitationally-bound systems that have relatively large kinematic contamination and thus are not so well distinguishable from chance-alignment pairs. One caveat of this approach is that when the gravitational anomaly is present, it can be somewhat underestimated if gravitationally-bound systems are appreciably removed. 

Considering the uncertainty of chance-alignment/flyby pairs in the PSS sample, we consider two extreme options that can bracket the possible range, as well as an intermediate option statistically matching the Newtonian modeling result by \cite{Pittordis:2025}. In one option to be referred to as `no flyby', the entire sample is used without removing any potential chance-alignment/flyby pairs (i.e., in this option it is assumed that the sample does not include any flyby). In the other extreme option to be referred to as `max flyby', we exclude all pairs that have $\mathcal{R}>0.1$ or are not included in the \cite{El-badry:2021} catalog (i.e., we exclude systems that are suspicious or do not have enough observational information). In the intermediate option to be referred to as the PSS-like option, we exclude pairs with $\tilde v > 5.5$ only. Figure~\ref{fig:f_flyby} shows these options. We note that the PSS-like option is overall closer to the `no flyby' option differing only at large $r_p$.

Another issue in using the PSS sample is how $f_{\rm trip}$ should be determined, as \cite{Pittordis:2025} did not use binaries with $r_p<1.25\,{\rm kau}$ to estimate their value of $f_{\rm trip}=0.185$. We will consider two options on $f_{\rm trip}$. In the first option, we simply take $f_{\rm trip}=0.185$ and check the result with it. In the second option, we take the approach of determining $f_{\rm trip}$ with binaries in the Newtonian regime that are available to us from the authors. As the PSS sample includes only binaries with measured RVs, the perspective effect is always taken into account.

Figure~\ref{fig:residual_7bins_PSSsample_ftripfixed} shows the results with the fixed value of $f_{\rm trip}=0.185$. Three features are obvious regardless of the option on flybys. First, the data and Newtonian prediction match well only in the bin centered on $x_0\approx -9.0$/$-9.5$ for the no/max flyby option. This means that the data in the high-acceleration Newtonian regime do not agree with the Newtonian prediction. This is particularly true for the max flyby case. This suggests that the \cite{Pittordis:2025} value of $f_{\rm trip}$ was overestimated. Second, the functional behavior of $\delta_{\rm obs-newt}$ is not flat (Newtonian) and clearly declines with $x_0$. Finally, the two bins with $x_0<-10$ show $\delta_{\rm obs-newt}>0$ with a high significance, even for $f_{\rm trip}=0.185$. These features provide evidence that Newtonian gravity does not work in the low-acceleration regime.

The results shown in Figure~\ref{fig:residual_7bins_PSSsample_ftripfixed} provide insight into understanding the conclusion that Newtonian gravity was preferred by \cite{Pittordis:2025} based on the same sample and the same value of $f_{\rm trip}$. Two points are relevant here. First, in the PSS-like flyby option (which is most relevant here), the bins in the range $-10\la x_0\la -8.5$ agree overall much better with Newton than the approximate numerical MOND model. Because \cite{Pittordis:2025} also used an approximate MOND model similar to that of \cite{Zonoozi:2021} (see Figure~\ref{fig:theory_prediction}), our result agrees overall with their conclusion. However, the realistic numerical model by \cite{Pflamm-Altenburg:2025} agrees with the data even better than Newton in the range $x_0\la -8.5$ probed by \cite{Pittordis:2025}. (This of course does not mean that the value of $f_{\rm trip}=0.185$ is correct because it causes a discrepancy at high accelerations $x_0> -8.5$.) Also, we note that the leftmost bin at $x_0\approx -11$ agrees much better with MOND regardless of which numerical MOND model is used. 

Second, \cite{Pittordis:2025} never tested Newton in an absolute sense, but only compared the relative performances of Newton and an inaccurate numerical representation of MOND, particularly in the transion regime $-10\la x_0 \la -9$, as acknowledged in the paper. This is why they obtained a stronger statistical preference of Newton in the transition regime (see Figure~11 of \cite{Pittordis:2025}). Thus, our results (particularly in the PSS-like flyby option) are consistent with the \cite{Pittordis:2025} results under similar conditions. More detailed and direct comparisons will be carried out below.

\begin{figure*}[tbh!]
    \centering
   \includegraphics[width=1.0\linewidth]{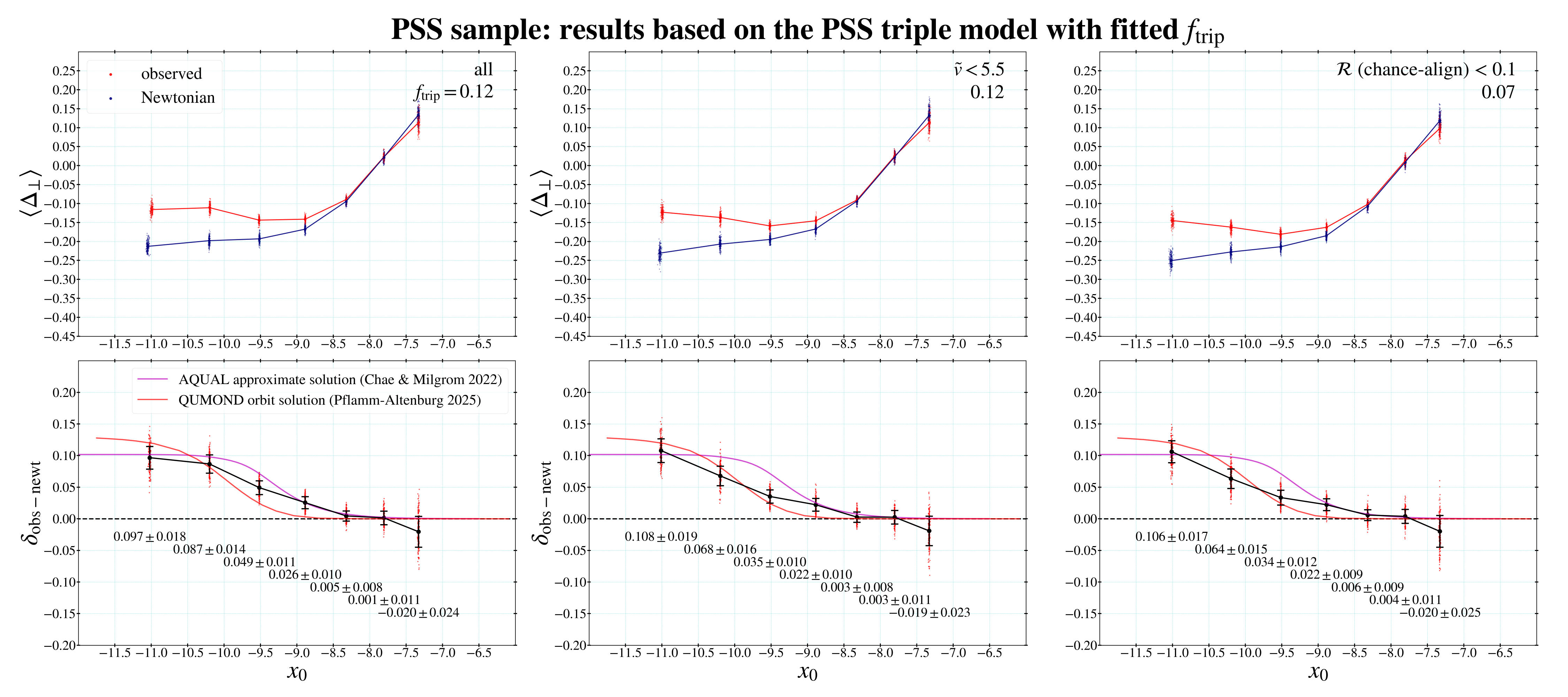}
    \caption{Same as Figure~\ref{fig:residual_7bins_PSSsample_ftripfixed} but for $f_{\rm trip}$ fitted with Newtonian regime data at $x_0\approx -8$.}
    \label{fig:residual_7bins_PSSsample_ftripfitted}
\end{figure*}

The declining behavior of $\delta_{\rm obs-newt}$ in the PSS sample strongly indicates a discrepancy with Newton and calls for the need to calibrate $f_{\rm trip}$ with Newtonian regime data. As Figure~\ref{fig:vpvt_rp} shows, the PSS sample is overall similar to the \cite{Chae:2023} sample in the $r_p$-$v_p$ and $r_p$-$\tilde v$ planes for $r_p\ga 0.3\,{\rm kau}$ except that the PSS sample includes some chance-alignment/flyby pairs at large $r_p$ ($>1\,{\rm kau}$). Thus, we apply the procedure of calibrating $f_{\rm trip}$ to the PSS sample as was done for the \cite{Chae:2023} sample.

Figure~\ref{fig:residual_7bins_PSSsample_ftripfitted} shows the results for the PSS sample with $f_{\rm trip}$ fitted using the bin at $x_0\approx -8.0$. We note the following features. First, the fitted values of $f_{\rm trip}=0.12$ (no/PSS-like flyby) and $0.07$ (max flyby), which bracket the likely value, are significantly lower than the \cite{Pittordis:2025} value of $0.185$. Comparison of Figure~\ref{fig:residual_7bins_PSSsample_ftripfitted} with Figure~\ref{fig:residual_7bins_PSSsample_ftripfixed} shows that the \cite{Pittordis:2025} value of $0.185$ is designed to bring the data into maximum agreement with Newton in the range $-9.5\la x_0 \la -9.0$, while our fitted values are designed to match Newton for $x_0\ga -8.5$. The problem with the \cite{Pittordis:2025} value is that it causes a tension with Newton for $x_0\ga -8.5$ where the binaries should obey Newton in any viable model of gravity. 

Second, the PSS sample clearly provides evidence for $\delta_{\rm obs-newt}>0$ with very high statistical significance in the bins of $x_0 \la -9.0$, in good agreement with the results for the \cite{Chae:2023} sample shown in Figure~\ref{fig:residual_7bins_Chaesample}. Moreover, the functional behavior of $\delta_{\rm obs-newt}>0$ matches approximately the numerical predictions of MOND models. The results with the PSS-like and max flyby options appear to favor the numerical prediction by \cite{Pflamm-Altenburg:2025} over that by \cite{ChaeMilgrom:2022}, although the distinction is not strong.

Third, the rightmost three bins with $x_0\ga -8.3$ follow the Newtonian flat line well within the estimated uncertainties. This is the desired result for binaries in the Newtonian regime. In other words, because the binaries in the Newtonian regime must obey Newtonian gravity, the same value of $f_{\rm trip}$ should work in any bin of $x_0$ as is the case here. This also indicates that the effective PSS triple model based on Equation~(\ref{eq:ainn_min}) is consistent with the PSS binaries of small separation ($r_p<1\,{\rm kau}$). 

\subsection{Auxiliary test with $\tilde v$ distributions} \label{sec:test_vtdist}

\begin{figure*}[tbh!]
    \centering
   \includegraphics[width=0.8\linewidth]{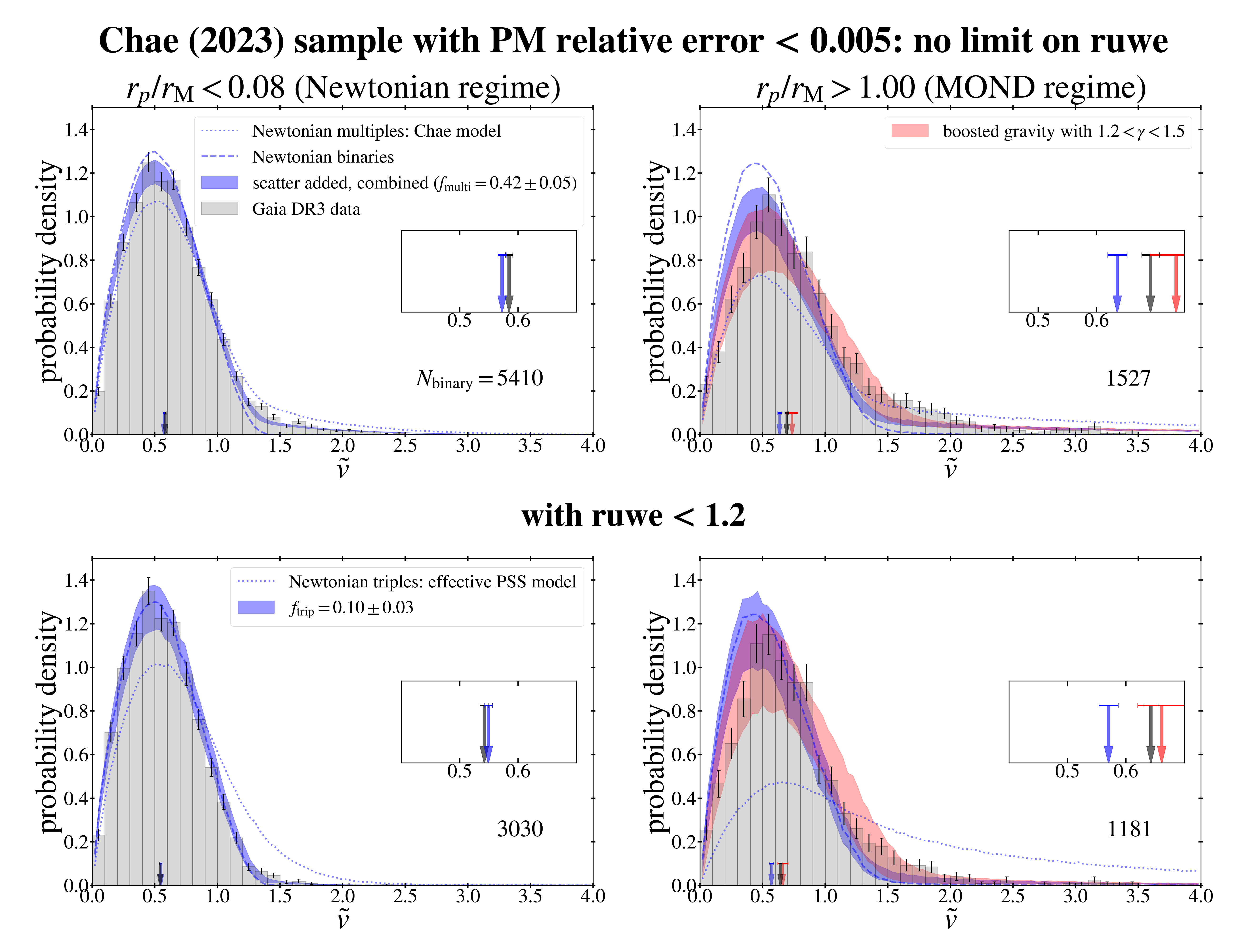}
    \caption{The observed distribution of $\tilde v$ is compared with the predictions of Newtonian gravity and a boosted gravity model to represent MOND gravity in the MOND (low-acceleration) regime. The left column is for binaries with $r_p<0.08 r_{\rm M}$, most of which must be in the Newtonian regime, while the right column is for $r_p > r_{\rm M}$ which is the MOND regime with internal acceleration $< a_0$. The top row is for the \cite{Chae:2023} sample without any limit on {\tt ruwe}. The \cite{Chae:2023} multiple-star model is used to represent hierarchical systems and $f_{\rm multi}=0.42\pm 0.05$ is taken from the acceleration-plane test results. The bottom row is for the sample with the limit {\tt ruwe} $<1.2$ and $f_{\rm trip}=0.10\pm 0.03$ determined with the PSS triple model from the acceleration-plane test. In calculating theoretical predictions of the $\tilde v$ distribution, we use the individual Bayesian information for each binary as described in the text. The arrows indicate the medians in the observed or theoretical distributions. The uncertainties of the histogram bars and the median in the observed distribution of $\tilde v$ are estimated from bootstrap resampling. }
    \label{fig:vtdist_Chaesample}
\end{figure*}

Here we analyze $\tilde v$ distributions as an auxiliary analysis given that the more rigorous and reliable acceleration-plane tests have already been performed above. \cite{Pittordis:2025} considered the distributions of $\tilde v$ in their sample to test Newtonian gravity and a control MOND model. They used only wide binaries with $r_p>1.25\,{\rm kau}$ noting that it was difficult to calibrate $f_{\rm trip}$ using the median of $\tilde v$ in a bin of small $r_p$ as binaries of smaller separations are less sensitive to inner close binaries. However, as Figure~\ref{fig:residual_3cols_Chaesample} above and Figure~22 of \cite{Chae:2024b} show, wide binaries of small separation have a clear sensitivity to $f_{\rm trip}$ or $f_{\rm multi}$. In other words, only a certain narrow range of $f_{\rm trip}$ or $f_{\rm multi}$ can be consistent with Newton in a high-acceleration bin. More importantly, the high-acceleration or small-$r_p$ bin is free of chance-alignment/flyby pairs, and thus $f_{\rm trip}$ can be reliably determined without worrying about degeneracy between triples and flybys.

Our analyses of $\tilde v$ distributions are different from \cite{Pittordis:2025} in several important ways. First, we use the values of $f_{\rm trip}$ fitted from the acceleration-plane analysis in Section~\ref{sec:test_acceleration}. The fitted values of $f_{\rm trip}$ for the considered samples are reliable because data and Newtonian prediction agree excellently in all bins of $x_0\ga -8.3$. 

Second, we use bins of normalized separation $r_p/r_{\rm M}$ instead of $r_p$, since the former takes into account individual masses of the binaries. Unlike \cite{Pittordis:2025}, we consider both a Newtonian regime $r_p/r_{\rm M}<0.08$ and a MOND regime $r_p/r_{\rm M}>1$. Because the 3D radius $r$ is always $>r_p$, the cut $r_p/r_{\rm M}>1$ ensures that all selected wide binaries have internal acceleration $<a_0$. However, the cut  $r_p/r_{\rm M}<0.08$, where $r_p/r_{\rm M}=0.08$ corresponds to $g_{\rm N}\approx 10^{-7.9}\,{\rm m}\,{\rm s}^{-2}$ in the median sense, may well include some wide binaries whose internal acceleration has $g_{\rm N}\la 10^{-8}$. This should matter little because we satisfy $g_{\rm N}\ga 10^{-8.3}\,{\rm m}\,{\rm s}^{-2}$ up to $r=2 r_p$ with the cut $r_p/r_{\rm M}<0.08$. We do not consider a transition regime because it is difficult to accurately represent the regime with a MOND gravity model, and exactly for that reason the recent tests by \cite{Banik:2024} and \cite{Pittordis:2025} suffered from an error in interpreting the transition regime data \citep{Pflamm-Altenburg:2025}. 

Third, we use individual Bayesian information of eccentricities from \cite{Hwang:2022} in calculating theoretical predictions of the $\tilde v$ distribution. While the individual Bayesian information covers a broad range of $e$ for each binary (see Figure~8 of \citealt{Chae:2023}), it is clearly more informative than a general power-law distribution. 

Fourth, we use a Monte Carlo (MC) approach to estimate the range of theoretical predictions. In one MC set, all orbit and inclination parameters are randomly sampled in the standard way as was done in Section~\ref{sec:WBTnature}. Then, we use $n$ ($n=50,\,100$, or $200$ depending on the numerical burden and sample size) MC sets to estimate the median probability and its uncertainty of occurrences for each bin of $\tilde v$, which will be compared with the observed distribution of $\tilde v$.

Finally, in testing gravity using the MOND regime data, we represent MOND gravity by a pseudo-Newtonian model with a gravity boost factor $\gamma$. This approximation is justified by the strong EFE due to the Galactic gravitational field as verified by numerical studies of MOND gravity (see Section~\ref{sec:mondgravity}).

\begin{figure*}[tbh!]
    \centering
   \includegraphics[width=0.75\linewidth]{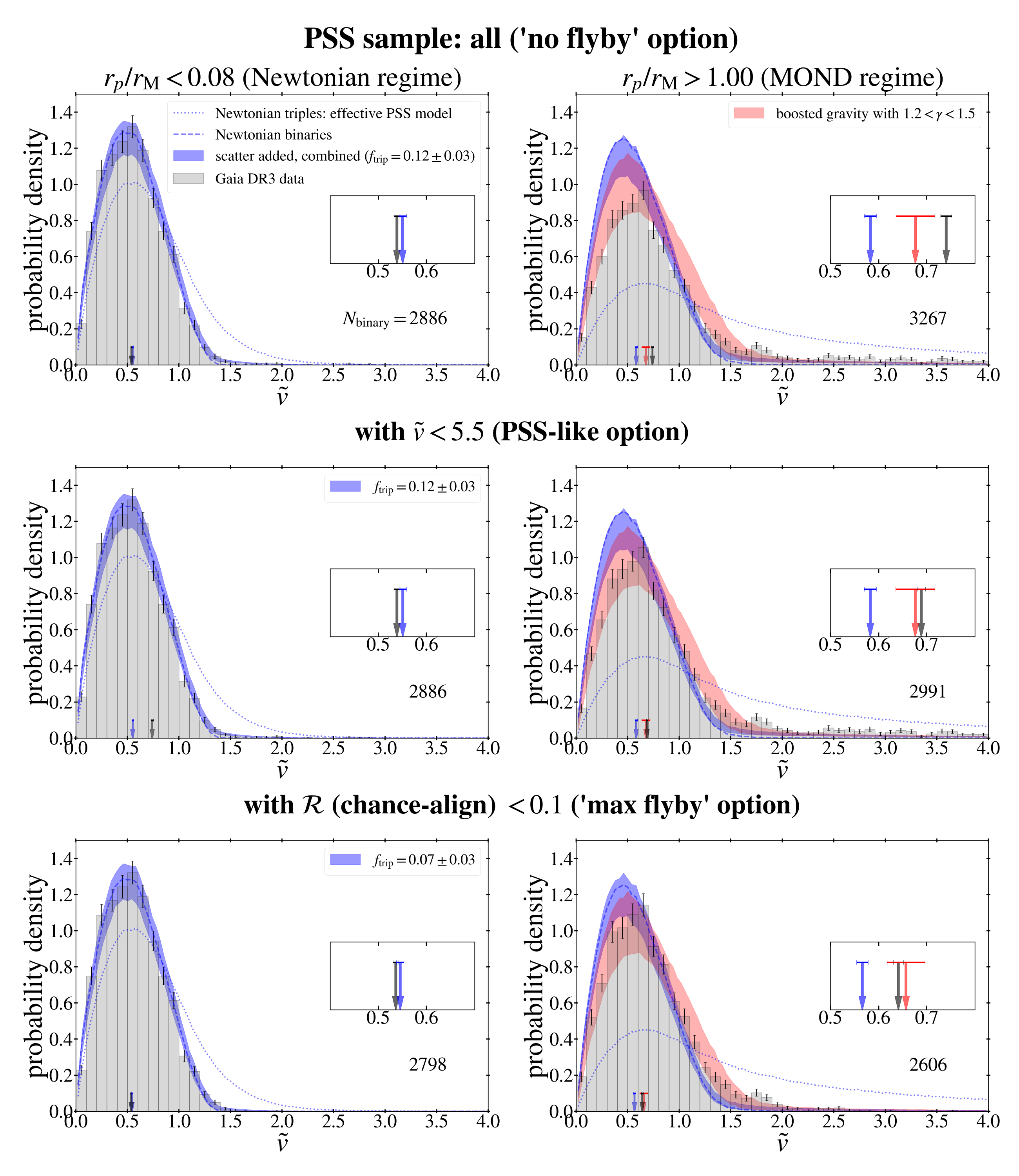}
    \caption{This figure shows $\tilde v$ distributions for the PSS sample in a similar format as Figure~\ref{fig:vtdist_Chaesample}. Since the PSS sample satisfies {\tt ruwe} $<1.2$, only the the PSS triple model is used. Regarding chance-alignment/flyby pairs included in the PSS sample, we consider three options in which none are removed (`no flyby' option), those with $\tilde v>5.5$ are removed (PSS-like option),  or all with the \cite{El-badry:2021} $\mathcal{R}>0.1$ are removed (`max flyby' option): see the text for the details. Considering that the PSS sample has a relatively large distance limit of 300~pc, we are excluding binaries with angular separation smaller than $2^{\prime\prime}$ for the subsamples in the Newtonian regime to ensure highest data qualities (though the angular cut has no impact). }
    \label{fig:vtdist_PSSsample}
\end{figure*}

We consider all three samples considered in this study (cf.\, Figure~\ref{fig:vpvt_rp}). In the case of the \cite{Banik:2024} sample, no acceleration-plane test was performed because it does not include any Newtonian-regime data, so we will assume a reasonable value of $f_{\rm trip}$ based on studies of the other samples. 

\begin{figure*}[tbh!]
    \centering
   \includegraphics[width=0.8\linewidth]{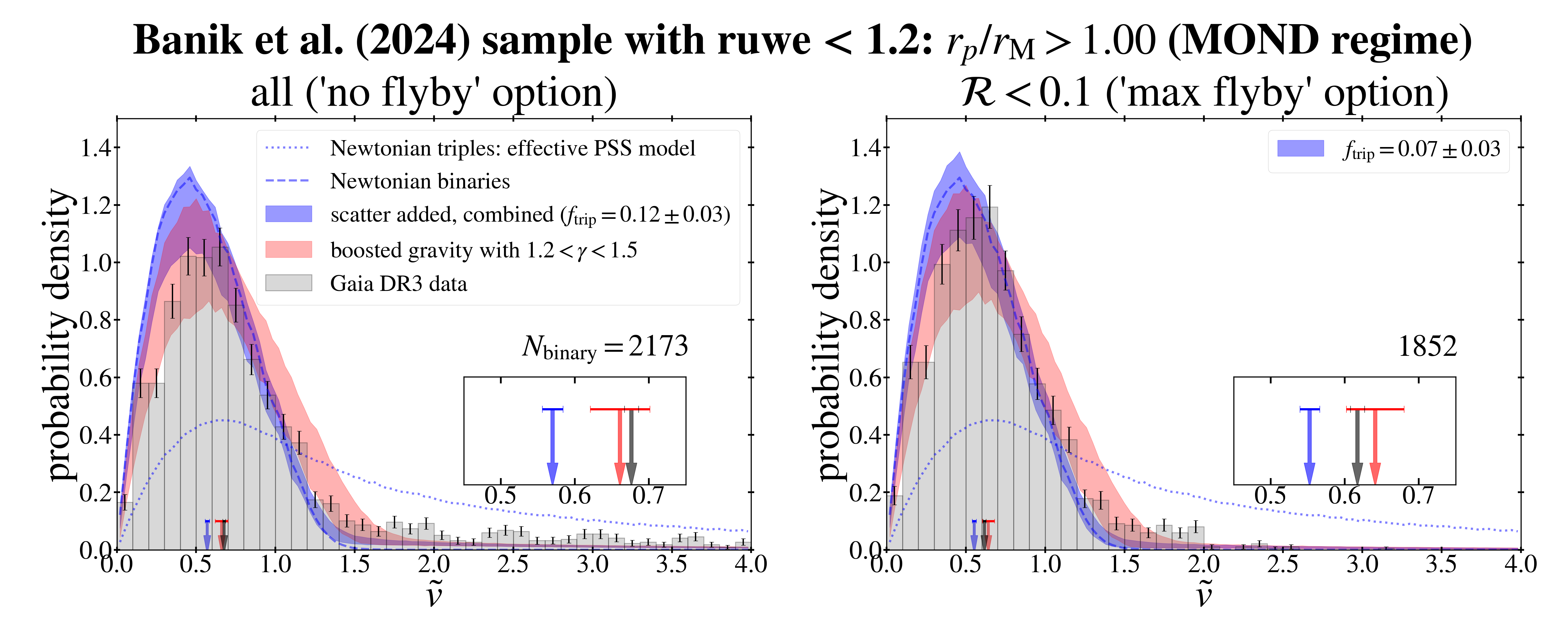}
    \caption{This figure shows $\tilde v$ distributions for the \cite{Banik:2024} sample with the limit {\tt ruwe} $<1.2$. The two panels are in the same format as those in the right column of Figure~\ref{fig:vtdist_PSSsample}. Since the \cite{Banik:2024} sample does not include small-$r_p$ data, we adopt $f_{\rm trip}$ values from the PSS sample with the same options. For the present sample, the `no flyby' option is equivalent to the PSS-like option because all pairs satisfy $\tilde v < 5$. }
    \label{fig:vtdist_Baniksample}
\end{figure*}

Figure~\ref{fig:vtdist_Chaesample} shows the $\tilde v$ distributions for the \cite{Chae:2023} sample. The uncertainties of the histogram bars and the median (indicated by the black arrow) are estimated from bootstrap samples (in Section~\ref{sec:nearby_WBs} we will show through simulations that the bootstrap error is reasonable). The upper row is for the entire sample without any limit on {\tt ruwe}, while the lower row is for the subsample with ${\tt ruwe} <1.2$. For the entire sample, the \cite{Chae:2023} multiple-star model is used, while for the subsample with ${\tt ruwe} <1.2$, the PSS triple model is used. In the Newtonian regime shown in the left column, the observed distributions and medians of $\tilde v$ are overall in agreement with the corresponding Newtonian predictions. The agreement is better for the subsample with ${\tt ruwe}<1.2$ and the PSS triple model. In detail, the general sample without the {\tt ruwe} cut shows some minor bump in the range $1.3\la \tilde v\la 1.7$, while the subsample with ${\tt ruwe}<1.2$ has a much weaker bump. In addition, the agreement between the observed median and the Newtonian prediction is better for the subsample with ${\tt ruwe}<1.2$. This may indicate that the subsample with ${\tt ruwe}<1.2$ and the PSS triple model are more reliable. 

In the MOND regime shown in the right column, the observed distributions and medians of $\tilde v$ clearly disagree with the corresponding Newtonian predictions in both cases. This disagreement is in stark contrast with the good agreement in the Newtonian regime. This result shows that Newtonian gravity breaks down in the low-acceleration regime in an \emph{absolute} sense, in line with the acceleration-plane test results. This is an important point regardless of whether wide binaries in the low-acceleration regime can be consistent with an existing nonstandard gravity model.

In the low-acceleration regime, we also test boosted gravity with the specific range $1.2<\gamma<1.5$ from numerical simulations of wide binary orbits (see Section~\ref{sec:mondgravity}). The boosted gravity model agrees much better with the data than the Newtonian model. In detail, the general sample without the {\tt ruwe} cut has an issue in the range $1.6\la\tilde v\la 2.0$, but the subsample with the {\tt ruwe} cut is fully consistent with the boosted gravity prediction with the PSS triple model. This indicates that some systems with ${\tt ruwe} > 1.2$ are contaminated and/or the \cite{Chae:2023} multiple-star model is not sufficiently precise.  

Figure~\ref{fig:vtdist_PSSsample} shows the $\tilde v$ distributions for the PSS sample. For this sample that satisfies {\tt ruwe} $<1.2$ by default, only the PSS triple model is used to calculate the kinematic contribution of hidden companions, and three options are considered regarding the removal of chance-alignment/flyby systems. In the Newtonian regime, the observed distribution and median of $\tilde v$ match well the Newtonian predictions with the values of $f_{\rm trip}$ from the acceleration-plane analysis in all three options. Unlike the \cite{Chae:2023} sample shown in Figure~\ref{fig:vtdist_Chaesample}, there is no noticeable bump near $\tilde v \approx 1.5$ in the Newtonian regime. The observed histogram is within the model uncertainty indicated by the blue band. Also, the observed median agrees well with predicted median in all cases. The good agreement in the Newtonian regime confirms the values of $f_{\rm trip}$ from the acceleration-plane analysis, since chance-alignment/flybys are negligible in the Newtonian regime.

In the MOND regime, the observed distribution cannot be described by the Newtonian model, in line with the result in Figure~\ref{fig:vtdist_Chaesample} for the \cite{Chae:2023} sample. The shape of the observed distribution is outside the range predicted by Newton. The Newtonian median is inconsistent with the observed median at a significance of $\approx 4.6\sigma$ (max flyby), $\approx 6.9\sigma$ (PSS-like flyby), or $\approx 9.1\sigma$ (no flyby).

The boosted gravity model performs much better than the Newtonian model in the MOND regime. In detail, the boosted gravity model is fully consistent with the data in the `max flyby' option both in shape and median. In the PSS-like flyby option, the predicted median agrees well with the observed median, but the shape does not match well for $\tilde v > 1.7$ as the observed distribution is somewhat above the predicted distribution. The result in the `no flyby' option is least consistent with the data. The predicted median is consistent with the observed median at $\approx 1.5\sigma$, and the predicted distribution shows some discrepancy with the observed distribution for $\tilde v \ga 1.7$. These issues indicate that chance-alignment/flyby pairs are nonnegligible in the low-acceleration regime of the PSS sample. 

Finally, Figure~\ref{fig:vtdist_Baniksample} shows the $\tilde v$ distributions in the MOND regime for the \cite{Banik:2024} sample with the limit {\tt ruwe} $<1.2$. Since the \cite{Banik:2024} sample includes chance-alignment/flyby pairs, we consider the two options of `no flyby' and `max fly'. Since we do not have any values of $f_{\rm trip}$ for the \cite{Banik:2024} sample, we simply take the fitted values of $f_{\rm trip}$ for the PSS sample in the two options of `no flyby' and `max flyby'. This seems reasonable as the fitted values of $f_{\rm trip}$ for the \cite{Chae:2023} sample and the PSS sample are similar in the same `max flyby' option. Clearly, the Newtonian model is discrepant with the observed distribution of $\tilde v$ in the \cite{Banik:2024} sample, in line with the results for the \cite{Chae:2023} sample and the PSS sample. The boosted gravity model performs much better than the Newtonian model in both options of flybys. However, there are issues in the range $\tilde v\ga 1.7$. In the case of the `max flyby' option, there is a minor bump near $1.8\la \tilde v \la 2.0$. In the case of the `no flyby' option, the observed distribution is somewhat above the predicted distribution for $\tilde v \ga 1.7$.

\subsection{Further comparison with previous results}\label{sec:compare_results}

For the PSS sample, \cite{Pittordis:2025} compared $\tilde v$ distributions in several ranges of $r_p$ (rather than $r_p/r_{\rm M}$) with the predictions of Newton and (their approximate/biased representation of) MOND gravity. Here we carry out additional analyses of $\tilde v$ distributions in specific ranges of $r_p$ so that their results can be more directly compared with our corresponding results. 

\cite{Pittordis:2025} considered ranges of $r_p$ from $1.25<r_p<1.77\,{\rm kau}$ to $14.1<r_p<20\,{\rm kau}$. We consider the lowest- and highest-$r_p$ ranges used by \cite{Pittordis:2025}. However, while we use the same range of $1.25<r_p<1.77\,{\rm kau}$, which includes up to $N_{\rm binary}=2833$ binaries, we consider a combined range of $10<r_p<20\,{\rm kau}$ (up to $1486$ binaries) since the range of $14.1<r_p<20\,{\rm kau}$ has relatively few ($\le 675$) binaries. We further consider a range of $r_p<0.7\,{\rm kau}$ with the constraint of angular separation $\Delta\theta>2^{\prime\prime}$, as binaries in the fully Newtonian regime as explained below. Here the angular separation constraint is recommended by observers\footnote{For example, Will Sutherland by private communication.} particularly because the PSS sample includes binaries up to distance of 300\,pc. However, only 3.7\% of the binaries in the range $r_p < 0.7\,{\rm kau}$ are excluded by the angular constraint and the exclusion has no impact on gravity tests. 

\begin{figure}[tbh!]
    \centering
   \includegraphics[width=0.9\linewidth]{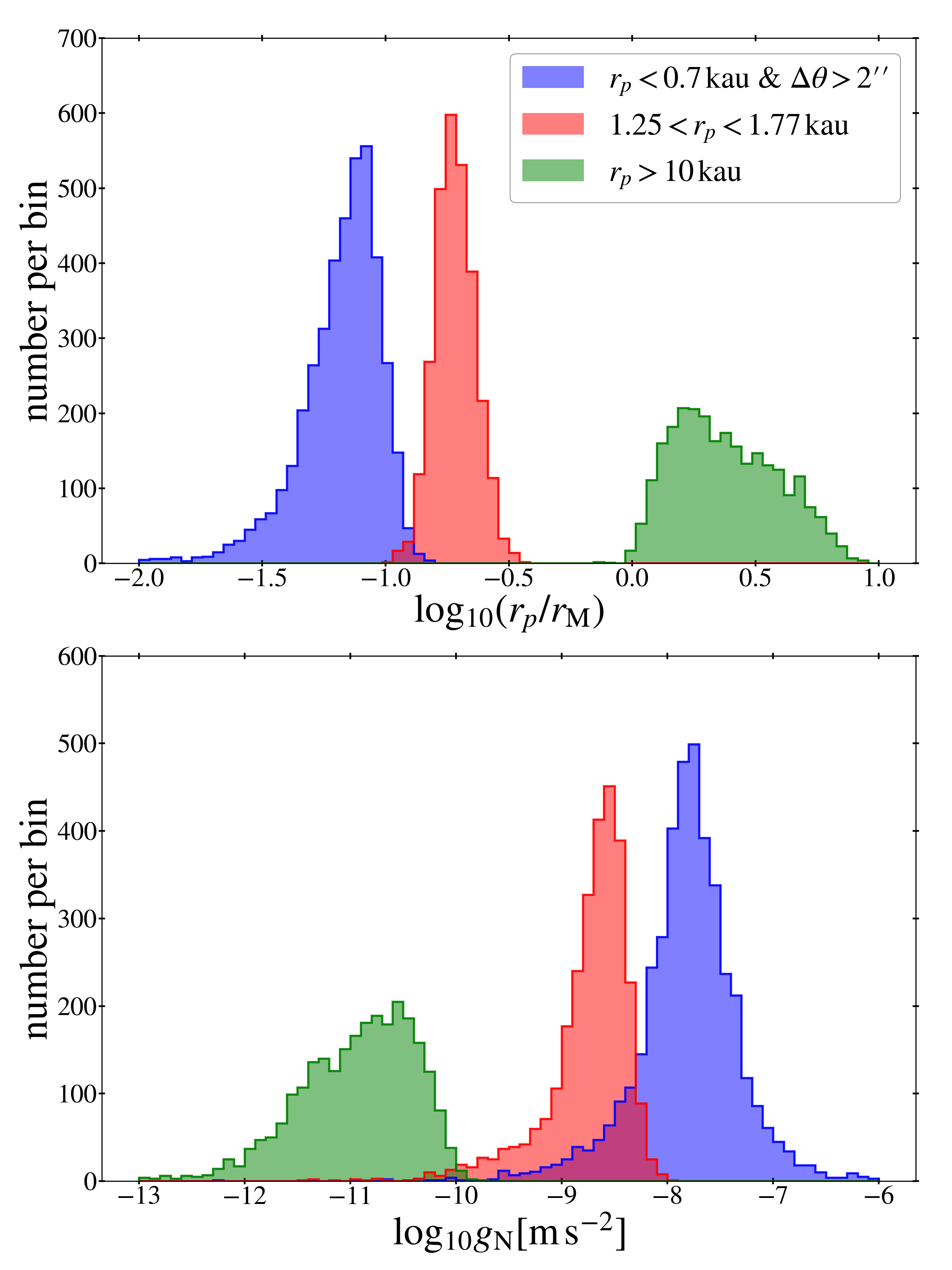}
    \caption{For wide binaries in the three ranges of $r_p$ considered in Section~\ref{sec:compare_results}, the distributions of $r_p/r_{\rm M}$ and $g_{\rm N}=G_{\rm N}M_{\rm tot}/r^2$ are shown. For binaries in the range $r_p < 0.7\,{\rm kau}$, we exclude systems whose angular separation is less than 2 arcseconds (3.7\%) to ensure highest possible data qualities, although this exclusion has negligible effects. }
    \label{fig:rp_rM}
\end{figure}

A fully Newtonian (i.e., sufficiently small $r_p$) range is valuable because chance-alignment/flybys are negligible and thus the range can be used to test the validity of $f_{\rm trip}$ determined from the acceleration-plane analysis. The lowest-$r_p$ range $1.25<r_p<1.77\,{\rm kau}$ used by \cite{Pittordis:2025} is not fully Newtonian because the 3D separation $r$ deprojected from $r_p$ can be in the transition or MOND regime (or, the internal acceleration $\la 10^{-9}\,{\rm m}\,{\rm s}^{-2}$). This can be explicitly seen by MC simulations taking random values of the orbit and orientation parameters. Figure~\ref{fig:rp_rM} shows the results from MC simulations for the three $r_p$ ranges. All binaries in the range $10<r_p<20\,{\rm kau}$ are completely in the MOND regime with $g_{\rm N}\la 10^{-10}\,{\rm m}\,{\rm s}^{-2}$. On the other hand, most cases in the range $r_p<0.7\,{\rm kau}$ satisfy $g_{\rm N}>10^{-9}\,{\rm m}\,{\rm s}^{-2}$ indicating that the range is statistically Newtonian containing only a tiny occurrence rate of transition-regime systems. However, the range $1.25<r_p<1.77\,{\rm kau}$ contains nonnegligible occurrence rate of $g_{\rm N}< 10^{-9}\,{\rm m}\,{\rm s}^{-2}$ showing that it is predominantly a mix of Newtonian and transition-regime systems, even including some MOND-regime systems. 

\begin{figure}[t!]
    \centering
   \includegraphics[width=0.9\linewidth]{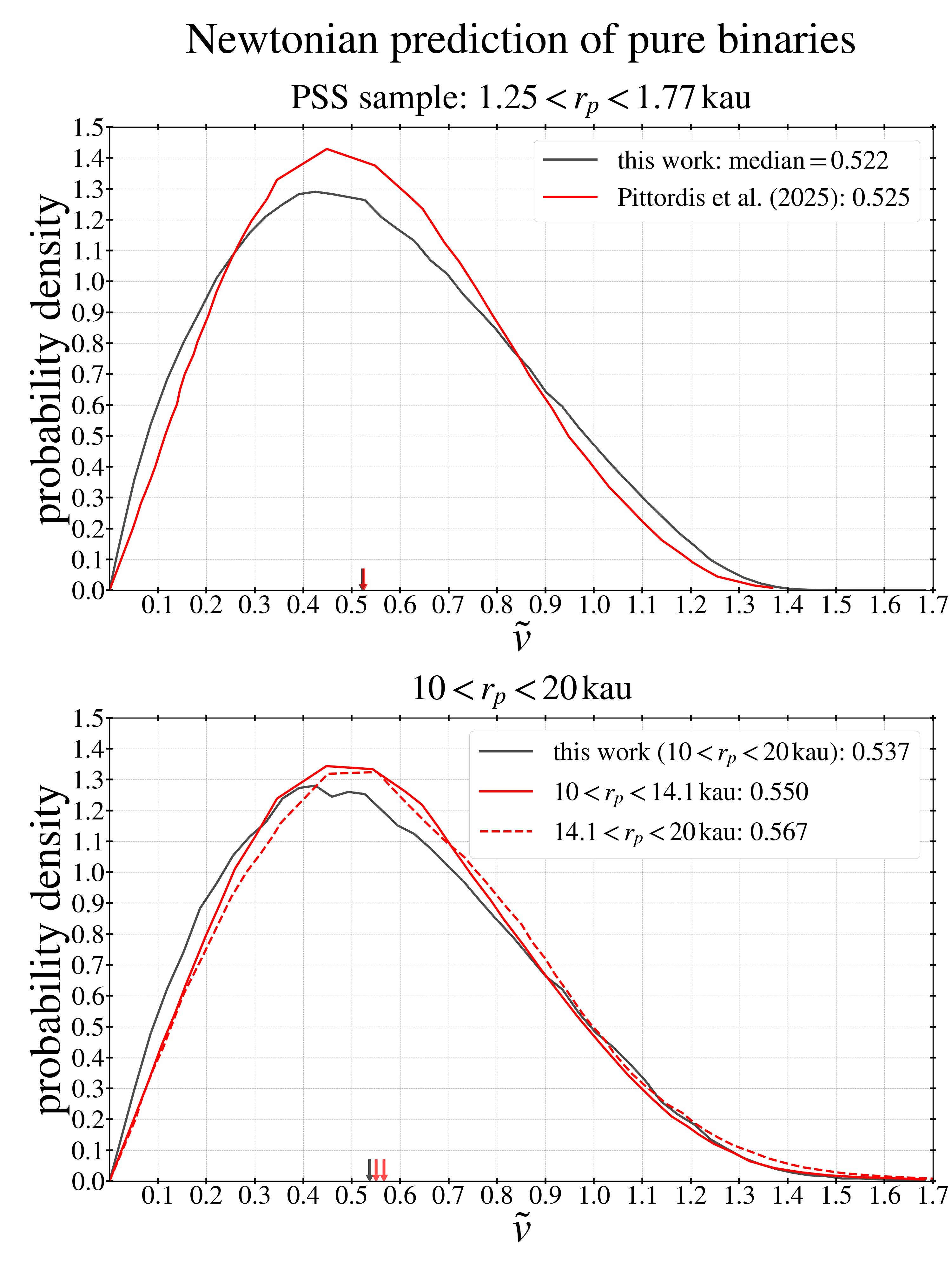}
    \caption{This figure compares Newtonian predictions of the $\tilde v$ distribution for pure binaries by this work with those by \cite{Pittordis:2025}. Because observational scatters are added to the theoretical predictions, ${\tilde v} > 1.4$ occur for wide binaries in the range $10<r_p<20\,{\rm kau}$. Because of differences in eccentricities and how scatters are added, our results and \cite{Pittordis:2025} results show some differences. }
    \label{fig:vtmedian_rp}
\end{figure}

The differences between the \cite{Pittordis:2025} analyses and our analyses have been described in Section~\ref{sec:test_vtdist}. Considering these differences, we first examine Newtonian predictions of the $\tilde v$ distribution for pure binaries by \cite{Pittordis:2025} and ourselves. In Figure~\ref{fig:vtmedian_rp}, the $\tilde v$ distributions of Newtonian pure binaries for $1.25<r_p<1.77\,{\rm kau}$ and $10<r_p<14.1\,{\rm kau}$/$14.1<r_p<20\,{\rm kau}$ from Figure~9 of \cite{Pittordis:2025} are compared with our predictions. For each range, our distribution and the \cite{Pittordis:2025} distribution have moderate differences in shape and/or median. In particular, our median $\tilde v$ for the range $10<r_p<20\,{\rm kau}$ is lower by $\approx 0.02$ than the corresponding one by \cite{Pittordis:2025}. This is due to two factors: (1) we use individual Bayesian information of eccentricities while \cite{Pittordis:2025} use the power-law distribution (Equation~(\ref{eq:powerlaw})) with $\alpha=1.3$; (2) we use observationally estimated uncertainties of $v_p$ and a mass uncertainty of 6.8\% to give scatter to $\tilde v$ while \cite{Pittordis:2025} use simulated measurement noise (see Section~3.2.2 of \cite{Pittordis:2025}). The difference in the range $10<r_p<20\,{\rm kau}$ have a subtle effect in gravity tests in the low-acceleration regime. We note that our calculations of Newtonian predictions have been verified in the Newtonian regime, but such a calibration step is missing in the \cite{Pittordis:2025} calculations.

\begin{figure*}[tbh!]
    \centering
   \includegraphics[width=1.0\linewidth]{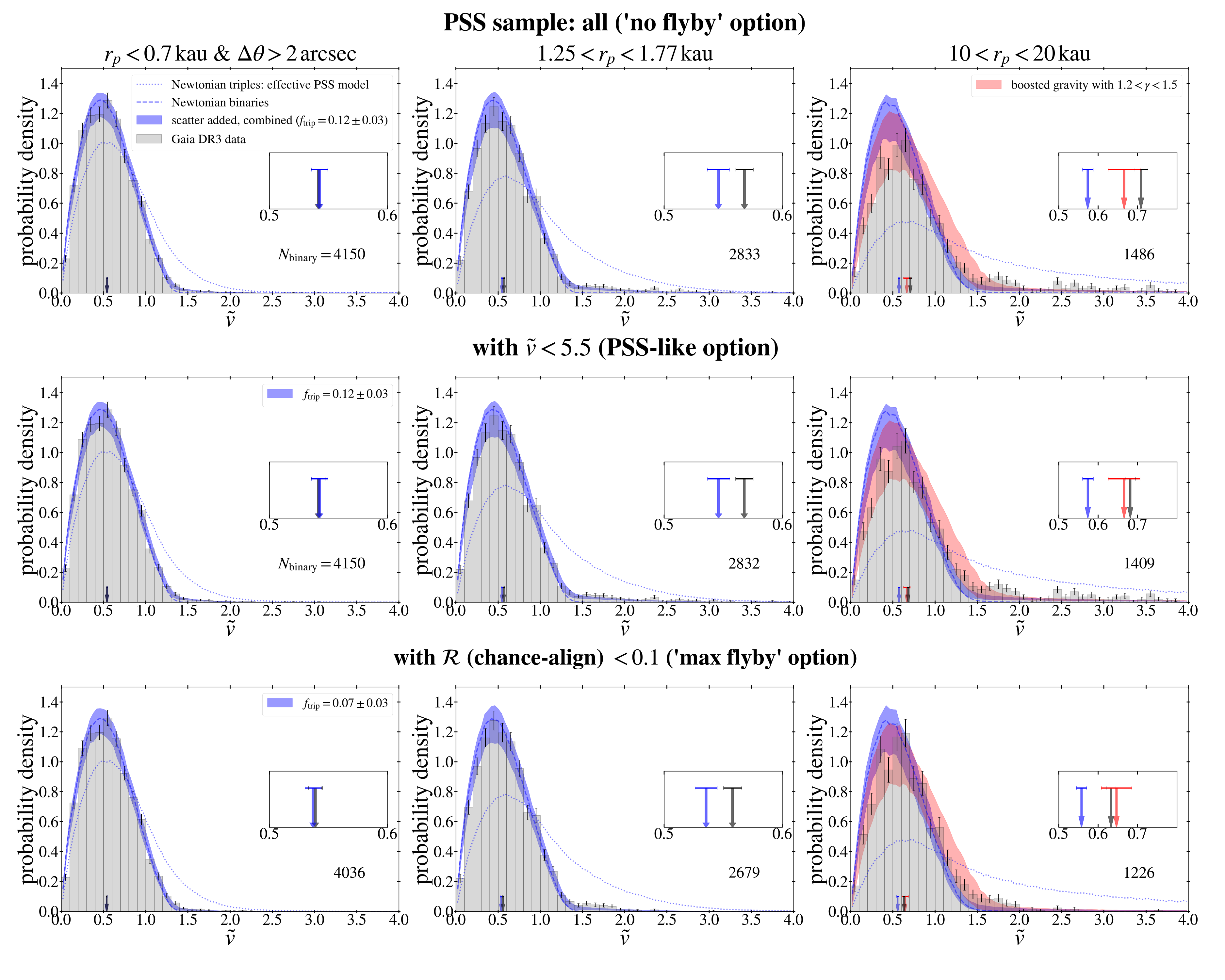}
    \caption{This figure is similar to Figure~\ref{fig:vtdist_PSSsample} but three ranges of $r_p$ are considered to compare more directly with the results by \cite{Pittordis:2025}. Note, however, that the fully Newtonian regime $r_p < 0.7\,{\rm kau}$ was not considered by \cite{Pittordis:2025}. }
    \label{fig:vtdist_PSSsample_rp}
\end{figure*}

Figure~\ref{fig:vtdist_PSSsample_rp} shows the results of testing gravity in the three ranges of $r_p$. Here we use our fitted values of $f_{\rm trip}$ from the acceleration-plane analysis in each of the `no flyby', PSS-like, and `max flyby' options. For the ranges of $r_p<0.7\,{\rm kau}$ and $1.25<r_p<1.77\,{\rm kau}$, only Newtonian gravity is tested. For the range of $10<r_p<20\,{\rm kau}$, both Newtonian and boosted gravities are tested. In the Newtonian regime of $r_p<0.7\,{\rm kau}$, the observed distribution of $\tilde v$ is in excellent agreement with the Newtonian prediction in both the `max flyby' and no/PSS-like flyby options. Because the sample size is sufficiently large ($N_{\rm binary}>4000$), the observed $\tilde v$ distribution is well defined, and both the shape and median of the distribution agree remarkably well with the Newtonian prediction, which is based on the most likely or reasonable input of eccentricities and $f_{\rm trip}$. This good agreement in the Newtonian regime is well consistent with the acceleration-plane test results shown in Figure~\ref{fig:residual_7bins_PSSsample_ftripfitted}. 

Unlike the range $r_p<0.7\,{\rm kau}$, the observed distribution of $\tilde v$ in the range $1.25<r_p<1.77\,{\rm kau}$ does not match well the Newtonian prediction in either of the `max flyby' and no/PSS-like flyby options. The observed median is about $2\sigma$ higher than the Newtonian prediction, and the distribution for $\tilde v > 1.5$ shows some discrepancy. This discrepancy in the median is unexpected from the viewpoint of Newtonian gravity because it demands that binaries in the range $1.25<r_p<1.77\,{\rm kau}$ must be gravitationally bound and satisfy Newtonian gravity. However, the discrepancy can be attributed to systems in the transition regime ($g_{\rm N}\la 10^{-9}\,{\rm m}\,{\rm s}^{-2}$) from the viewpoint of MOND gravity.

The low-acceleration regime with $10<r_p<20\,{\rm kau}$ shown in the right column of Figure~\ref{fig:vtdist_PSSsample_rp} reveals that Newtonian gravity is $>4\sigma$ discrepant with the observed median in any option of flyby assumptions. Also, the shape of the observed distribution is clearly discrepant with the Newtonian distribution. In contrast, the prediction of boosted gravity with a gravity boost range of $1.2<\gamma<1.5$ (or a velocity boost range of $1.10< \gamma_v <1.22$) matches the observed distribution. In detail, the predicted median matches well the observed median in both of the PSS-like and max flyby options. In the `max flyby' option, the shape of the observed distribution is well within the predicted shape range. This good agreement reaffirms the results for the PSS sample shown in Figure~\ref{fig:residual_7bins_PSSsample_ftripfitted} and Figure~\ref{fig:vtdist_PSSsample}.

\begin{figure*}[tbh!]
    \centering
   \includegraphics[width=1.0\linewidth]{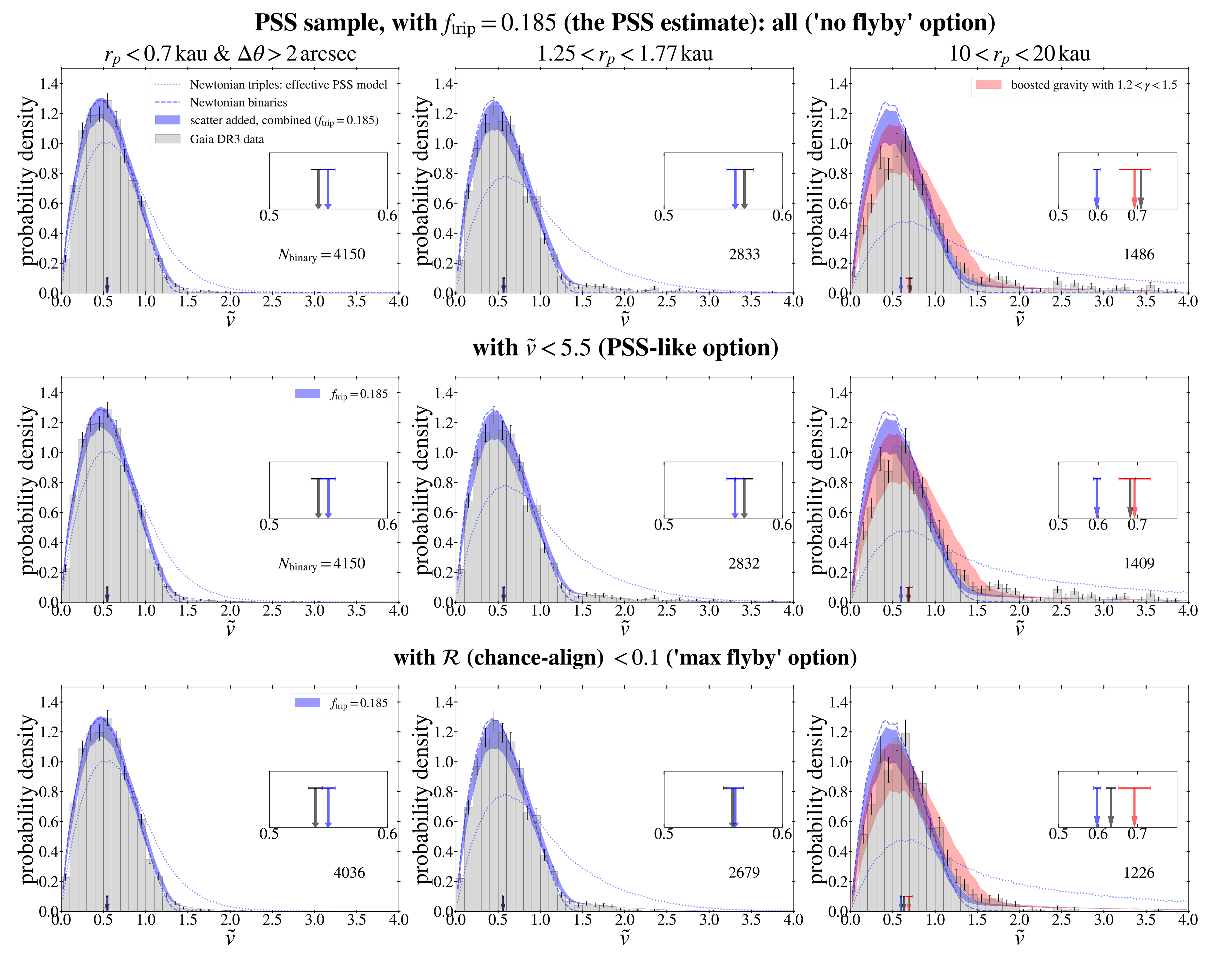}
    \caption{same as Figure~\ref{fig:vtdist_PSSsample_rp} but for a fixed value of $f_{\rm trip}=0.185$ from \cite{Pittordis:2025}. }
    \label{fig:vtdist_PSSsample_rp_ftrip185}
\end{figure*}

\cite{Pittordis:2025} obtained $f_{\rm trip}=0.185$ without considering a Newtonian regime and claimed that Newtonian gravity was moderately preferred over MOND gravity in the range $10<r_p<20\,{\rm kau}$ and more strongly preferred in other ranges of $r_p$ including the range $1.25<r_p<1.77\,{\rm kau}$. To test the validity of their claim more directly, we consider $f_{\rm trip}=0.185$ and the results are shown in Figure~\ref{fig:vtdist_PSSsample_rp_ftrip185}. In this test, the `max flyby' option and the PSS-like flyby (or `no flyby') option give qualitatively different results. 

In the `max flyby' option of Figure~\ref{fig:vtdist_PSSsample_rp_ftrip185}, the Newtonian prediction matches well the observed distribution in the range $1.25<r_p<1.77\,{\rm kau}$. However, in the MOND regime, the observed median is between the Newtonian prediction and the boosted gravity prediction, being somewhat closer to the Newtonian prediction in terms of absolute difference. At face value, these results appear to be qualitatively consistent with the \cite{Pittordis:2025} results. However, the observed median of $\tilde v$ is $2.3\sigma$ away from the Newtonian median in the MOND regime. Thus, in an absolute sense, Newtonian gravity has some tension in the MOND regime even with $f_{\rm trip}=0.185$. Moreover, in contrast to the case of Figure~\ref{fig:vtdist_PSSsample_rp}, in the Newtonian regime of $r_p<0.7\,{\rm kau}$ Newtonian gravity starts to deviate from the observed distribution of $\tilde v$: the histogram bars near $\tilde v =1.5$ are lower than the Newtonian band, and the Newtonian median of $\tilde v$ starts to deviate from the observed median. If a higher value of $f_{\rm trip}$ is considered to make the Newtonian median fully agree with the observed median in the MOND regime, then the discrepancy of Newton in the Newtonian regime will get clearer.

In the PSS-like or no flyby option of Figure~\ref{fig:vtdist_PSSsample_rp_ftrip185}, tension of Newtonian gravity is slightly lessened in the range $r_p<0.7\,{\rm kau}$. However, in the MOND regime of $10<r_p<20\,{\rm kau}$, Newtonian gravity has a worsened tension of $> 5\sigma$. In contrast, MOND/boosted gravity agrees well with the observed distribution in terms of the median, although there is some tension in the shape of the distribution at large $\tilde v$. 

Therefore, in an absolute sense, Newtonian gravity has a significant tension with the data in the MOND regime regardless of whether our fitted value of $f_{\rm trip}$ or the \cite{Pittordis:2025} value is used. In contrast, MOND gravity represented by a boosted gravity is well consistent with the data. This is consistent with the acceleration-plane test results shown in Figure~\ref{fig:residual_7bins_PSSsample_ftripfixed} and Figure~\ref{fig:residual_7bins_PSSsample_ftripfitted} where the bins with $x_0 \la -10$ are clearly discrepant with Newton regardless of the choice of $f_{\rm trip}$. This conclusion is inconsistent with the \cite{Pittordis:2025} claim that Newton was preferred over their MOND model. This disagreement can be attributed to several factors. 

First, \cite{Pittordis:2025} did not consider the fully Newtonian regime of $r_p<0.7\,{\rm kau}$, so they could not check whether their value of $f_{\rm trip}$ started to cause tension with Newton. Second, our prediction of the $\tilde v$ distribution for Newtonian pure binaries is somewhat different from the \cite{Pittordis:2025} prediction as shown in Figure~\ref{fig:vtmedian_rp} due to differences in eccentricities and how the scatters of $\tilde v$ are added. Third, we fix $f_{\rm flyby}$ for the data and consider three options of $f_{\rm flyby}$ including the two extreme options, while \cite{Pittordis:2025} fit $f_{\rm flyby}$ separately for each gravity model resulting in a factor of 4 different values for Newton and MOND models. Fourth, our representation of MOND gravity covers a broad range of gravity boost factor of $1.2<\gamma<1.5$ motivated from realistic two-body orbit simulations by \cite{Pflamm-Altenburg:2025}, while \cite{Pittordis:2025} use a model based on approximate numerical results. Fifth, \cite{Pittordis:2025} did not test Newtonian gravity in an absolute sense but tested only relative merits of Newton and their MOND model by calculating $\chi^2$ of the two models. Finally, our fitted value of $f_{\rm trip}$ from the acceleration plane analysis is significantly lower than the value used by \cite{Pittordis:2025} that was not calibrated with the Newtonian-regime data free of chance-alignment/flybys.

Comparison of the left columns of Figure~\ref{fig:vtdist_PSSsample_rp} and Figure~\ref{fig:vtdist_PSSsample_rp_ftrip185} shows that a fully Newtonian regime with small separation has a sensitivity to $f_{\rm trip}$. A sample within such a range of $r_p$ is effectively free of chance-alignment/flyby pairs, so that the observed $\tilde v$ distribution is (almost) entirely determined by pure binaries and triples. Thus, a specific value of $f_{\rm trip}$ can be tested, or $f_{\rm trip}$ can be determined by comparing the observed distribution with the predicted distribution as a  function of $f_{\rm trip}$. This determination is certainly more reliable than fitting both $f_{\rm trip}$ and $f_{\rm flyby}$ simultaneously using a sample of large separations that contains flybys because the latter approach suffers from an error due to the degeneracy between the two parameters. A factor of 4 difference in the fitted values of $f_{\rm flyby}$ for the Newton and MOND models by \cite{Pittordis:2025} may provide such an example of the degeneracy issue.

\cite{Pittordis:2025} raised criticisms against previous studies that detected MOND effects. They found that a significant proportion of apparent binaries with hidden tertiaries have $\tilde v > 1.5$ and this proportion increases with $r_p$ (see Figure~\ref{fig:vt_triple_PSS}). In other words, apparent binaries with hidden tertiaries tend to have larger $\tilde v$ for larger $r_p$. Thus, they claimed that this had led to an increase in $\tilde v$ for larger $r_p$, which was misinterpreted as the MOND signal.

\begin{figure}[h!]
    \centering
   \includegraphics[width=1.0\linewidth]{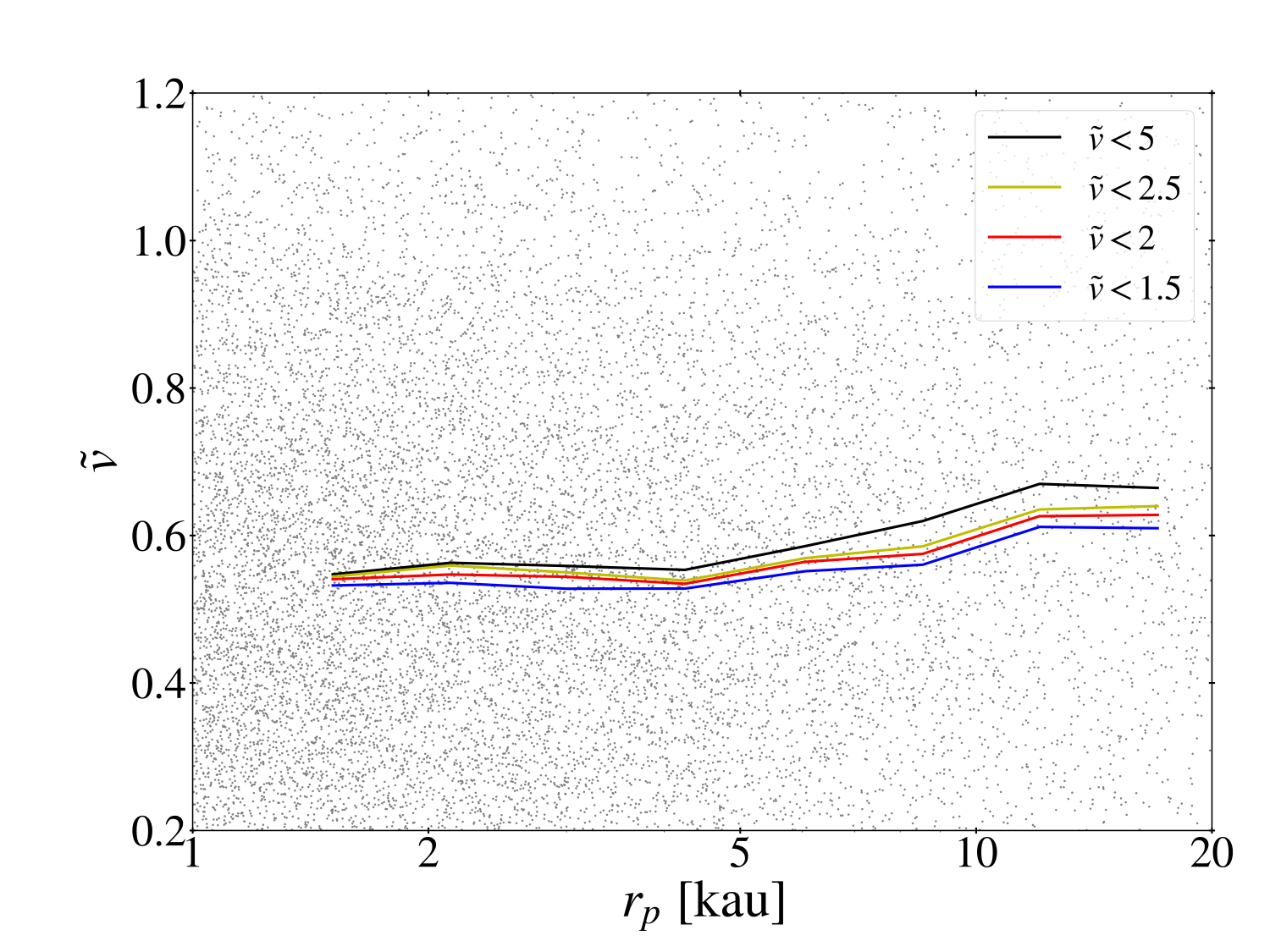}
    \caption{The median of $\tilde v$ increases with $r_p$, even for subsamples with $\tilde v<1.5,2,2.5,$ and $5$.
    }
    \label{fig:vt_rp_PSS}
\end{figure}

However, as we saw in Figure~\ref{fig:vt_scaling_3samples_vtmax15}, the median of $\tilde v$ has an increasing behavior, even with the limit $\tilde v<1.5$. This is apparent not only when plotting $\tilde v$ as a function of $r_p/r_{\rm M}$, but also when plotting it as a function of $r_p$, as shown in Figure~\ref{fig:vt_rp_PSS}. In this figure, for a direct comparison, we chose our bins to correspond to the $r_p$ ranges of the triple model in Figure~8 and Table~2 of \cite{Pittordis:2025}. In the case of the PSS sample, the median of $\tilde v$ increases by 0.078 from 0.532 to 0.610 for $\tilde v<1.5$, and by 0.087 from 0.541 to 0.628 for $\tilde v<2$. 

\begin{figure*}[tbh!]
    \centering
   \includegraphics[width=1.0\linewidth]{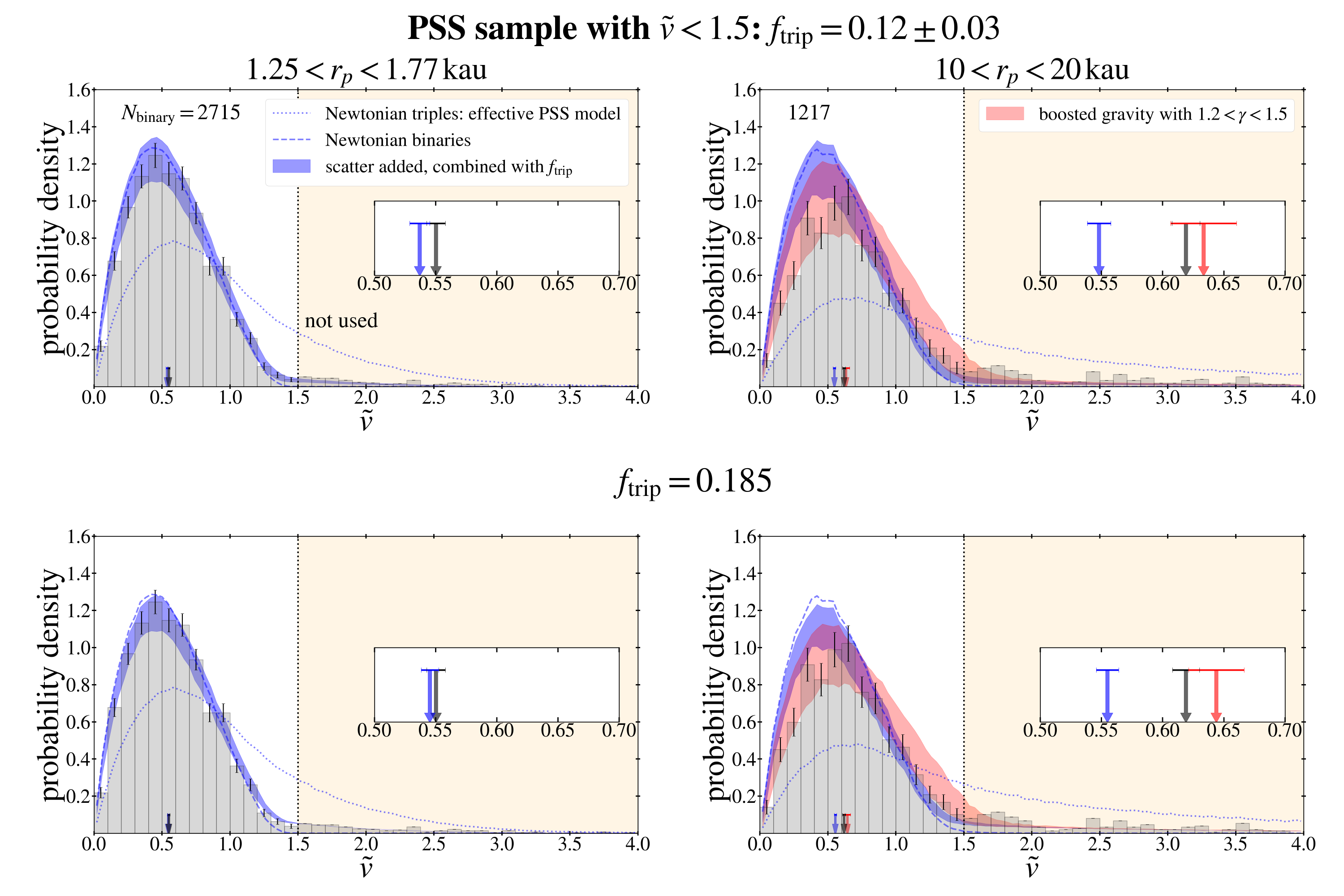}
    \caption{This figure shows results of the $\tilde v$-distribution test with the limit $\tilde v<1.5$ for the PSS sample. Two ranges of $r_p$ used in Figure~\ref{fig:vtdist_PSSsample_rp} are considered. The upper row shows the results with our fitted value of $f_{\rm trip}$ while the lower row shows the results with the \cite{Pittordis:2025} value.}
    \label{fig:vtdist_PSSsample_rp_vtmax15}
\end{figure*}

Figure~\ref{fig:vtdist_PSSsample_rp_vtmax15} shows the results of the $\tilde v$-distribution test with the limit $\tilde v < 1.5$. Since chance-alignment/flyby pairs are negligible with $\tilde v < 1.5$, $f_{\rm trip}$ is the only unknown. We consider two cases of $f_{\rm trip}=0.12\pm0.03$ (our fit) and $0.185$ (the \cite{Pittordis:2025} fit). In the range of $10<r_p<20\,{\rm kau}$, Newton is strongly discrepant with the data, while the boosted gravity agrees well with it, in both cases of $f_{\rm trip}$. Notably, the increase of $f_{\rm trip}$ does not resolve the discrepancy of Newton at all. Comparison of the lower right panel of Figure~\ref{fig:vtdist_PSSsample_rp_vtmax15} with the bottom right panel of Figure~\ref{fig:vtdist_PSSsample_rp_ftrip185} indicates that the discrepancy is weaker in the latter case probably because some wide binaries with $\tilde v$ (see Figure~\ref{fig:vt_mass_hist}) were removed from the sample with $\mathcal{R}<0.1$ just because $\mathcal{R}$ values are not available for them. In the following section, we will carry out extensive analyses of highest-quality samples satisfying $\tilde v<1.5$.

\cite{Pittordis:2025} also noted that Chae's sample had been derived from \cite{El-badry:2021}, which contained a cutoff $v_p\leq 2.1\,{\rm  km}\,{\rm s}^{-1}(r_p/1\,{\rm kau})^{-0.5}$, or equivalently $\tilde v\leq 2.24/\sqrt{M_{\rm tot}}$. They argued that ``the extended tail at $\tilde v>2$ is almost entirely removed from the Chae sample'' and ``greatly reduces the leverage on $f_{\rm trip}$.'' A few comments are warranted here. First, due to scatter in the data, the cut is not a hard one as shown in Figure~1 of \cite{Chae:2024b}. In particular, the claim that $\tilde v>2$ is almost entirely removed is exaggerated as can be clearly seen from Figure~\ref{fig:vpvt_rp} and Figure~\ref{fig:vtdist_Chaesample}. Second, the distribution of $\tilde v$ for triples is limited to $\tilde v \la 2$ (see the left column of Figure~\ref{fig:vtdist_PSSsample}) in the Newtonian regime of sufficiently small $r_p$ where $f_{\rm trip}$ can be reliably calibrated without the concern of chance-alignment/flybys. Finally, the cut would tend to not create but remove MOND effects in a study of the median trend of $\tilde v$ with $r_p$, as excluding $\tilde v$ beyond a fixed threshold would more affect data in the MOND regime, depressing $\tilde v$ more. Thus, MOND effects would be slightly enhanced, rather than suppressed, without the cut.

\section{Testing Gravity with Nearby Highest-quality Samples}\label{sec:nearby_WBs}

\cite{Cookson:2026} have discussed various quality cuts including the Banik cut, a {\tt ruwe} cut, a CMD cut, a requirement on Gaia's {\tt ipd\_frac\_multi\_peak} parameter, and a tight distance limit such as $<130\,{\rm pc}$ (or $<150\,{\rm pc}$) to mitigate the role of contaminated data in gravity tests. They also emphasized correcting measured velocities for the perspective effect. They proposed a quality framework for wide binary data in statistical tests with sky-projected velocities. Based on a small sample of wide binaries satisfying all their quality cuts, \cite{Cookson:2026} claimed that there was no evidence for the low-acceleration gravitational anomaly, in contrast to a number of recent results based on various samples and methodologies including 3D velocity analysis. Here we first discuss obvious problems in their analysis and results, and then carry out our own analysis based on a much larger sample satisfying all their quality cuts.

\cite{Cookson:2026} investigated the medians of $\tilde v$ in bins of $r_p/r_{\rm M}$. Their sample has a total number of binaries from 1216 to 1421 (their Table~3) including just 61 binaries satisfying $r_p/r_{\rm M}>1$ (fully MOND/low-acceleration regime). We identify the following problems in the analysis, the results, and the claim. 

First, their sample size is not large enough to achieve even a $2\sigma$ distinction between Newton and boosted gravity with $\gamma=1.4$ in the regime of $r_p/r_{\rm M}>1$. To check this, we carry out mock observations of wide binaries satisfying $r_p/r_{\rm M}>1$ as a function of the number of wide binaries. For this, we use wide binaries from the PSS sample satisfying $10<r_p<20\,{\rm kau}$ and $d<150\,{\rm pc}$, and produce mock wide binaries satisfying pseudo-Newtonian orbits in boosted gravity with $\gamma=1.4$ by taking random values of the orbit and orientation parameters as usual. We perform $2\times 10^6$ mock observations of $N_{\rm binary}$ wide binaries and obtain the predicted distribution of median $\tilde v$. 

\begin{figure}[tbh!]
    \centering
   \includegraphics[width=1.0\linewidth]{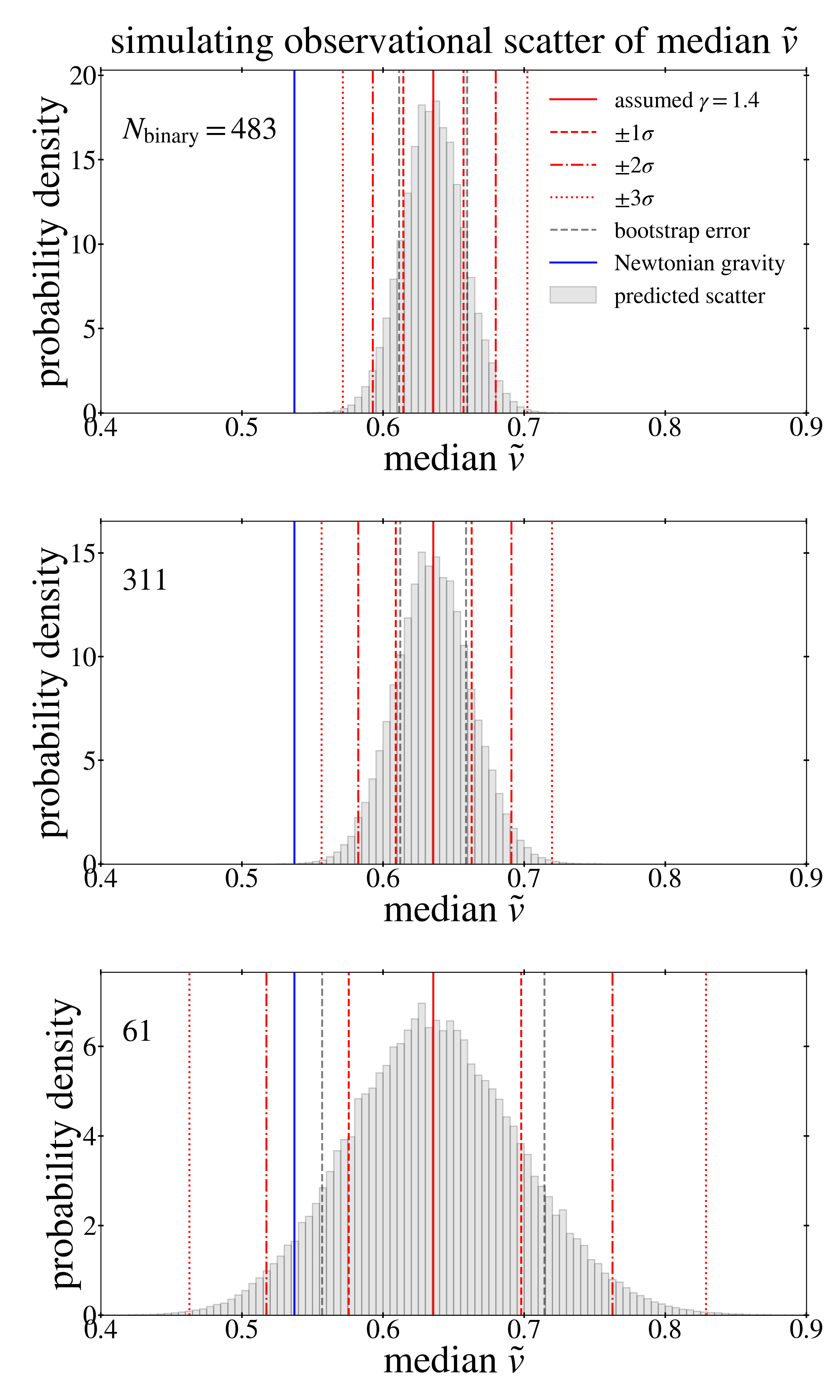}
    \caption{Each histogram represents the distribution of median $\tilde v$ in $2\times 10^6$ random observations of $N_{\rm binary}$ mock wide binaries obeying pseudo-Newtonian gravity with a gravity boost factor of $\gamma=1.4$ (so, a velocity boost of $\sqrt{1.4}$). Mock wide binaries are produced using actual wide binaries in the low-acceleration regime (satisfying $10<r_p<20\,{\rm kau}$) taken from the PSS sample. When $N_{\rm binary}=483$ (311), Newton would be outside $4.67\sigma$ ($3.73\sigma$) confidence limit. Thus, in these cases, Newton and a boosted gravity are expected to be discriminated with a high confidence. However, when $N_{\rm binary}=61$ (the case of \cite{Cookson:2026}), Newton would be just outside the $1.66\sigma$ confidence limit, far below the $3\sigma$ rule of thumb. In this case, Newton and the assumed boosted gravity cannot be distinguished meaningfully due to the small sample size. We note that a bootstrap error estimated from one mock observation is similar to $1\sigma$ from the distribution. This property will be used in estimating the uncertainty in statistical analyses of real samples. }
    \label{fig:vtmedian_simulation}
\end{figure}

Figure~\ref{fig:vtmedian_simulation} shows the results for three cases of $N_{\rm binary}=483$, $311$, and $61$, which match the sizes of the samples to be used below, or used by \cite{Cookson:2026}. When $N_{\rm binary}=61$, Newtonian gravity would occur within $2\sigma$ of the uncertainty, meaning that such a small sample of 2D velocity data cannot meaningfully discriminate between boosted gravity and Newton. In other words, out of a large sample of wide binaries obeying boosted gravity, it is not difficult to select a sample of 61 wide binaries with $10<r_p<20\,{\rm kau}$ whose median is Newtonian. This will be explicitly demonstrated below with the PSS sample. Thus, it is even in principle not possible to claim meaningful evidence for or against Newton or MOND with the \cite{Cookson:2026} sample. Their claim of ``no evidence for MOND'' is not justified based on basic statistics even without considering other issues below. However, when $N_{\rm binary}=483$ or $311$, the data can clearly distinguish between Newton and boosted gravity such as $\gamma=1.4$, with a statistical significance well above $3\sigma$. Such samples will be considered below.

\begin{figure}[tbh!]
    \centering
   \includegraphics[width=1.0\linewidth]{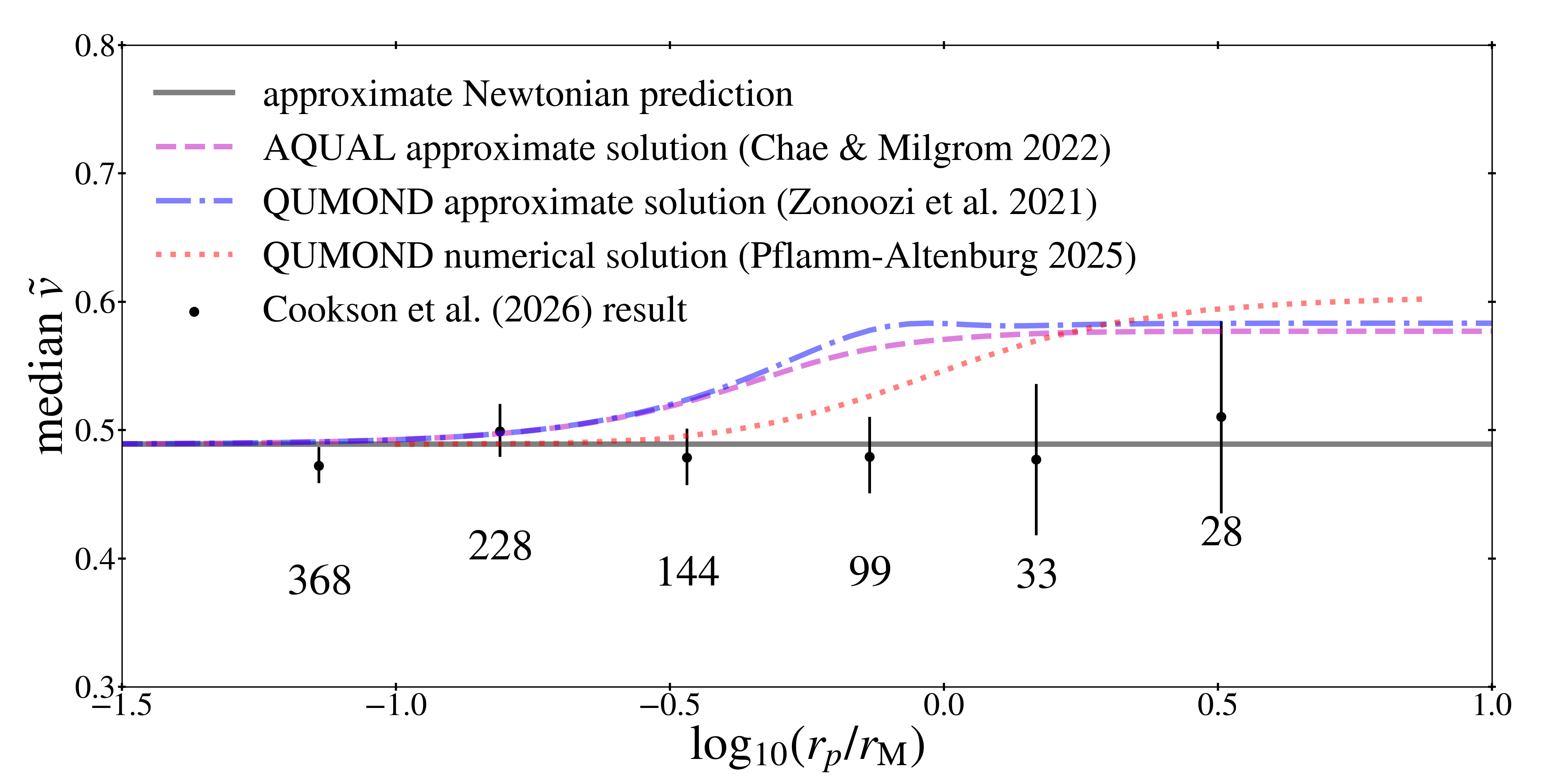}
    \caption{The profile of median $\tilde v$ from Figure~8 of \cite{Cookson:2026} is compared with the latest numerical prediction of QUMOND by \cite{Pflamm-Altenburg:2025} as well as past approximate numerical predictions. }
    \label{fig:vtprofile_Cookson}
\end{figure}

Second, \cite{Cookson:2026} used only the approximate numerical prediction of MOND presented by \cite{Zonoozi:2021}, although more recently \cite{Pflamm-Altenburg:2025} presented realistic wide binary orbit solutions. Figure~\ref{fig:vtprofile_Cookson} reproduces the profile of median $\tilde v$ presented in Figure~8 of \cite{Cookson:2026} including the numerical prediction by \cite{Pflamm-Altenburg:2025}. In the two bins of the transition regime ($-0.6\la \log_{10} (r_p/r_{\rm M}) \la 0$), the \cite{Pflamm-Altenburg:2025} prediction is clearly different from \cite{Zonoozi:2021} and much closer to Newton, while they are similar in the last two bins with $\log_{10} (r_p/r_{\rm M}) > 1$. In addition to the low number statistics mentioned above, consideration of the more realistic numerical prediction of MOND gravity makes the Newton-MOND distinction more difficult with the \cite{Cookson:2026} sample.  

\begin{figure}[tbh!]
    \centering
   \includegraphics[width=1.0\linewidth]{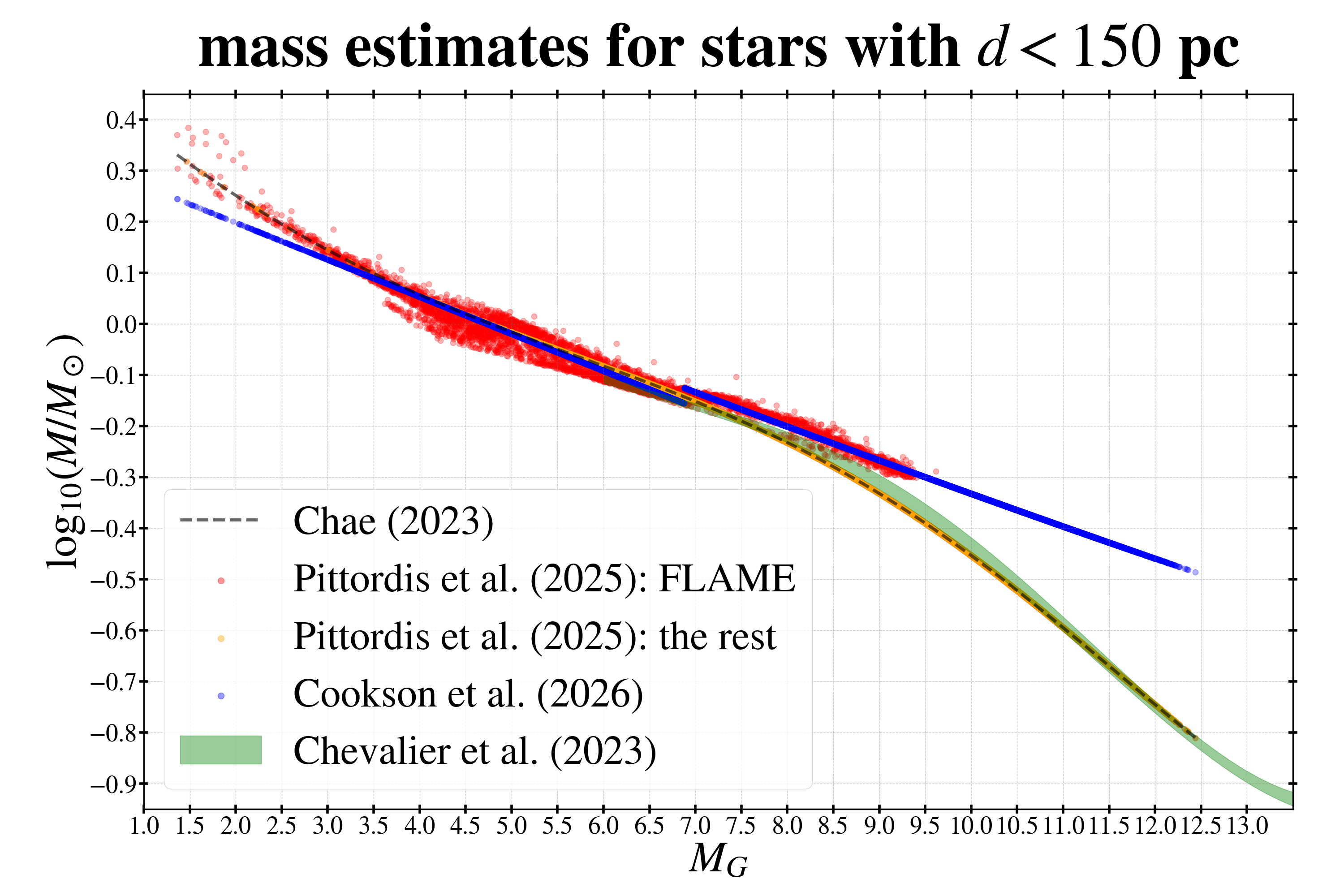}
    \caption{Various mass estimates of stars are compared for a subsample taken from the PSS sample \citep{Pittordis:2025}. For 36\% of the stars, Gaia DR3 FLAME masses are available; they are represented by red dots. The rest (64\%) are represented by orange dots that follow the relation derived by \cite{Chae:2023} based on \cite{PecautMamajek:2013}. The relation derived by \cite{Chevalier:2023} based on Gaia DR3 photometric and spectroscopic data along with literature data agrees well with the relation derived by \cite{Chae:2023}. The linear relation used by \cite{Cookson:2026} deviates from these relations. }
    \label{fig:mass_compare}
\end{figure}

Third, \cite{Cookson:2026} used too simplified a luminosity-mass relation for a large portion of their sample for which Gaia FLAME masses are not available. In a sample of stars, FLAME masses are available only for some fraction of the stars with $>0.5M_{\odot}$ and the fraction depends on how the sample is defined. Figure~\ref{fig:mass_compare} shows various mass estimates for a subsample with $d<150\,{\rm pc}$ taken from the PSS sample. In this example, FLAME masses are not available for 64\% of the stars. Since the \cite{Cookson:2026} sample covers the range $4<M_G<14$, FLAME masses are not available for the majority of their stars (the fraction of missing FLAME masses is 81\% as we read the paper). They used the linear relation between $M_G$ and $\log_{10}(M/M_\odot)$ (with a slight adjustment of the $y$ intercept for $M<0.7M_\odot$) shown in Figure~\ref{fig:mass_compare}. However, FLAME masses are less accurate near $0.5M_\odot$, so it is incorrect to use the linear relation for $M\la 0.5M_\odot$ based on the extrapolation of FLAME masses to large $M_G$. A curvature in the magnitude-$\log({\rm mass})$ relation for low-mass stars near $M_G=9$ is a well-established empirical fact (e.g., \citealt{PecautMamajek:2013,Chevalier:2023}; see also Figure~7 of \cite{Chae:2023} and references therein). Thus, for stars without FLAME masses, the relation derived by \cite{Chae:2023} based on \cite{PecautMamajek:2013} or the relation (though for a limited $M_G$ range) derived by \cite{Chevalier:2023} (or something similar), which are consistent with each other, should be used. The linear relation used by \cite{Cookson:2026} significantly overestimates masses for stars with $M_G \ga 9$.

Also, they used an incorrect benchmark for the median $\tilde v$ of pure Newtonian binaries as already pointed out in Figure~\ref{fig:vtmedian_alpha}. Their median value of $\tilde v =0.489$ is outside the range for any plausible distribution of eccentricities. We suspect that they used such a low value to be consistent with their overestimated masses for the significant portion of their sample without FLAME masses. The quantity $\tilde v$ is independent of mass, so it can be robustly predicted by Newtonian gravity for the given distribution of eccentricities. On the other hand, the measurement of $\tilde v$ depends on the estimated mass due to the factor $v_c$ (Equation~(\ref{eq:vc})). Thus, systematically overestimated (underestimated) masses will result in underestimated (overestimated) $\tilde v$.

Although systematically biased masses do not affect the profile of $\tilde v$ with respect to $r_p/r_{\rm M}$, correct masses are still important to verify required self-consistency. If a stringently selected sample consists of pure binaries in a statistical sense (meaning that a small number of contaminated cases do not significantly affect the median), the measured median should agree with the Newtonian prediction in the Newtonian high-acceleration regime for a reasonable distribution of eccentricities. Correct masses are essential for this important check. In the case of \cite{Cookson:2026}, this self-consistency check is missing.

Finally, \cite{Cookson:2026} did not calculate the Newtonian prediction on the $\tilde v$ profile for their sample but simply assumed that a flat line was the Newtonian prediction. Newtonian gravity will predict a flat profile of $\langle\tilde v\rangle (r_p/r_{\rm M})$ (remember that the notation $\langle\cdots\rangle$ is sometimes used to refer to median in this paper) only when eccentricity is independent of $r_p/r_{\rm M}$. As observational studies show that eccentricity increases with $r_p$ from the fully Newtonian regime (small separation) to the fully MOND regime (large separation), Newton predicts a mildly declining profile from a small separation through the transition regime. Moreover, the Banik cut can make the profile further decline in general, although it is less important for a sample with a distance limit such as $d<150\,{\rm pc}$ (see Figure~\ref{fig:vperr_vterr_distance}). However, the measurement error of $\tilde v$ at large separation is larger because the relative error of $v_p$ becomes relatively larger due to the smaller $v_p$. Its effect will be in the opposite direction. In principle, it is necessary to calculate the Newtonian prediction specific to the particular sample under consideration, taking into account all these factors. 

\begin{figure}[tbh!]
    \centering
   \includegraphics[width=1.0\linewidth]{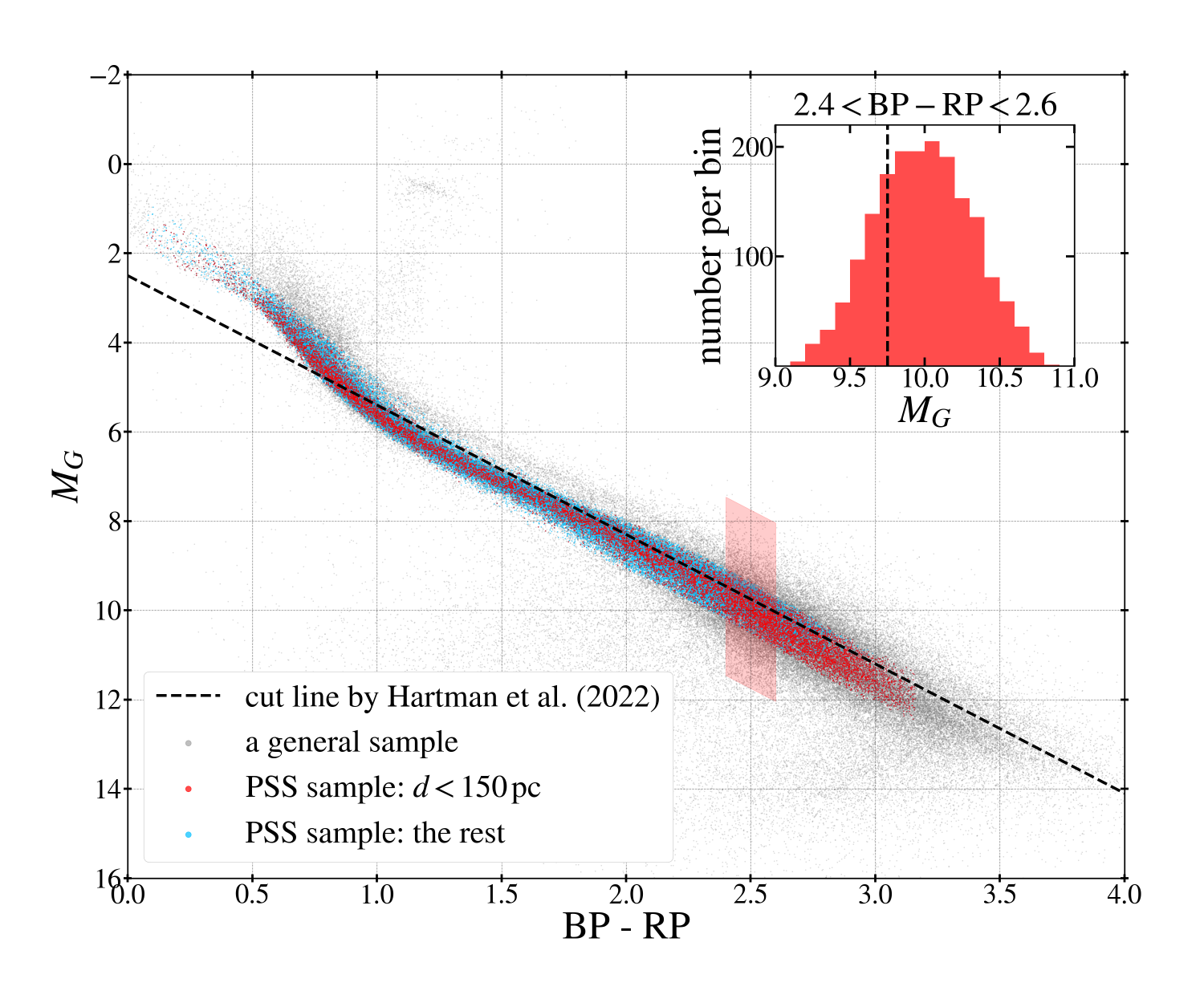}
    \caption{This figure shows a color-magnitude diagram for stars using the Gaia DR3 color BP$-$RP and absolute magnitude $M_G$. The stars with $d<150\,{\rm pc}$ from the PSS sample follow a narrow diagonal band. Stars with ${\rm BP-RP}>1.0$ are mostly under the arbitrary cut line suggested by \cite{Hartman:2022}. However, for ${\rm BP-RP}\ga 1.8$ the cut removes a higher-luminosity portion of a normal distribution at a fixed ${\rm BP-RP}$ as the inset exhibits. The other stars with $d>150\,{\rm pc}$ from the PSS sample are similar but have larger scatters due to larger measurement errors at larger distances.  }
    \label{fig:CM_PSS}
\end{figure}

Now we perform an analysis of $\langle\tilde v\rangle(r_p/r_{\rm M})$ based on a much larger sample than the \cite{Cookson:2026} sample that satisfies all their quality cuts and the distance limit $d<150\,{\rm pc}$. Our analysis of $\langle\tilde v\rangle(r_p/r_{\rm M})$ is mainly motivated to illustrate the effect of the sample size and the unbiased masses, compared with the analysis of \cite{Cookson:2026}. But, we will also consider the acceleration-plane test and the $\tilde v$ distribution (histogram) test for the sample with $d<150\,{\rm pc}$. We select wide binaries with $d<150\,{\rm pc}$ from the PSS sample. The member stars follow a narrow band in the CMD as shown in Figure~\ref{fig:CM_PSS}. These stars satisfy the `Lobster cut' introduced by \cite{Hartman:2022} to remove unresolved doubles. Stars selected by \cite{Cookson:2026} based mainly on ${\tt ruwe}<1.25$ are strictly under the arbitrary cut line introduced by \cite{Hartman:2022} for ${\rm BP-RP}\ga 1.0$. This cut line is intended to separate unresolved overluminous binaries from single stars. We note that this cut has two limitations: (1) it is not applicable for stars with ${\rm BP-RP}< 1.0$; (2) it can distort the luminosity distribution of single stars at a given ${\rm BP-RP}$ by removing some intrinsically luminous stars as shown by the inset of Figure~\ref{fig:CM_PSS}. The majority of the stars selected with ${\tt ruwe}<1.2$ are also under the cut line for ${\rm BP-RP}\ga 1.0$ but not as strictly as the \cite{Cookson:2026} sample. As our primary choice, we will not apply the \cite{Hartman:2022} cut, but will also consider the subsample that strictly satisfies it. As for masses, we will use the PSS masses, which consist of FLAME masses for 36\% and masses from \cite{Chae:2023} magnitude-mass relation for the rest.

\begin{figure*}[tbh!]
    \centering
   \includegraphics[width=1.0\linewidth]{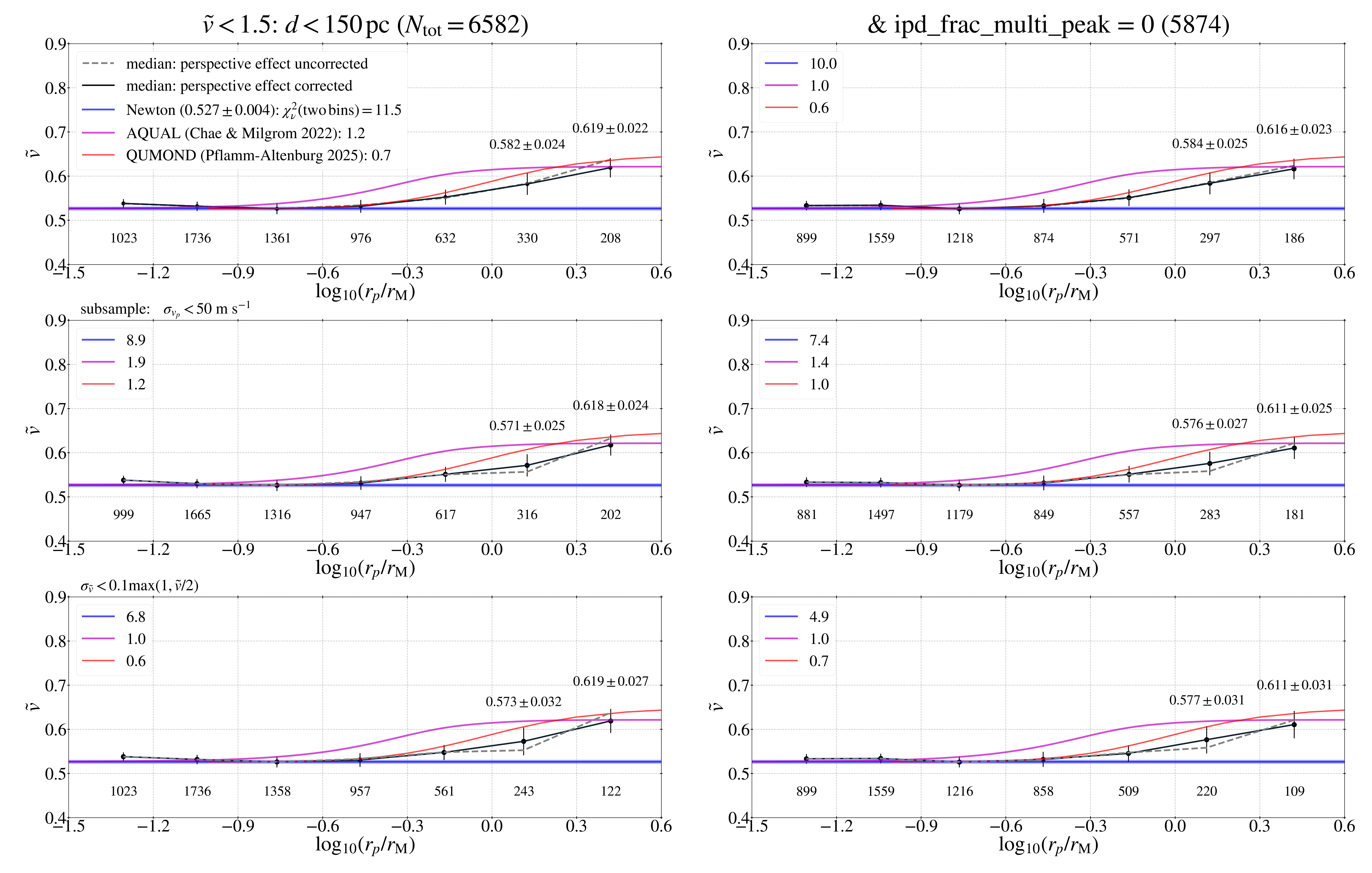}
    \caption{Profile of median $\tilde v$ in the sample satisfying $\tilde v <1.5$ and $d<150\,{\rm pc}$ is shown and compared with theoretical predictions of Newton and MOND models. The Newtonian prediction is calculated for all wide binaries in the sample based on individual eccentricities from \cite{Hwang:2022}. Two MOND models are from Figure~\ref{fig:theory_prediction}. The statistic $\chi^2_\nu$ ($\chi^2$ per degree of freedom) is calculated for the two bins in the MOND regime. In the Newtonian regime with $\log_{10}(r_p/r_{\rm M})<-0.6$, the measured medians agree well with the Newtonian prediction. }
    \label{fig:vt_scaling_PSS_vtmax15}
\end{figure*}

Figure~\ref{fig:vt_scaling_PSS_vtmax15} shows the measured medians of $\tilde v$ in bins of $\log_{10}(r_p/r_{\rm M})$ in the main sample with $d<150\,{\rm pc}$. As in \cite{Cookson:2026}, we require $\tilde v < 1.5$ to maximally remove triples (and other multiple-star systems) in a statistical sense. The left column presents the results for all wide binaries including those with ${\tt ipd\_frac\_multi\_peak = 1, 2}$, while the right column is for ${\tt ipd\_frac\_multi\_peak = 0}$ only. This sample with ${\tt ipd\_frac\_multi\_peak = 0}$ includes 481 wide binaries with $r_p/r_{\rm M}>1$, which are 8 times as many as those in the \cite{Cookson:2026} sample. In the three bins with $\log_{10}(r_p/r_{\rm M})<-0.6$, the medians agree well with the Newtonian prediction for pure binaries with individual eccentricities taken from \cite{Hwang:2022}. This result is different from that for the entire PSS sample with $\tilde v < 1.5$ shown in the right column of Figure~\ref{fig:vt_scaling_3samples_vtmax15}, where the measured medians are shifted upward by a small amount from the Newtonian prediction. This indicates that data for $d>150\,{\rm pc}$ make the difference, as will be shown below. 

\begin{figure*}[tbh!]
    \centering
   \includegraphics[width=1.0\linewidth]{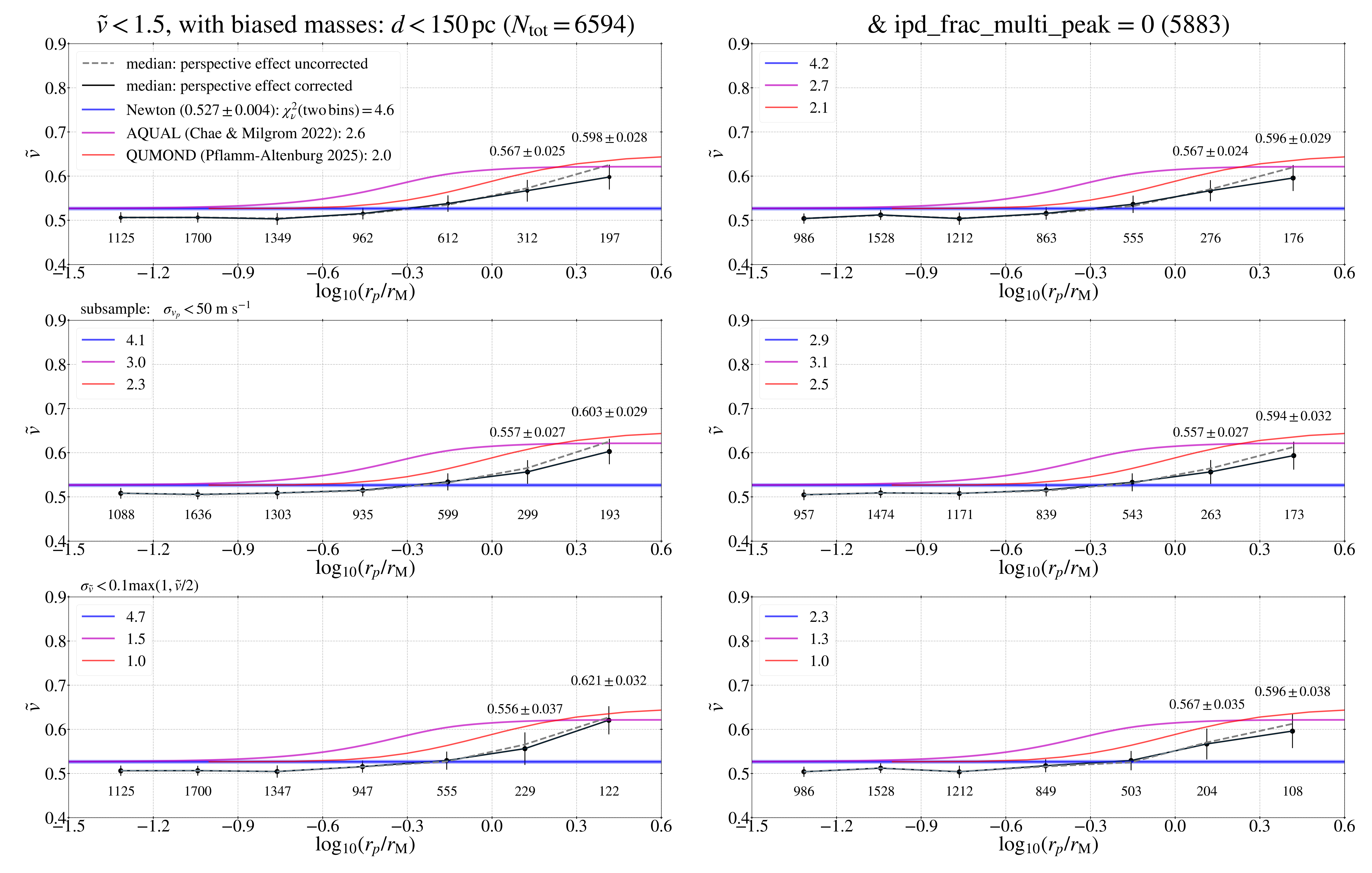}
    \caption{Similar to Figure~\ref{fig:vt_scaling_PSS_vtmax15} but with biased masses based on the linear relation between $M_G$ and $\log_{10}(M/M_\odot)$ used by \cite{Cookson:2026}.}
    \label{fig:vt_scaling_PSS_vtmax15_linearML}
\end{figure*}

The results shown in Figure~\ref{fig:vt_scaling_PSS_vtmax15} can be contrasted with the results of \cite{Cookson:2026} (see Figure~\ref{fig:vtprofile_Cookson} for their typical result) as follows. First, our values of median $\tilde v$ naturally match the Newtonian prediction in the Newtonian regime with no adjustment or fine-tuning. Our Newtonian values are higher by $\approx 0.04$ than the Newtonian benchmark value of $0.489$ and the biased medians assumed or derived by \cite{Cookson:2026}. This is largely due to the biased masses used by \cite{Cookson:2026}, but also in part to the difference in the sample definition in the CM diagram. Figure~\ref{fig:vt_scaling_PSS_vtmax15_linearML} shows the results with the biased masses of \cite{Cookson:2026}. The ``measured'' medians of $\tilde v$ in the Newtonian regime are now clearly lower than the Newtonian prediction. Nevertheless, the shape of the profile is unchanged from Figure~\ref{fig:vt_scaling_PSS_vtmax15}.

Second, the medians of $\tilde v$ in the last three bins with $\log_{10}(r_p/r_{\rm M})>-0.3$, which correspond to the transition and MOND regimes, exhibit a characteristic trend of deviations from the Newtonian prediction. These deviations are reliable because the sample meets the required calibration in the Newtonian regime. Considering the theoretical uncertainty in the transition regime (see Section~\ref{sec:mondgravity}), let us first focus on testing theoretical predictions in the MOND regime with $\log_{10}(r_p/r_{\rm M})>0$. For the two bins in the MOND regime, the reduced $\chi^2$ value of $\chi^2_\nu = 11.5$ or $10.0$ (with $\nu=2$) for Newton means a survival probability of $P_c = 1.0\times 10^{-5}$ or $4.5\times 10^{-5}$ (i.e., excluding Newton at $> 4\sigma$). The subsamples with $\sigma_{v_p}<50\,{\rm m}\,{\rm s}^{-1}$ give qualitatively similar results with slightly reduced statistical power. The subsamples with the Banik cut give also similar results with further reduced statistical power. However, the latter results need not be counted because of the bias of the Banik cut described in Sections~\ref{sec:WBTnature} and \ref{sec:tildev}. 

\begin{figure}[tbh!]
    \centering
   \includegraphics[width=1.0\linewidth]{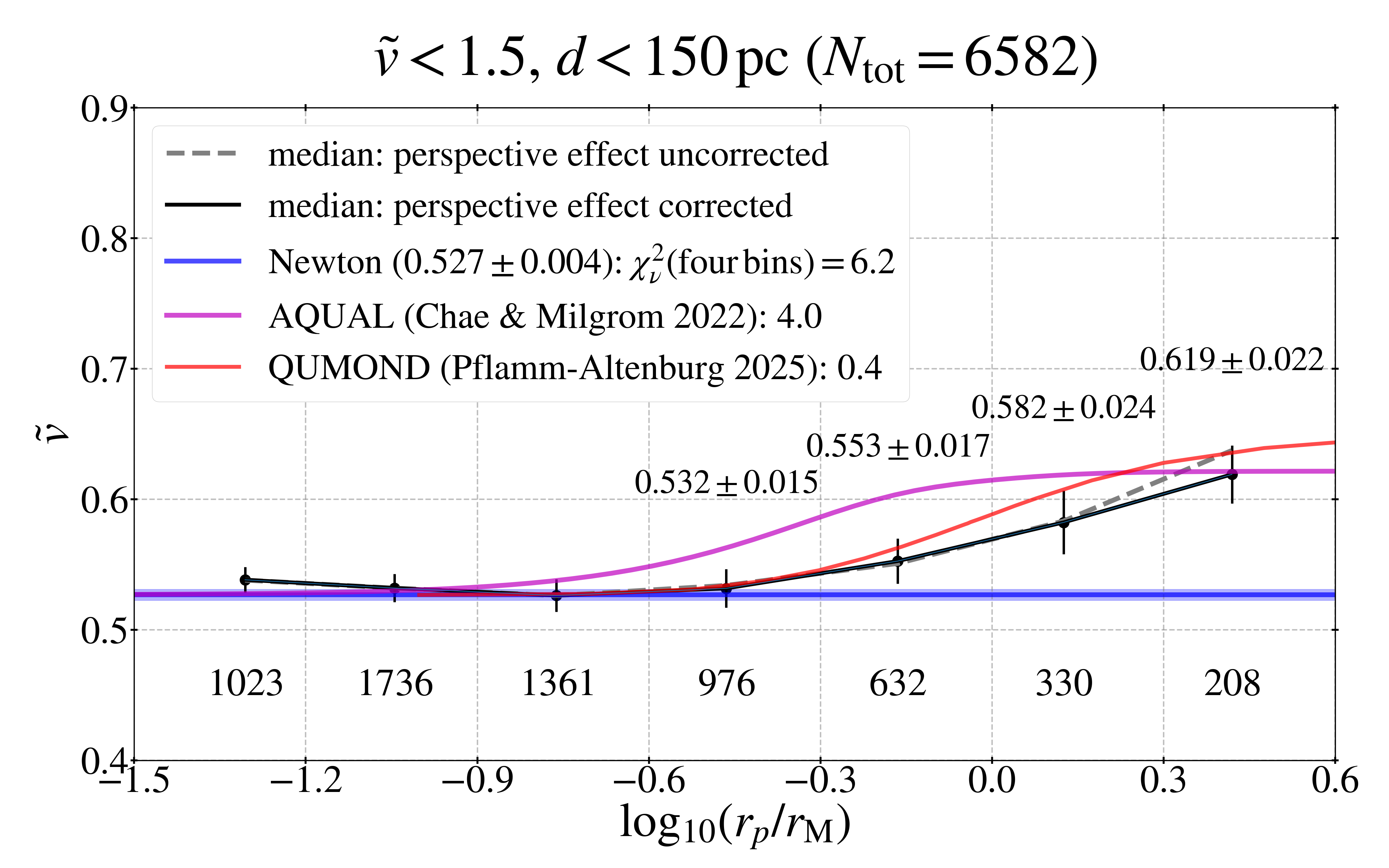}
    \caption{Same as the upper left panel of Figure~\ref{fig:vt_scaling_PSS_vtmax15} but $\chi^2_\nu$ calculated with four bins in the transition and MOND regimes.}
    \label{fig:vt_scaling_PSS_vtmax15_chi2dof4}
\end{figure}

Third, in the MOND regime existing MOND predictions converge (for similar MOND transition functions) and agree remarkably well with the data with $\chi^2_\nu\la 1$. In the transition regime ($-0.6\la\log_{10}(r_p/r_{\rm M})\la 0$) existing numerical MOND predictions cover a broad range (see Section~\ref{sec:mondgravity}) and the data appear to discriminate them. For statistical testing of the MOND predictions including the transition regime, we calculate $\chi^2_\nu$ for four bins in the range $-0.6<\log_{10}(r_p/r_{\rm M})< 0.6$. The result for the main sample is shown in Figure~\ref{fig:vt_scaling_PSS_vtmax15_chi2dof4}. The realistic numerical QUMOND prediction by \cite{Pflamm-Altenburg:2025} agrees well with the observed profile of median $\tilde v$ with $\chi^2_\nu= 0.4$. However, the approximate numerical AQUAL prediction by \cite{ChaeMilgrom:2022} has a significant tension in the transition regime with $\chi^2_\nu=4.0$ (note that the approximate QUMOND prediction presented by \cite{Zonoozi:2021} is slightly worse and not considered here), although it is acceptable in the MOND regime. These results are in line with the results from the acceleration-plane test for the general sample without the constraint $\tilde v < 1.5$, but the distinction between realistic and approximate numerical MOND solutions appears to be clearer with the sample shown in Figure~\ref{fig:vt_scaling_PSS_vtmax15}. Nevertheless, even the approximate numerical MOND solution is relatively better than Newton because of the clear velocity boost in the MOND regime.

Finally, we note that the correction of the perspective effect or the requirement of ${\tt ipd\_frac\_multi\_peak} = 0$ has negligible effects on the low-acceleration gravitational anomaly, contrary to the concerns raised by \cite{Cookson:2026}. This is not surprising. As for the perspective effect, the results shown in Figure~\ref{fig:vt_scaling_PSS_vtmax15} indicate that it can be only a minor issue in the last bin of $0.3<\log_{10}(r_p/r_{\rm M})<0.6$ (or for $15\la r_p \la 30\,{\rm kau}$ for typical stars in the sample), but the low-acceleration gravitational anomaly has already been clear in the data with $7\la r_p\la 15\,{\rm kau}$ (corresponding to the penultimate bin in Figure~\ref{fig:vt_scaling_PSS_vtmax15}) from various large samples (e.g., \citealt{Chae:2023,Chae:2024a,Chae:2024b,Yoonetal:2025}). Moreover, the correction of the perspective effect for $\log_{10}(r_p/r_{\rm M})>0.3$ does not reduce the measured velocities enough to remove the anomaly, rather it tends to make the anomaly more consistent with the MOND-type velocity boost (see \cite{Yoonetal:2025} for further details). As for ${\tt ipd\_frac\_multi\_peak} = 0$, it is already met by 89\% of the binaries in the PSS sample, and the rest have $1$ and $2$ that can be consistent with single stars rather than unresolved doubles.

\begin{figure*}[tbh!]
    \centering
   \includegraphics[width=1.0\linewidth]{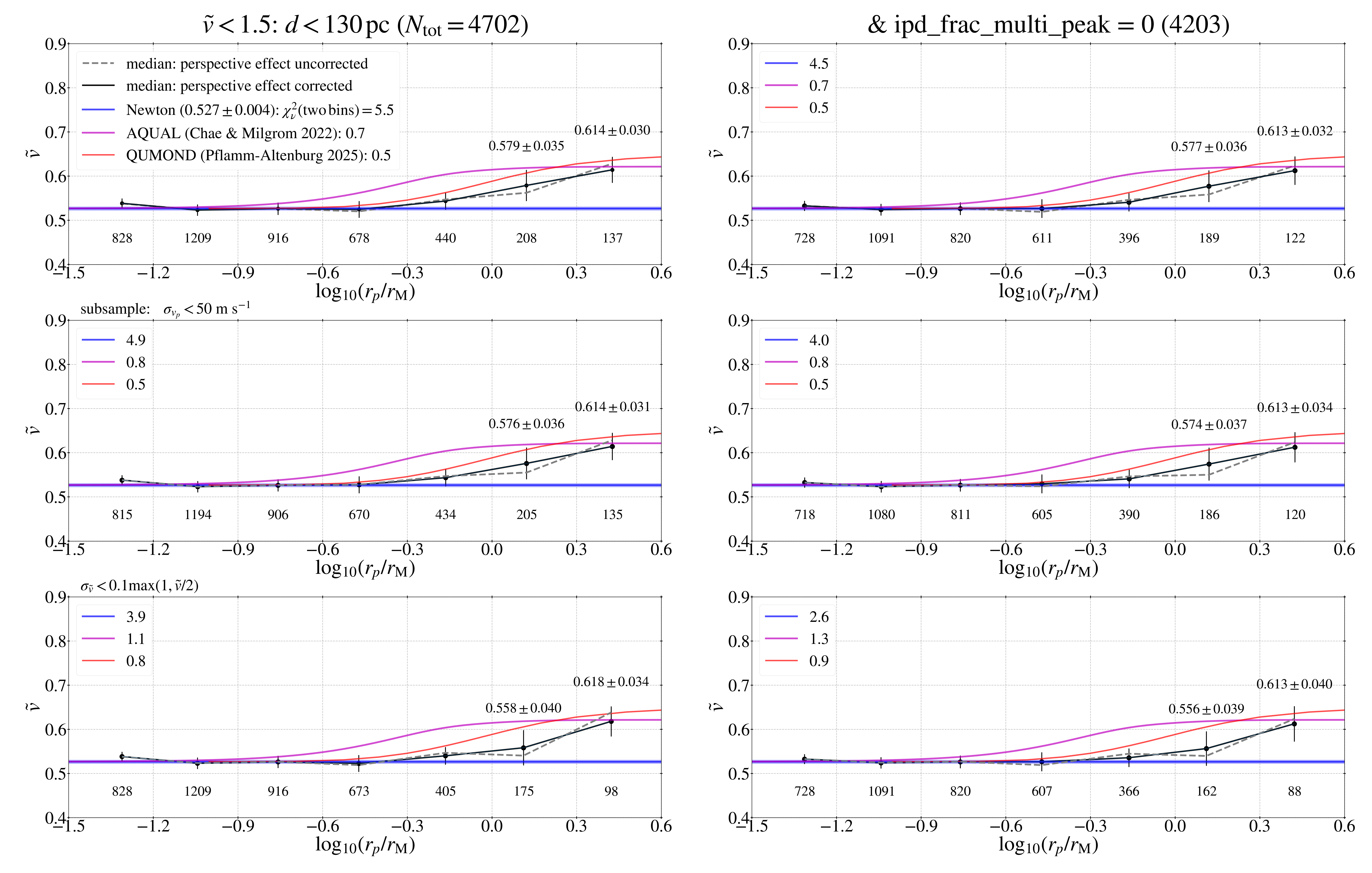}
    \caption{Similar to Figure~\ref{fig:vt_scaling_PSS_vtmax15} but for a smaller sample with $d<130\,{\rm pc}$.}
    \label{fig:vt_scaling_PSS_vtmax15_dmax130}
\end{figure*}

\begin{figure}[tbh!]
    \centering
   \includegraphics[width=1.0\linewidth]{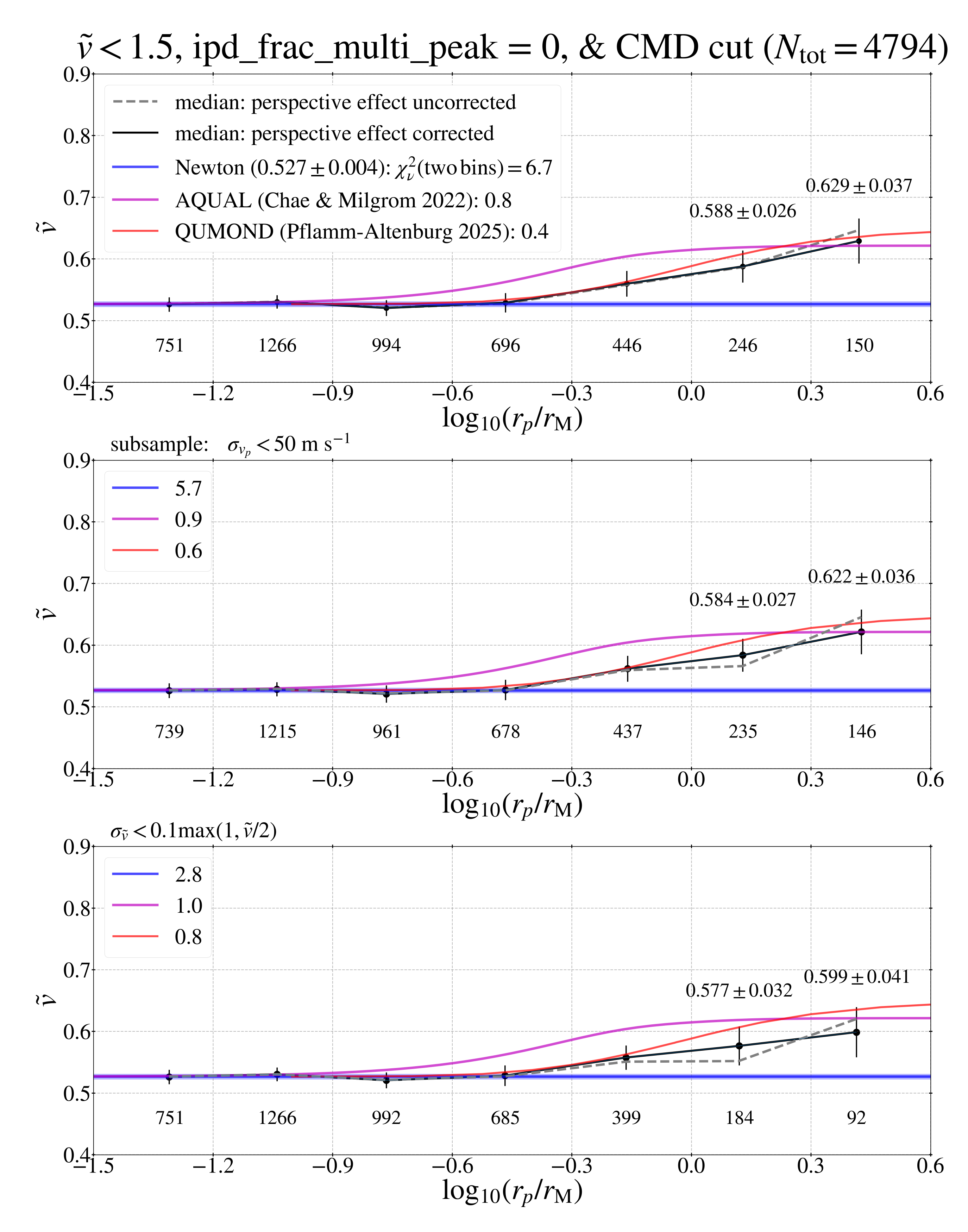}
    \caption{Similar to the right column of Figure~\ref{fig:vt_scaling_PSS_vtmax15} but for a smaller sample with the CMD cut for ${\rm BP-RP}>1$ shown in Figure~\ref{fig:CM_PSS}. }
    \label{fig:vt_scaling_PSS_vtmax15_CMDcut}
\end{figure}

If smaller samples with $d<130\,{\rm pc}$ and/or the \cite{Hartman:2022} CMD cut line are used, qualitatively similar results are obtained with reduced statistical power, as shown in Figures~\ref{fig:vt_scaling_PSS_vtmax15_dmax130} and \ref{fig:vt_scaling_PSS_vtmax15_CMDcut}. Statistical power can be maximally reduced by imposing the Banik cut on sufficiently small samples. 

\begin{figure}[tbh!]
    \centering
   \includegraphics[width=1.0\linewidth]{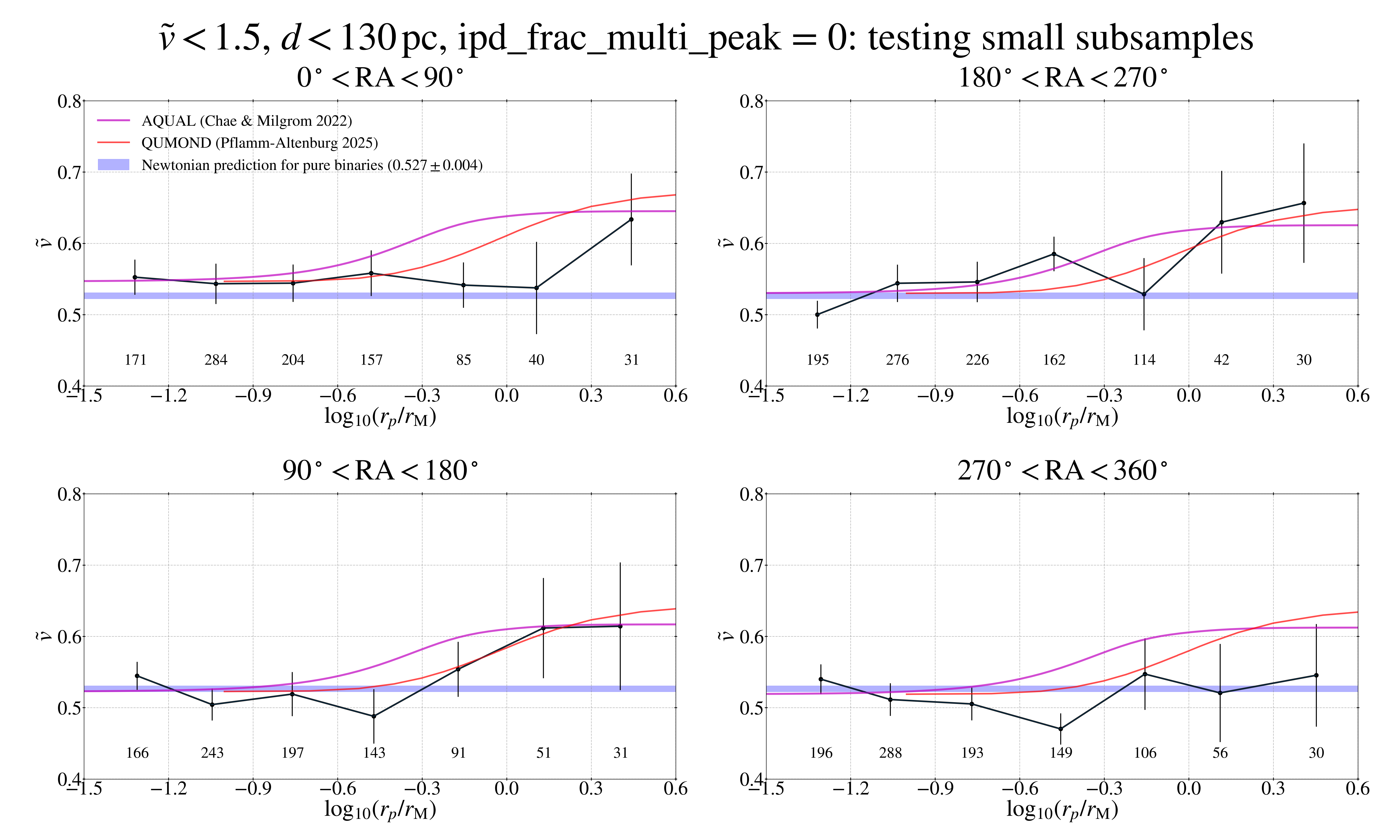}
    \caption{Each panel shows the profile of median $\tilde v$ for a subsample belonging to a quadrant in the R.A. space of the sample shown in the upper right panel of Figure~\ref{fig:vt_scaling_PSS_vtmax15_dmax130}. Each subsample is designed to contain a small number of wide binaries in the MOND regime with $r_p/r_{\rm M}>1$ (last two bins) as in \cite{Cookson:2026}.}
    \label{fig:vt_scaling_vtmax15_test}
\end{figure}

The flat profile of median $\tilde v$ obtained by \cite{Cookson:2026} is mainly a random fluctuation due to low statistics with some help from the Banik cut and the CMD cut. To check this, we consider four small subsamples, each of which includes $<90$ wide binaries in the MOND regime, similar in number to the \cite{Cookson:2026} sample. These samples are obtained by dividing the sample shown in the upper right panel of Figure~\ref{fig:vt_scaling_PSS_vtmax15_dmax130} (satisfying $\tilde v <1.5$, $d<130\,{\rm pc}$, and ${\tt ipd\_frac\_multi\_peak} = 0$) into quadrants in the R.A. space. Figure~\ref{fig:vt_scaling_vtmax15_test} shows the profiles of median $\tilde v$ in the small samples. Each profile has large uncertainties as in \cite{Cookson:2026}. The four results can be classified into three categories. The two samples in $90^\circ<{\rm R.A.}<180^\circ$ or $180^\circ<{\rm R.A.}<270^\circ$ prefer MOND over Newton, one sample in $270^\circ<{\rm R.A.}<360^\circ$ prefers Newton agreeing with the \cite{Cookson:2026} result, and the remaining sample in $0^\circ<{\rm R.A.}<90^\circ$ represents a mixed case (the last bin prefers MOND, but the penultimate bin prefers Newton). This exercise demonstrates that a small sample cannot be used to distinguish between Newton and MOND. If all four samples are combined, the numerical MOND model by \cite{Pflamm-Altenburg:2025} is preferred over Newton with $\chi^2_\nu = 0.5$ versus $4.5$ for the two bins with $\log_{10}(r_p/r_{\rm M})>0$, as shown in the upper right panel of Figure~\ref{fig:vt_scaling_PSS_vtmax15_dmax130}.

\begin{figure}[tbh!]
    \centering
   \includegraphics[width=1.0\linewidth]{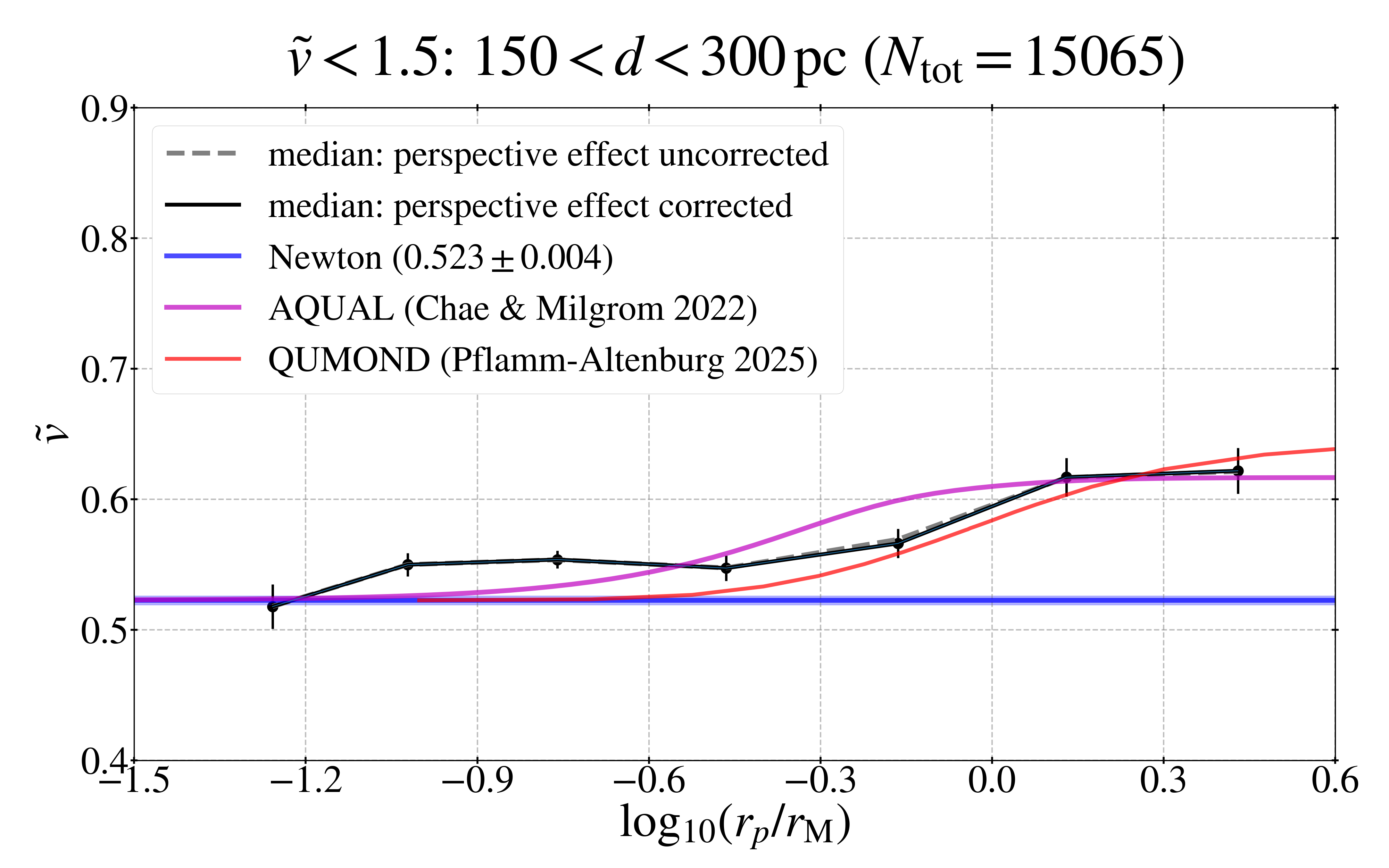}
    \caption{Similar to the upper left panel of Figure~\ref{fig:vt_scaling_PSS_vtmax15} for the sample with $150<d<300\,{\rm pc}$. }
    \label{fig:vt_scaling_vtmax15_d150300}
\end{figure}

The difference in the Newtonian regime between the total PSS sample with the limit $\tilde v < 1.5$ (the right column of Figure~\ref{fig:vt_scaling_3samples_vtmax15}) and its subsample with $d<150\,{\rm pc}$ (Figure~\ref{fig:vt_scaling_PSS_vtmax15}) indicates that data of wide binaries with $d>150\,{\rm pc}$ are different from those with $d<150\,{\rm pc}$ when the same constraint $\tilde v < 1.5$ is imposed. Figure~\ref{fig:vt_scaling_vtmax15_d150300} shows the profile of median $\tilde v$ in the subsample with $150<d<300\,{\rm pc}$. Unlike the subsample with $d<150\,{\rm pc}$, medians of $\tilde v$ in the Newtonian regime are clearly above the Newtonian prediction for pure binaries although the same constraint $\tilde v < 1.5$ is imposed. This may indicate that hierarchical systems containing unresolved inner binaries have been less effectively removed in wide binaries at larger distances by the observational cuts including $\tilde v < 1.5$. It may also be partially due to the larger scatter in $\tilde v$ arising from larger data uncertainties at larger distances.  

\begin{figure}[tbh!]
    \centering
   \includegraphics[width=1.0\linewidth]{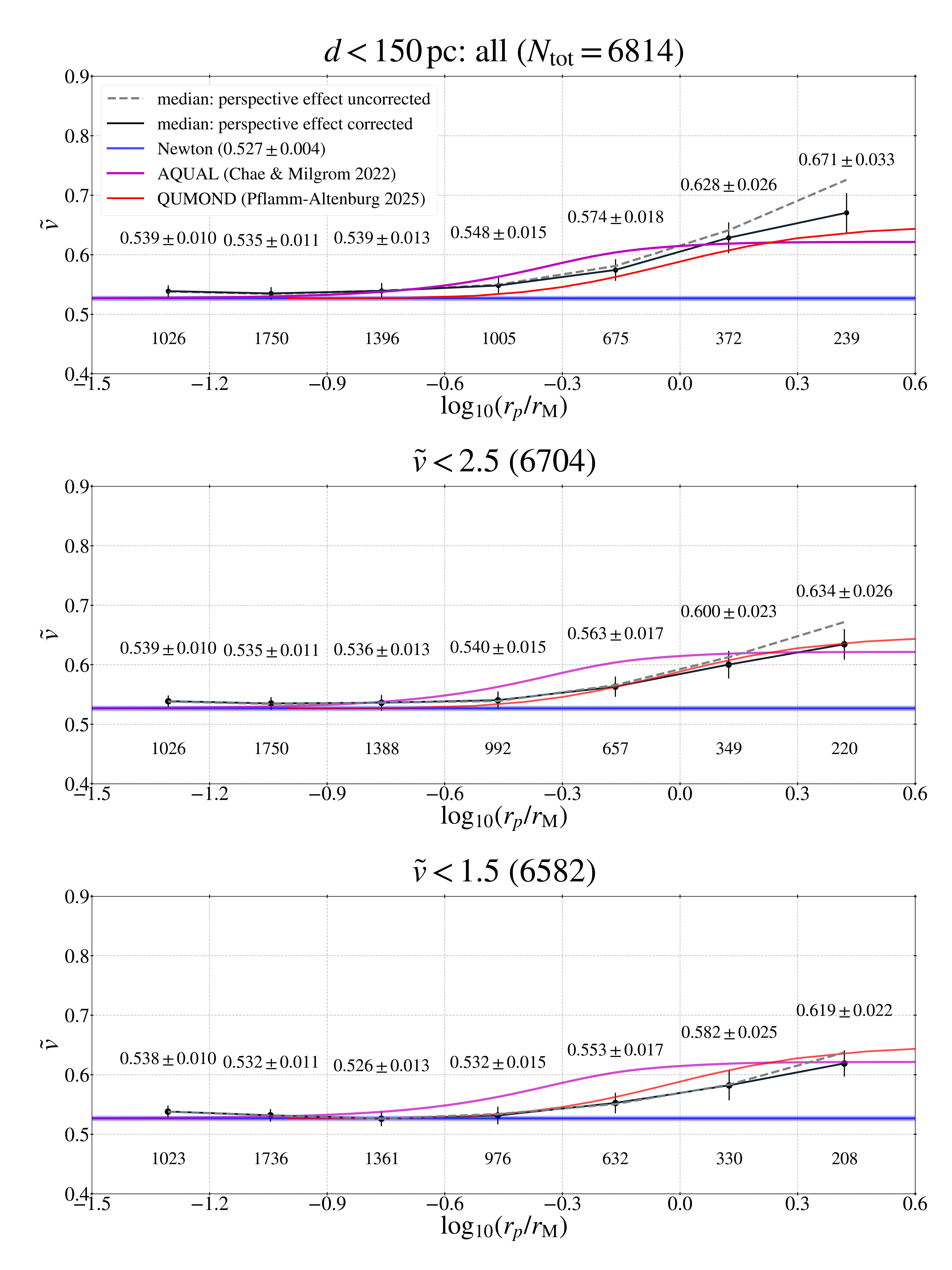}
    \caption{Top panel shows the profile of median $\tilde v$ for all wide binaries within $150\,{\rm pc}$ from the Sun. The other panels show the profiles in the subsamples with upper limits on $\tilde v$. }
    \label{fig:vt_scaling_d150_vtmaxvaried}
\end{figure}

\begin{figure}[tbh!]
    \centering
   \includegraphics[width=1.0\linewidth]{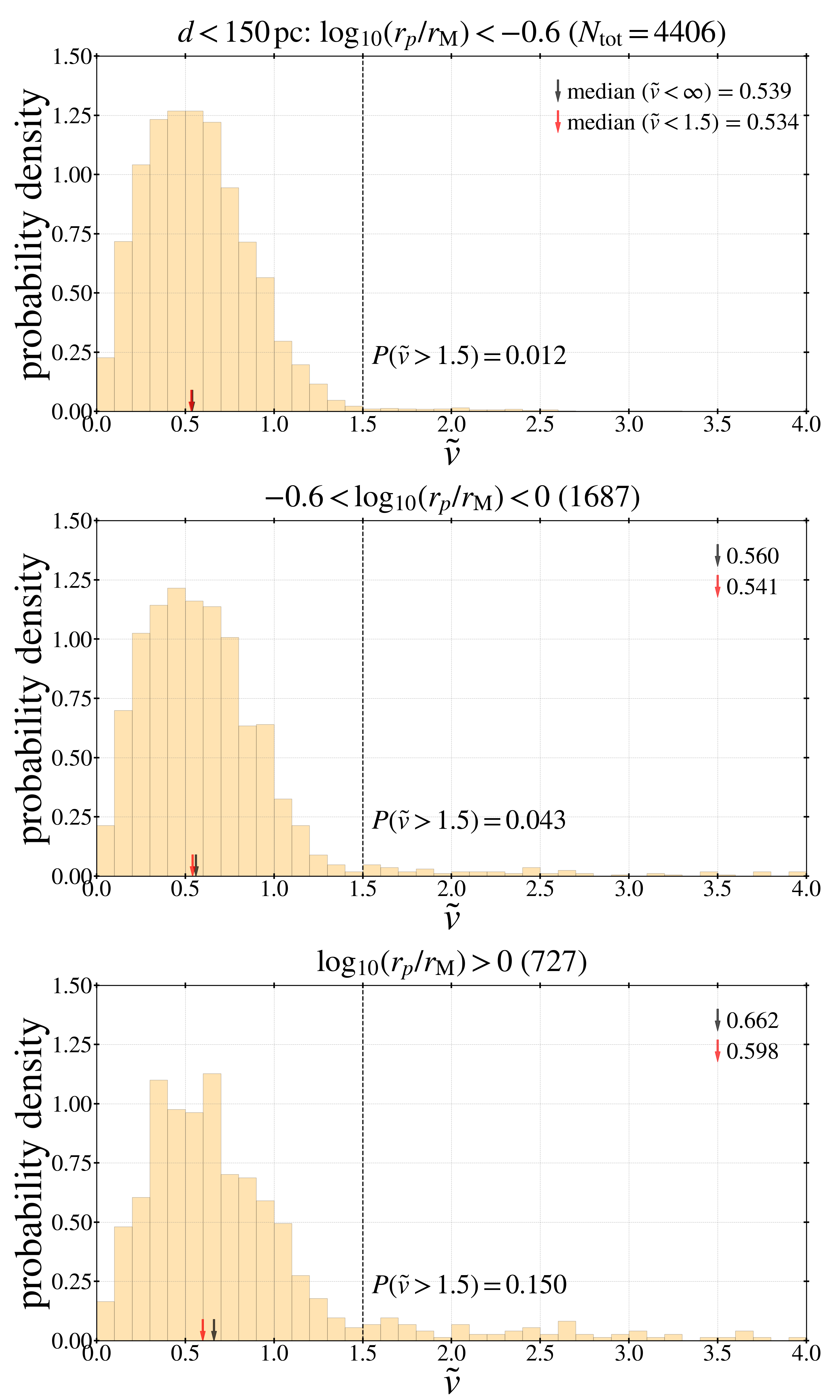}
    \caption{This figure shows the normalized histogram (or, probability distribution) of $\tilde v$ in three regimes of the PSS sample with $d<150\,{\rm pc}$. The upper limit of $\tilde v =1.5$ used in this section is indicated by the vertical dashed line, and the fraction outside the limit is indicated near the line. }
    \label{fig:vthist_d150}
\end{figure}

So far we have only analyzed the profile of median $\tilde v$ as a function of $r_p/r_{\rm M}$. We now consider the acceleration-plane test and the $\tilde v$ distribution for the main sample with $d<150\,{\rm pc}$. For the acceleration-plane test, we use the whole sample without the cut $\tilde v <1.5$ so that $f_{\rm trip}$ can be determined with the effective PSS triple model. Let us first examine the property of the whole sample with $d<150\,{\rm pc}$. Figure~\ref{fig:vt_scaling_d150_vtmaxvaried} shows the profile of median $\tilde v$ for wide binaries of the whole sample, and illustrates how it is varied if an upper limit on $\tilde v$ is imposed. In the three bins with $\log_{10}(r_p/r_{\rm M})<-0.6$, the measured medians are consistent with the Newtonian prediction for pure binaries even without an upper limit on $\tilde v$. This indicates that this sample contains a negligible fraction of triples, since chance-alignment/flyby is already negligible in these bins. This can be seen in the histogram of $\tilde v$ displayed in the top panel of Figure~\ref{fig:vthist_d150}. The distribution includes only 1.2\% of wide binaries with $\tilde v >1.5$ (also 1.4\% even with $\tilde v >1.4$). Thus, those systems with $\tilde v>1.4$ have negligible effects on the median. 

As we move to the transition ($-0.6<\log_{10}(r_p/r_{\rm M})<0$) and MOND/low-acceleration ($\log_{10}(r_p/r_{\rm M})>0$) regimes, the fraction with $\tilde v > 1.5$ increases dramatically to 4.3\% and 15.0\%. Because the triple fraction is negligibly small from the Newtonian regime inference and should not depend much on $r_p/r_{\rm M}$, this dramatic increase must be due to chance-alignment/flyby pairs consistent with the trend for the entire PSS sample with $d<300\,{\rm pc}$ shown in the right panel of Figure~\ref{fig:f_flyby}. 

\begin{figure*}[tbh!]
    \centering
   \includegraphics[width=1.0\linewidth]{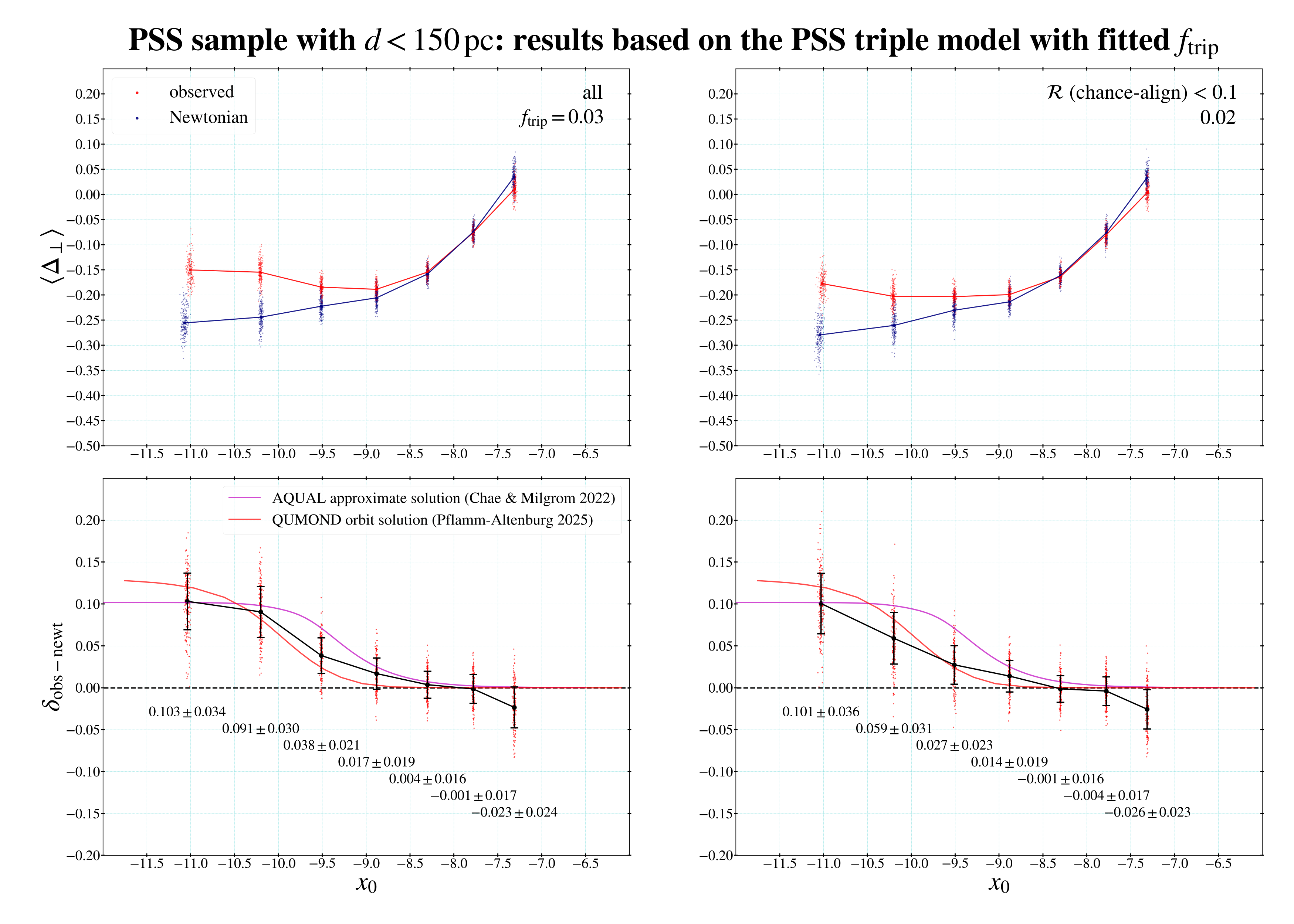}
    \caption{Similar to Figure~\ref{fig:residual_7bins_PSSsample_ftripfitted} but for the subsample with $d<150\,{\rm pc}$. The case $\tilde v < 5.5$ is not considered here as it is indistinguishable from the case of all binaries. }
    \label{fig:residual_7bins_PSSsample_d150}
\end{figure*}

Figure~\ref{fig:residual_7bins_PSSsample_d150} shows the acceleration-plane test results for the PSS sample with $d<150\,{\rm pc}$. As expected from Figure~\ref{fig:vt_scaling_d150_vtmaxvaried}, the fitted value of $f_{\rm trip}$ is close to zero, regardless of whether all binaries or binaries with $\mathcal{R}<0.1$ are used. The inferred values of $\delta_{\rm obs-newt}$ are consistent with those for the entire PSS sample shown in Figure~\ref{fig:residual_7bins_PSSsample_ftripfitted}, but the fitted values of $f_{\rm trip}$ for $d<150\,{\rm pc}$ are dramatically lower. The values of $\delta_{\rm obs-newt}$ agree well with the QUMOND numerical prediction by \cite{Pflamm-Altenburg:2025}, consistent with the profile of median $\tilde v$ shown in Figure~\ref{fig:vt_scaling_PSS_vtmax15}.

\begin{figure*}[tbh!]
    \centering
   \includegraphics[width=1.0\linewidth]{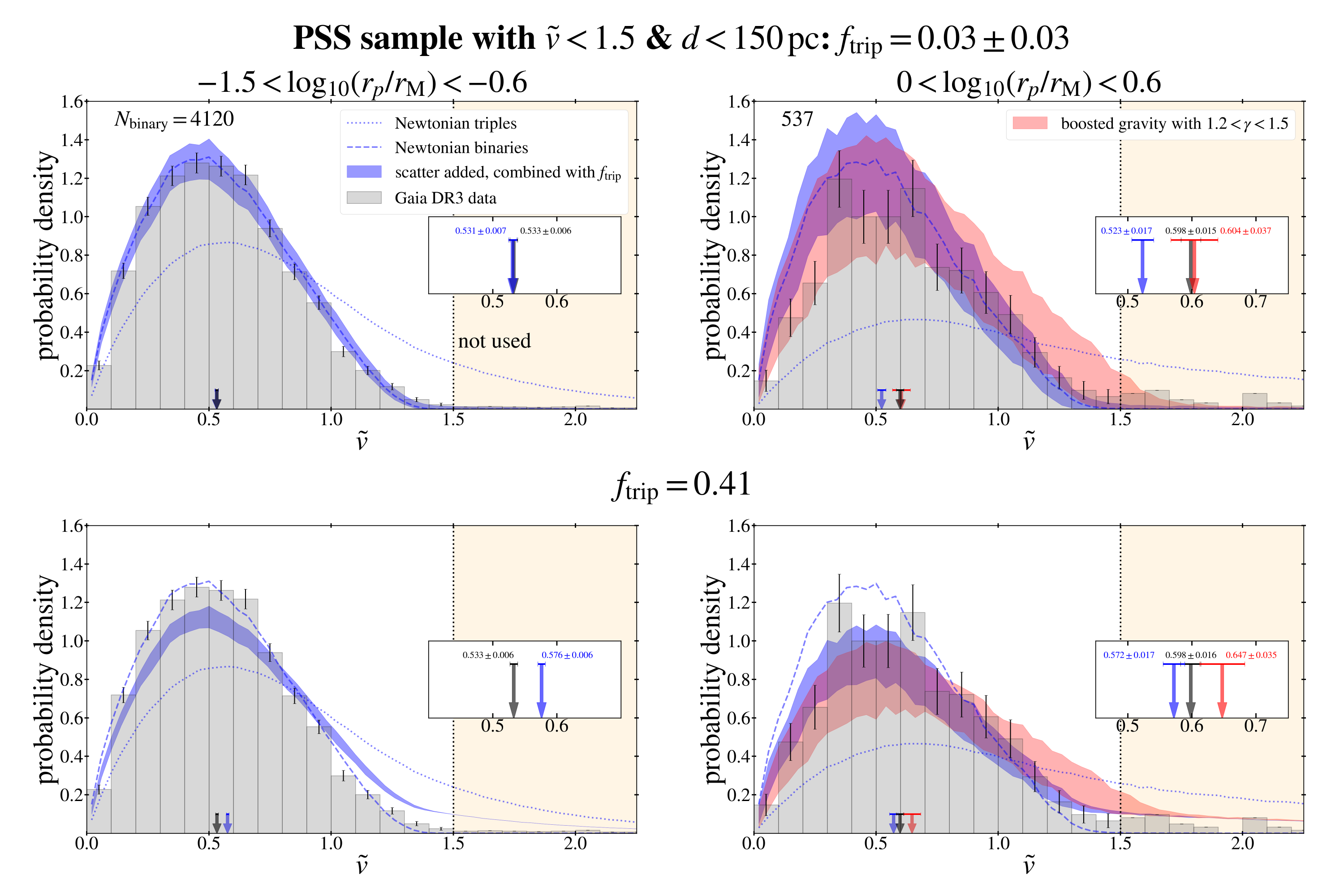}
    \caption{Similar to Figure~\ref{fig:vtdist_PSSsample} but for the subsample with $d<150\,{\rm pc}$ and $\tilde v < 1.5$. Here we consider the two specific ranges of $\log_{10}(r_p/r_{\rm M})$ indicated in the upper row that cover respectively the Newtonian and MOND regimes of Figure~\ref{fig:vt_scaling_PSS_vtmax15}. The upper row shows the results with $f_{\rm trip}$ fitted from the acceleration-plane test (see Figure~\ref{fig:residual_7bins_PSSsample_d150}), while the lower row shows the results with an arbitrary value of $f_{\rm trip}$ that makes the observed median agree with the Newtonian prediction in the MOND regime up to the combined error.  }
    \label{fig:vtdist_PSSsample_d150}
\end{figure*}

Next, we consider the $\tilde v$ distribution for the sample with $d<150\,{\rm pc}$ as was done for the entire PSS sample in Section~\ref{sec:test_vtdist}. In this test, we are particularly interested in whether the velocity boost evident in the MOND regime of the $\tilde v$ profile (Figure~\ref{fig:vt_scaling_PSS_vtmax15}) can be attributed to triples through a high value of $f_{\rm trip}$. Thus, we consider the sample with the limit $\tilde v < 1.5$ to be (almost) completely free of chance-alignment/flyby pairs. Figure~\ref{fig:vtdist_PSSsample_d150} shows the test results for the two specific ranges of $-1.5<\log_{10}(r_p/r_{\rm M})<-0.6$ and $0<\log_{10}(r_p/r_{\rm M})<0.6$. With $f_{\rm trip}=0.03\pm 0.03$ from the acceleration-plane analysis (the upper row), the observed distribution of $\tilde v$ in $-1.5<\log_{10}(r_p/r_{\rm M})<-0.6$ agrees excellently with the Newtonian prediction both in the shape and the median. In the MOND regime with $0<\log_{10}(r_p/r_{\rm M})<0.6$, the Newtonian prediction is clearly discrepant with the observed distribution, as the observed histogram is not contained in the predicted band and the observed median is clearly shifted from the predicted median. The statistical significance of the discrepancy in the median is about $\approx 3.2\sigma$, which is lower than in the $\chi^2_\nu$ test for the two bins in the profile of $\tilde v$ (Figure~\ref{fig:vt_scaling_PSS_vtmax15}) because here we include the uncertainty of the Newtonian prediction. In contrast to the Newtonian prediction, the prediction of boosted gravity matches well the observed distribution in line with various results above.

In the lower row of Figure~\ref{fig:vtdist_PSSsample_d150}, we test whether Newtonian gravity can be somehow consistent with the data in the MOND regime by varying $f_{\rm trip}$ based on the PSS triple model. We consider an arbitrary high value of $f_{\rm trip}=0.41$ that resolves the discrepancy of the Newtonian median in the MOND regime up to the combined uncertainty of the measured median and the Newtonian prediction. In this example, while the desired agreement is forced in the MOND regime, the Newtonian prediction in the Newtonian regime becomes clearly discrepant with the observed distribution both in the shape and the median. This test demonstrates that a high fraction of apparent binaries with hidden tertiaries (and other hierarchical systems) cannot bring the data into agreement with Newton, as was also demonstrated in the acceleration test in this and previous studies (see, e.g., Figure~\ref{fig:residual_3cols_Chaesample}). 

As we have demonstrated above, the \cite{Cookson:2026} quality framework does not remove the low-acceleration gravitational anomaly when a much larger sample than the \cite{Cookson:2026} sample is used. We also note the following misleading aspects of their quality framework and giving scores to samples used in wide binary gravity tests summarized in their Table 4.

First, data cuts required for a wide binary sample depend on the statistical method employed to test gravity. For example, cuts such as a {\tt ruwe} cut, a CMD cut, a {\tt ipd\_frac\_multi\_peak} cut, or a $\tilde v$ cut, etc, designed to remove apparent binaries that are actually hierarchical systems, are irrelevant for the method that models and takes into account the effects of hidden stars. What is important is to model their effects properly and calibrate $f_{\rm multi}$ or $f_{\rm trip}$ with the Newtonian-regime data. If modeling is carried out correctly, gravity test results should agree regardless of the sample, the method, and the implied value of $f_{\rm multi}$/ $f_{\rm trip}$ as demonstrated in various previous studies and in this study. In particular, the acceleration-plane test introduced by \cite{Chae:2023} and updated in subsequent studies does not require any quality cuts based on $\tilde v$ because PM uncertainties are taken directly into account \citep{Chae:2024b}. 

Second, their score system is misleading and inconsistent, not only because samples selected for different tests are compared based on the same quality cuts as pointed out above, but also because a crucial criterion/requirement and a minor one are treated equally with the same value of 1. 

Third, \cite{Cookson:2026} advocated the Banik cut. But, as shown in this work, the Banik cut biases the statistical properties of the sample relevant to gravity tests, by selectively removing higher $\tilde v$ values at larger $r_p$ that are based on sufficiently precise $v_p$.

Fourth, the importance of correcting for the perspective effect was misfocused on criticizing the valid results on the breakdown of standard gravity at low acceleration. As demonstrated in \cite{Yoonetal:2025}, which was published prior to \cite{Cookson:2026} but not included in their Table 4, and throughout this work, correcting for the perspective effect does not affect the presence of the gravitational anomaly because the perspective effect is minor or negligible for $r_p \la 20\,{\rm kau}$ and the evidence for the anomaly is already clear even without considering the data in which the perspective effect matters. The right focus should be on the nature of gravity in the low-acceleration limit at $r_p > 20\,{\rm kau}$. This work and \cite{Yoonetal:2025} based on sufficiently large samples show that correcting for the perspective effect at $r_p > 20\,{\rm kau}$ does not remove the gravity boost but makes it similar to the one present in the range $7\la r_p \la 20\,{\rm kau}$, bringing the scaling of $\gamma$ with $r_p$ into a better agreement with the MOND-type gravitational anomaly under the EFE. 

Finally, because of the above points, evaluating various recent publications across the board, as given at the last row of Table~4, is misleading and nonsensical. It is hardly based on sound judgment on the proper application of a specific wide binary gravity test to specific data.

\section{Summary, Conclusions, and Prospects} \label{sec:conclusion}

Since the release of Gaia DR3 \citep{Gaia:2023}, three independent dedicated groups have performed statistical gravity tests based on sky-plane 2D velocities data for the past three years. Two groups led by K.-H. Chae and X. Hernandez have independently obtained consistent results showing a gravitational anomaly with $\gamma\approx 1.4-1.5$ at low accelerations (Chae's group: \citealt{Chae:2023,Chae:2024a,Chae:2024b,Yoonetal:2025}, Hernandez's group: \citealt{Hernandez:2023,HernandezChae:2024,Hernandez:2024b,HernandezKroupa:2025}). Moreover, orbit modeling based on accurate 3D velocities has started producing results consistent with the results from statistical analyses of 2D velocities \citep{Chae:2025,Chae:2026,Chaeetal:2026}. However, one group led or collaborated by W. Sutherland has recently argued that there was no gravitational anomaly at low acceleration \citep{Banik:2024,Pittordis:2025,Cookson:2026}. Although 3D orbit modeling is providing more direct evidence, it is still important for statistical analyses of 2D velocities to reach a consensus. Thus, we have undertaken this project to revisit all the issues raised by those studies that found no gravitational anomaly, bearing in mind that the same Gaia DR3 data cannot imply different gravities.

The issues raised recently fall broadly into two categories: data quality control and how to deal with apparent binaries that are hiding a tertiary star (and more). \cite{Banik:2024} have emphasized a quality cut based on the error of $\tilde v$ (the Banik cut given by $\sigma_{\tilde v}<0.1$) while \cite{Cookson:2026} emphasized a whole set of quality control including the Banik cut, a {\tt ruwe} cut, a distance limit, correcting for the perspective effect, a CMD cut, a {\tt ipd\_frac\_multi\_peak} cut, etc. \cite{Pittordis:2025} emphasized a more realistic modeling of hierarchical systems focusing on triple systems with the limit ${\tt ruwe}<1.2$. We have thoroughly investigated whether the Banik cut does not bias gravity tests relying on $\tilde v$ while only improving data quality through analyses of a large number of mock binaries with controlled errors of $v_p$ (the 2D sky-plane scalar velocity) and $M_{\rm tot}$ (the total mass of the binary system) (Section~\ref{sec:WBTnature}). We then investigated the statistical properties of the three samples from the recent literature \citep{Chae:2023,Banik:2024,Pittordis:2025} with the Banik cut and other cuts, in particular a cut based on the absolute error of $v_p$ (Section~\ref{sec:tildev}). 

Regarding the Banik cut, a cut based on $v_p$ error, and the \cite{Cookson:2026} quality framework, we find or note the following:
\begin{itemize}
    \item We have demonstrated that the Banik cut introduces a bias in the scaling of $\tilde v$ with $r_p$ even for sufficiently precise $v_p$ because it selectively removes higher $\tilde v$ at larger $r_p$ (Figure~\ref{fig:vpvt_illustration_M14_vperr35}). 
    \item For the same precision of $v_p$, the Banik-cut-introduced bias is stronger for a binary of lower mass (Figure~\ref{fig:vpvt_illustration_M10_vperr35}), so the Banik cut selectively removes systems of lower mass (Figure~\ref{fig:mass_vp_Banik}), as first noticed by \cite{Chae:2024b}.
    \item The cut based on the error of $v_p$ such as $\sigma_{v_p}<50\,{\rm m}\,{\rm s}^{-1}$ does not introduce a bias in gravity tests, and the majority of wide binaries in the existing samples already satisfy $\sigma_{v_p}<50\,{\rm m}\,{\rm s}^{-1}$ or similar.
    \item The acceleration-plane test does not require the Banik cut, a $\sigma_{v_p}$ cut, or in particular the \cite{Cookson:2026} quality framework because the uncertainties of PM components can be directly taken into account \citep{Chae:2024b}.
    \item The $\tilde v$ distribution (histogram) test in ranges of $r_p/r_{\rm M}$ (or $r_p$) does not require the Banik cut, a $\sigma_{v_p}$ cut, or the \cite{Cookson:2026} quality framework because scatters due to $v_p$ errors can be used to derive the uncertainties of the distribution (or histogram bars) and the median. 
    \item The \cite{Cookson:2026} quality framework includes many criteria that are not relevant to certain tests under certain conditions. For example, a statistical analysis taking into account hidden hierarchical systems does not require specific criteria needed to remove hierarchical systems. For a sample consisting mainly of wide binaries with separation $<15\,{\rm kau}$ (or $<20\,{\rm kau}$), the perspective effect is minor or negligible and so its correction is insignificant.  
\end{itemize}

We have implemented the PSS triple model in the acceleration-plane test applicable to samples with ${\tt ruwe}<1.2$. For the \cite{Chae:2023} and PSS samples, the acceleration-plane test was used to calibrate $f_{\rm trip}$ using the Newtonian-regime data. Then, through the acceleration-plane test and the $\tilde v$ distribution test with the calibrated values of $f_{\rm trip}$, we have obtained the following results on gravity at low acceleration (Sections~\ref{sec:test_acceleration} and \ref{sec:test_vtdist}):
\begin{itemize}
    \item For the \cite{Chae:2023} sample, the fitted value of $f_{\rm trip}=0.10\pm0.03$ with ${\tt ruwe}<1.2$ is much lower than $f_{\rm multi}=0.42\pm0.05$ for the whole sample without the {\tt ruwe} cut and based on the \cite{Chae:2023} multiple-star model.
    \item For the PSS sample, the fitted values of $f_{\rm trip}=0.12\pm0.03$ or $0.07\pm0.03$ (depending on the assumption on chance-alignment/flyby pairs present in the sample) are lower than the value of $\approx 0.2$ estimated by \cite{Pittordis:2025} through fitting of the data in transition and MOND regimes excluding the data in a fully Newtonian regime. 
    \item The inferred behaviors of the gravitational anomaly parameter $\delta_{\rm obs-newt}[=(\log_{10}\gamma_g)/\sqrt{2}]$ (cf.\, Figure~\ref{fig:residual_3cols_Chaesample}) on the plane spanned by Newtonian ($g_{\rm N}$) and empirical accelerations are very similar for the two samples. For $g_{\rm N}\la 10^{-9}\,{\rm m}\,{\rm s}^{-2}$, $\delta_{\rm obs-newt} > 0$ is evident with very high statistical significance (with combined statistical significance of $>5\sigma$), consistent with numerous previous results obtained with the \cite{Chae:2023} multiple-star model.
    \item The detailed behaviors of $\delta_{\rm obs-newt}$ in the transition regime ($10^{-10}\la g_{\rm N}\la 10^{-9}\,{\rm m}\,{\rm s}^{-2}$) better agree with realistic numerical QUMOND solutions of wide binary orbits \citep{Pflamm-Altenburg:2025} than approximate numerical solutions \citep{Banik:2018,ChaeMilgrom:2022}. 
    \item For the PSS sample with $f_{\rm trip}\approx 0.2$ from \cite{Pittordis:2025}, we obtain $\delta_{\rm obs-newt} < 0$ in the Newtonian regime showing that the PSS value is overestimated (Figure~\ref{fig:residual_7bins_PSSsample_ftripfixed}). Even with this uncalibrated value of $f_{\rm trip}$, the global behavior of $\delta_{\rm obs-newt}$ for a broad dynamic range from highest to lowest acceleration shows a characteristic trend similar to the MOND prediction save an overall vertical shift. The \cite{Pittordis:2025} preference for Newton in the regime $g_{\rm N}\la 10^{-9}\,{\rm m}\,{\rm s}^{-2}$ was a consequence of the incorrect value of $f_{\rm trip}$ and a limited dynamics range.
    \item The $\tilde v$-distribution test was carried out for the \cite{Chae:2023}, \cite{Banik:2024}, and PSS samples. In the case of the \cite{Banik:2024} sample, the value of $f_{\rm trip}$ from the PSS sample was used because it does not have the Newtonian-regime data. The results show that in the Newtonian regime with $r_p/r_{\rm M}<0.08$, the observed distribution matches excellently the Newtonian prediction. However, in the MOND regime of $r_p/r_{\rm M}>1$, the observed distribution is inconsistent with the Newtonian prediction with high statistical significance (in some cases with $>5\sigma$) but is well consistent the prediction of boosted gravity with the ranage of $1.2<\gamma<1.5$ from numerical simulations of \cite{Pflamm-Altenburg:2025}. 
    \item The $\tilde v$-distribution test was also performed with the limit $\tilde v < 1.5$ to be free of the uncertainty due to chance-alignment/flyby pairs. The results (Figure~\ref{fig:vtdist_PSSsample_rp_vtmax15}) show that the gravitational anomaly in the MOND regime is present with high statistical significance regardless of the assumed value (taken from this work or \cite{Pittordis:2025}) of $f_{\rm trip}$ which is the only unknown in this specific test.
\end{itemize}

We have also considered a highest-quality sample with $d<150\,{\rm pc}$, ${\tt ruwe}<1.2$, a ``Lobster'' cut in the CMD, and $\tilde v < 1.5$ taken from \cite{Pittordis:2025}. This sample is free of chance-alignment/flyby pairs, and the acceleration-plane test shows that the fraction of hierarchical systems is quite low with $f_{\rm trip}=0.03\pm0.03$. For this sample, we perform three different tests, i.e., the test of the median-$\tilde v$ profile as well as the acceleration-plane test and the $\tilde v$-distribution test. We find the following:
\begin{itemize}
    \item For a sample that satisfies the data quality framework of \cite{Cookson:2026} with correct stellar masses and the correct CMD cut, median $\tilde v$ agrees with the Newtonian prediction ($ 0.527\pm 0.004$) for the sample up to the measurement uncertainties for $\log_{10}(r_p/r_{\rm M}) < -0.6$, indicating that hierarchical systems are minimal and not significant statistically. We note that \cite{Cookson:2026} have an incorrect median of $\tilde v =0.486$ because they use an incorrect magnitude-mass relation.
    \item We find that the profile of median $\tilde v$ is not flat but rises with $r_p/r_{\rm M}$ when a sufficiently large sample is used. We show that the \cite{Cookson:2026} result of the flat profile is mainly a consequence of random fluctuation (Figure~\ref{fig:vt_scaling_vtmax15_test}) due to the small number ($N=61$) of wide binaries in the MOND regime  $r_p/r_{\rm M}>1$.  
    \item We find that the profile of median $\tilde v$ matches well the prediction of realistic QUMOND simulations by \cite{Pflamm-Altenburg:2025} but has tension with the prediction of approximate numerical AQUAL/QUMOND solutions due to the difference in the transition regime.
    \item We find that the acceleration-plane test for this sample gives results consistent with the profile of median $\tilde v$ and the acceleration-plane test results (Section~\ref{sec:test_acceleration}) for the much larger general samples.
    \item We find that the $\tilde v$-distribution test returns results (Figure~\ref{fig:vtdist_PSSsample_d150}) consistent with the profile of median $\tilde v$ and the acceleration-plane test results. In particular, we demonstrate that the discrepancy of standard gravity in the MOND regime cannot be resolved by increasing $f_{\rm trip}$ because that will make the distribution discrepant with Newton in the Newtonian regime. 
\end{itemize}

In conclusion, we have thoroughly investigated all the issues raised recently to challenge the low-acceleration gravitational anomaly and performed extensive gravity tests including the median-$\tilde v$-profile test, the $\tilde v$-distribution test, and the acceleration-plane test. Our key findings are:
\begin{enumerate}
    \item Implementation of the \cite{Pittordis:2025} triple model in the acceleration-plane test and the $\tilde v$-distribution test does not remove gravitational anomalies in the low-acceleration regime as long as a sufficiently broad dynamic range is used and the fraction of hierarchical systems is calibrated with the Newtonian-regime data free of chance-alignment/flyby pairs.
    \item Implementation of the \cite{Cookson:2026} data quality framework in the median-$\tilde v$-profile test does not remove gravitational anomalies in the low-acceleration regime as long as a sufficiently large sample is used to ensure $>3\sigma$ statistical significance.
    \item All three gravity tests based on various samples give consistent results that agree well with the prediction by realistic numerical solutions of wide binary orbits by \cite{Pflamm-Altenburg:2025} but show some tensions with approximate numerical solutions. Although the distinction is not conclusive from the present studies, this is the first result of its kind in wide binary gravity tests. 
\end{enumerate}

Although direct orbit modeling with accurate 3D velocities of wide binaries is likely to play more and more important roles in the future \citep{Chae:2025,Saglia:2025,Chae:2026,Chaeetal:2026}, gravity tests based on a large number of 2D velocities are also promising with the upcoming release of Gaia DR4 in December 2026, which will provide dramatically larger samples with significantly improved data qualities. Further refined modeling of hierarchical systems than the \cite{Pittordis:2025} triple model are also likely to be important, as the triple model by \cite{Pittordis:2025} improved over earlier models indicates a distinction among existing numerical predictions of MOND. It is interesting and encouraging that refined statistical analyses appear to better agree with more accurate numerical orbit solutions. 

\section*{acknowledgments}
We are deeply indebted to Will Sutherland and Charalambos Pittordis for providing us with their numerical data for their triple model and their full sample including the Newtonian-regime wide binaries, and for discussions. We also thank Indranil Banik for sharing the sample used in \cite{Banik:2024}. We thank Jan Pflamm-Altenburg for providing the numerical data for his wide binary orbit solutions. This work was supported by the National Research Foundation of Korea (NRF) grant funded by the Korea government (MSIT) (RS-2026-25492976).

\bibliography{ms}{}

\begin{thebibliography}{}
\expandafter\ifx\csname natexlab\endcsname\relax\def\natexlab#1{#1}\fi
\providecommand{\url}[1]{\href{#1}{#1}}
\providecommand{\dodoi}[1]{doi:~\href{http://doi.org/#1}{\nolinkurl{#1}}}
\providecommand{\doeprint}[1]{\href{http://ascl.net/#1}{\nolinkurl{http://ascl.net/#1}}}
\providecommand{\doarXiv}[1]{\href{https://arxiv.org/abs/#1}{\nolinkurl{https://arxiv.org/abs/#1}}}

\bibitem[{I. {Banik} {et~al.}(2024){Banik}, {Pittordis}, {Sutherland}, {Famaey}, {Ibata}, {Mieske}, \& {Zhao}}]{Banik:2024}
{Banik}, I., {Pittordis}, C., {Sutherland}, W., {et~al.} 2024, \bibinfo{title}{{Strong constraints on the gravitational law from Gaia DR3 wide binaries},} \mnras, 527, 4573, \dodoi{10.1093/mnras/stad3393}

\bibitem[{I. {Banik} \& H. {Zhao}(2018){Banik} \& {Zhao}}]{Banik:2018}
{Banik}, I., \& {Zhao}, H. 2018, \bibinfo{title}{{Testing gravity with wide binary stars like {\ensuremath{\alpha}} Centauri},} \mnras, 480, 2660, \dodoi{10.1093/mnras/sty2007}

\bibitem[{J. {Bekenstein} \& M. {Milgrom}(1984){Bekenstein} \& {Milgrom}}]{BekensteinMilgrom:1984}
{Bekenstein}, J., \& {Milgrom}, M. 1984, \bibinfo{title}{{Does the missing mass problem signal the breakdown of Newtonian gravity?},} \apj, 286, 7, \dodoi{10.1086/162570}

\bibitem[{K.-H. {Chae}(2023){Chae}}]{Chae:2023}
{Chae}, K.-H. 2023, \bibinfo{title}{{Breakdown of the Newton-Einstein Standard Gravity at Low Acceleration in Internal Dynamics of Wide Binary Stars},} \apj, 952, 128, \dodoi{10.3847/1538-4357/ace101}

\bibitem[{K.-H. {Chae}(2024{\natexlab{a}}){Chae}}]{Chae:2024a}
{Chae}, K.-H. 2024{\natexlab{a}}, \bibinfo{title}{{Robust Evidence for the Breakdown of Standard Gravity at Low Acceleration from Statistically Pure Binaries Free of Hidden Companions},} \apj, 960, 114, \dodoi{10.3847/1538-4357/ad0ed5}

\bibitem[{K.-H. {Chae}(2024{\natexlab{b}}){Chae}}]{Chae:2024b}
{Chae}, K.-H. 2024{\natexlab{b}}, \bibinfo{title}{{Measurements of the Low-acceleration Gravitational Anomaly from the Normalized Velocity Profile of Gaia Wide Binary Stars and Statistical Testing of Newtonian and Milgromian Theories},} \apj, 972, 186, \dodoi{10.3847/1538-4357/ad61e9}

\bibitem[{K.-H. {Chae}(2025){Chae}}]{Chae:2025}
{Chae}, K.-H. 2025, \bibinfo{title}{{Low-acceleration Gravitational Anomaly from Bayesian 3D Modeling of Wide Binary Orbits: Methodology and Results with Gaia Data Release 3},} \apj, 985, 210, \dodoi{10.3847/1538-4357/adce09}

\bibitem[{K.-H. {Chae}(2026){Chae}}]{Chae:2026}
{Chae}, K.-H. 2026, \bibinfo{title}{{Bayesian Inference of Gravity through Realistic 3D Modeling of Wide Binary Orbits: General Algorithm and a Pilot Study with HARPS Radial Velocities},} \apjl, 998, L43, \dodoi{10.3847/2041-8213/ae40ef}

\bibitem[{K.-H. {Chae} {et~al.}(2019){Chae}, {Bernardi}, {Sheth}, \& {Gong}}]{Chae:2019}
{Chae}, K.-H., {Bernardi}, M., {Sheth}, R.~K., \& {Gong}, I.-T. 2019, \bibinfo{title}{{Radial Acceleration Relation between Baryons and Dark or Phantom Matter in the Supercritical Acceleration Regime of Nearly Spherical Galaxies},} \apj, 877, 18, \dodoi{10.3847/1538-4357/ab18f8}

\bibitem[{K.-H. {Chae} {et~al.}(2026){Chae}, {Lee}, {Hernandez}, {Orlov}, {Lim}, {Turnshek}, \& {Lee}}]{Chaeetal:2026}
{Chae}, K.-H., {Lee}, B.-C., {Hernandez}, X., {et~al.} 2026, \bibinfo{title}{{Detection of Gravitational Anomaly at Low Acceleration from a Highest-quality Sample of 36 Wide Binaries with Accurate 3D Velocities},} arXiv e-prints, arXiv:2601.21728.
\newblock \doarXiv{2601.21728}

\bibitem[{K.-H. {Chae} \& M. {Milgrom}(2022){Chae} \& {Milgrom}}]{ChaeMilgrom:2022}
{Chae}, K.-H., \& {Milgrom}, M. 2022, \bibinfo{title}{{Numerical Solutions of the External Field Effect on the Radial Acceleration in Disk Galaxies},} \apj, 928, 24, \dodoi{10.3847/1538-4357/ac5405}

\bibitem[{S. {Chevalier} {et~al.}(2023){Chevalier}, {Babusiaux}, {Merle}, \& {Arenou}}]{Chevalier:2023}
{Chevalier}, S., {Babusiaux}, C., {Merle}, T., \& {Arenou}, F. 2023, \bibinfo{title}{{Binary masses and luminosities with Gaia DR3},} \aap, 678, A19, \dodoi{10.1051/0004-6361/202347111}

\bibitem[{S.~A. {Cookson} {et~al.}(2026){Cookson}, {Banik}, {El-Badry}, {Sutherland}, {Penoyre}, {Pittordis}, \& {Clarke}}]{Cookson:2026}
{Cookson}, S.~A., {Banik}, I., {El-Badry}, K., {et~al.} 2026, \bibinfo{title}{{A quality framework for testing gravity with wide binaries: no evidence for MOND},} \mnras, 547, stag342, \dodoi{10.1093/mnras/stag342}

\bibitem[{K. {El-Badry} {et~al.}(2021){El-Badry}, {Rix}, \& {Heintz}}]{El-badry:2021}
{El-Badry}, K., {Rix}, H.-W., \& {Heintz}, T.~M. 2021, \bibinfo{title}{{A million binaries from Gaia eDR3: sample selection and validation of Gaia parallax uncertainties},} \mnras, 506, 2269, \dodoi{10.1093/mnras/stab323}

\bibitem[{B. {Famaey} \& J. {Binney}(2005){Famaey} \& {Binney}}]{Famaey:2005}
{Famaey}, B., \& {Binney}, J. 2005, \bibinfo{title}{{Modified Newtonian dynamics in the Milky Way},} \mnras, 363, 603, \dodoi{10.1111/j.1365-2966.2005.09474.x}

\bibitem[{Z.~D. {Hartman} {et~al.}(2022){Hartman}, {L{\'e}pine}, \& {Medan}}]{Hartman:2022}
{Hartman}, Z.~D., {L{\'e}pine}, S., \& {Medan}, I. 2022, \bibinfo{title}{{Vetting the ``Lobster'' Diagram: Searching for Unseen Companions in Wide Binaries Using NASA Space Exoplanet Missions},} \apj, 934, 72, \dodoi{10.3847/1538-4357/ac72a0}

\bibitem[{X. {Hernandez}(2023){Hernandez}}]{Hernandez:2023}
{Hernandez}, X. 2023, \bibinfo{title}{{Internal kinematics of Gaia DR3 wide binaries: anomalous behaviour in the low acceleration regime},} \mnras, 525, 1401, \dodoi{10.1093/mnras/stad2306}

\bibitem[{X. {Hernandez} {et~al.}(2024{\natexlab{a}}){Hernandez}, {Chae}, \& {Aguayo-Ortiz}}]{HernandezChae:2024}
{Hernandez}, X., {Chae}, K.-H., \& {Aguayo-Ortiz}, A. 2024{\natexlab{a}}, \bibinfo{title}{{A critical review of recent Gaia wide binary gravity tests},} \mnras, 533, 729, \dodoi{10.1093/mnras/stae1823}

\bibitem[{X. {Hernandez} \& P. {Kroupa}(2025){Hernandez} \& {Kroupa}}]{HernandezKroupa:2025}
{Hernandez}, X., \& {Kroupa}, P. 2025, \bibinfo{title}{{A recent confirmation of the wide binary gravitational anomaly},} \mnras, 537, 2925, \dodoi{10.1093/mnras/staf210}

\bibitem[{X. {Hernandez} {et~al.}(2024{\natexlab{b}}){Hernandez}, {Verteletskyi}, {Nasser}, \& {Aguayo-Ortiz}}]{Hernandez:2024b}
{Hernandez}, X., {Verteletskyi}, V., {Nasser}, L., \& {Aguayo-Ortiz}, A. 2024{\natexlab{b}}, \bibinfo{title}{{Statistical analysis of the gravitational anomaly in Gaia wide binaries},} \mnras, 528, 4720, \dodoi{10.1093/mnras/stad3446}

\bibitem[{H.-C. {Hwang} {et~al.}(2022){Hwang}, {Ting}, \& {Zakamska}}]{Hwang:2022}
{Hwang}, H.-C., {Ting}, Y.-S., \& {Zakamska}, N.~L. 2022, \bibinfo{title}{{The eccentricity distribution of wide binaries and their individual measurements},} \mnras, 512, 3383, \dodoi{10.1093/mnras/stac675}

\bibitem[{V.~V. {Makarov}(2026){Makarov}}]{Makarov:2026}
{Makarov}, V.~V. 2026, \bibinfo{title}{{Distributions of Wide Binary Stars in Theory and in Gaia Data. III. Orbital Momenta, Masses, and Manifestations of MOND},} \aj, 171, 79, \dodoi{10.3847/1538-3881/ae2757}

\bibitem[{S.~S. {McGaugh}(2008){McGaugh}}]{McGaugh:2008}
{McGaugh}, S.~S. 2008, \bibinfo{title}{{Milky Way Mass Models and MOND},} \apj, 683, 137, \dodoi{10.1086/589148}

\bibitem[{S.~S. {McGaugh} {et~al.}(2016){McGaugh}, {Lelli}, \& {Schombert}}]{McGaugh:2016}
{McGaugh}, S.~S., {Lelli}, F., \& {Schombert}, J.~M. 2016, \bibinfo{title}{{Radial Acceleration Relation in Rotationally Supported Galaxies},} \prl, 117, 201101, \dodoi{10.1103/PhysRevLett.117.201101}

\bibitem[{M. {Milgrom}(1983){Milgrom}}]{Milgrom:1983}
{Milgrom}, M. 1983, \bibinfo{title}{{A modification of the Newtonian dynamics as a possible alternative to the hidden mass hypothesis.},} \apj, 270, 365, \dodoi{10.1086/161130}

\bibitem[{M. {Milgrom}(2010){Milgrom}}]{Milgrom:2010}
{Milgrom}, M. 2010, \bibinfo{title}{{Quasi-linear formulation of MOND},} \mnras, 403, 886, \dodoi{10.1111/j.1365-2966.2009.16184.x}

\bibitem[{M.~J. {Pecaut} \& E.~E. {Mamajek}(2013){Pecaut} \& {Mamajek}}]{PecautMamajek:2013}
{Pecaut}, M.~J., \& {Mamajek}, E.~E. 2013, \bibinfo{title}{{Intrinsic Colors, Temperatures, and Bolometric Corrections of Pre-main-sequence Stars},} \apjs, 208, 9, \dodoi{10.1088/0067-0049/208/1/9}

\bibitem[{J. {Pflamm-Altenburg}(2025){Pflamm-Altenburg}}]{Pflamm-Altenburg:2025}
{Pflamm-Altenburg}, J. 2025, \bibinfo{title}{{Numerical solutions of the complete two-body system in QUMOND},} \aap, 703, A68, \dodoi{10.1051/0004-6361/202555656}

\bibitem[{C. {Pittordis} \& W. {Sutherland}(2019){Pittordis} \& {Sutherland}}]{Pittordis:2019}
{Pittordis}, C., \& {Sutherland}, W. 2019, \bibinfo{title}{{Testing modified gravity with wide binaries in Gaia DR2},} \mnras, 488, 4740, \dodoi{10.1093/mnras/stz1898}

\bibitem[{C. {Pittordis} {et~al.}(2025){Pittordis}, {Sutherland}, \& {Shepherd}}]{Pittordis:2025}
{Pittordis}, C., {Sutherland}, W., \& {Shepherd}, P. 2025, \bibinfo{title}{{Wide Binaries from GAIA DR3 : testing GR vs MOND with realistic triple modelling},} The Open Journal of Astrophysics, 8, 109, \dodoi{10.33232/001c.142887}

\bibitem[{R. {Saglia} {et~al.}(2025){Saglia}, {Pasquini}, {Patat}, {Ludwig}, {Giribaldi}, {Leao}, {de Medeiros}, \& {Murphy}}]{Saglia:2025}
{Saglia}, R., {Pasquini}, L., {Patat}, F., {et~al.} 2025, \bibinfo{title}{{Testing gravity with wide binaries: 3D velocities and distances of wide binaries from Gaia and HARPS},} \aap, 699, A151, \dodoi{10.1051/0004-6361/202555115}

\bibitem[{E.~J. {Shaya} \& R.~P. {Olling}(2011){Shaya} \& {Olling}}]{Shaya:2011}
{Shaya}, E.~J., \& {Olling}, R.~P. 2011, \bibinfo{title}{{Very Wide Binaries and Other Comoving Stellar Companions: A Bayesian Analysis of the Hipparcos Catalogue},} \apjs, 192, 2, \dodoi{10.1088/0067-0049/192/1/2}

\bibitem[{A. {Vallenari} {et~al.}(2023){Vallenari}, {Brown}, {Prusti}, {de Bruijne}, {Arenou}, {Babusiaux}, {Biermann}, {Creevey}, {Ducourant}, {Evans}, {Eyer}, {Guerra}, {Hutton}, {Jordi}, {Klioner}, {Lammers}, {Lindegren}, {Luri}, {Mignard}, {Panem}, {Pourbaix}, {Randich}, {Sartoretti}, {Soubiran}, {Tanga}, {Walton}, {Bailer-Jones}, {Bastian}, {Drimmel}, {Jansen}, {Katz}, {Lattanzi}, {van Leeuwen}, {Bakker}, {Cacciari}, {Casta{\~n}eda}, {De Angeli}, {Fabricius}, {Fouesneau}, {Fr{\'e}mat}, {Galluccio}, {Guerrier}, {Heiter}, {Masana}, {Messineo}, {Mowlavi}, {Nicolas}, {Nienartowicz}, {Pailler}, {Panuzzo}, {Riclet}, {Roux}, {Seabroke}, {Sordo}, {Th{\'e}venin}, {Gracia-Abril}, {Portell}, {Teyssier}, {Altmann}, {Andrae}, {Audard}, {Bellas-Velidis}, {Benson}, {Berthier}, {Blomme}, {Burgess}, {Busonero}, {Busso}, {C{\'a}novas}, {Carry}, {Cellino}, {Cheek}, {Clementini}, {Damerdji}, {Davidson}, {de Teodoro}, {Nu{\~n}ez Campos}, {Delchambre}, {Dell'Oro}, {Esquej}, {Fern{\'a}ndez-Hern{\'a}ndez}, {Fraile}, {Garabato},
  {Garc{\'\i}a-Lario}, {Gosset}, {Haigron}, {Halbwachs}, {Hambly}, {Harrison}, {Hern{\'a}ndez}, {Hestroffer}, {Hodgkin}, {Holl}, {Jan{\ss}en}, {Jevardat de Fombelle}, {Jordan}, {Krone-Martins}, {Lanzafame}, {L{\"o}ffler}, {Marchal}, {Marrese}, {Moitinho}, {Muinonen}, {Osborne}, {Pancino}, {Pauwels}, {Recio-Blanco}, {Reyl{\'e}}, {Riello}, {Rimoldini}, {Roegiers}, {Rybizki}, {Sarro}, {Siopis}, {Smith}, {Sozzetti}, {Utrilla}, {van Leeuwen}, {Abbas}, {{\'A}brah{\'a}m}, {Abreu Aramburu}, {Aerts}, {Aguado}, {Ajaj}, {Aldea-Montero}, {Altavilla}, {{\'A}lvarez}, {Alves}, {Anders}, {Anderson}, {Anglada Varela}, {Antoja}, {Baines}, {Baker}, {Balaguer-N{\'u}{\~n}ez}, {Balbinot}, {Balog}, {Barache}, {Barbato}, {Barros}, {Barstow}, {Bartolom{\'e}}, {Bassilana}, {Bauchet}, {Becciani}, {Bellazzini}, {Berihuete}, {Bernet}, {Bertone}, {Bianchi}, {Binnenfeld}, {Blanco-Cuaresma}, {Blazere}, {Boch}, {Bombrun}, {Bossini}, {Bouquillon}, {Bragaglia}, {Bramante}, {Breedt}, {Bressan}, {Brouillet}, {Brugaletta}, {Bucciarelli},
  {Burlacu}, {Butkevich}, {Buzzi}, {Caffau}, {Cancelliere}, {Cantat-Gaudin}, {Carballo}, {Carlucci}, {Carnerero}, {Carrasco}, {Casamiquela}, {Castellani}, {Castro-Ginard}, {Chaoul}, {Charlot}, {Chemin}, {Chiaramida}, {Chiavassa}, {Chornay}, {Comoretto}, {Contursi}, {Cooper}, {Cornez}, {Cowell}, {Crifo}, {Cropper}, {Crosta}, {Crowley}, {Dafonte}, {Dapergolas}, {David}, {David}, {de Laverny}, {De Luise}, \& {De March}}]{Gaia:2023}
{Vallenari}, A., {Brown}, A.~G.~A., {Prusti}, T., {et~al.} 2023, \bibinfo{title}{{Gaia Data Release 3. Summary of the content and survey properties},} \aap, 674, A1, \dodoi{10.1051/0004-6361/202243940}

\bibitem[{Y. {Yoon} {et~al.}(2025){Yoon}, {Tian}, \& {Chae}}]{Yoonetal:2025}
{Yoon}, Y., {Tian}, Y., \& {Chae}, K.-H. 2025, \bibinfo{title}{{Probing the Nature of Gravity in the Low-acceleration Limit: Wide Binaries of Extreme Separations with Perspective Effects},} \apj, 992, 102, \dodoi{10.3847/1538-4357/ae0190}

\bibitem[{A.~H. {Zonoozi} {et~al.}(2021){Zonoozi}, {Lieberz}, {Banik}, {Haghi}, \& {Kroupa}}]{Zonoozi:2021}
{Zonoozi}, A.~H., {Lieberz}, P., {Banik}, I., {Haghi}, H., \& {Kroupa}, P. 2021, \bibinfo{title}{{The Kennicutt-Schmidt law and the main sequence of galaxies in Newtonian and milgromian dynamics},} \mnras, 506, 5468, \dodoi{10.1093/mnras/stab2068}

\end{thebibliography}
\bibliographystyle{aasjournalv7}

\end{document}